\documentclass[fleqn,usenatbib]{mnras}
\usepackage{fix-cm}
\usepackage{newtxtext,newtxmath}

\usepackage[T1]{fontenc}
\usepackage{subfiles}
\usepackage{xcolor}
\usepackage{xcolor}
\DeclareRobustCommand{\VAN}[3]{#2}
\let\VANthebibliography\thebibliography
\def\thebibliography{\DeclareRobustCommand{\VAN}[3]{##3}\VANthebibliography}

\usepackage{graphicx}	
\usepackage{amsmath}	

\title[LFT3 Mission]{The Lunar Farside Transients and Technology Telescope (LFT3) Mission}

\author[D. R. DeBoer et al.]{
David R. DeBoer,$^{1,2}$\thanks{E-mail: dave.deboer@physics.ox.ac.uk}
Charlie K. Ashe,$^{3}$
Owen A. Johnson,$^{3,2}$
Evan F. Keane,$^{3}$
Andrew C. Lesh,$^{4}$
\newauthor
Richard Lynch,$^{5}$
Ella J. Marshall,$^{6}$
Steve Prabu,$^{1}$
An\v{z}e Slosar,$^{7}$
Chenoa D. Tremblay,$^{2,8,9}$
Jake D. Turner,$^{10}$
\newauthor
Karl F. Warnick,$^{11}$
Andrew P. V. Siemion,$^{1,2,8}$
Jamie Drew,$^{1,12}$
S. Pete Worden$^{12}$
\\
$^{1}$Sub-department of Astrophysics, Department of Physics, Keble Road, University of Oxford, OX1-3RH, UK\\
$^{2}$Radio Astronomy Lab, University of California, 501 Campbell Hall, Berkeley 94720, USA\\
$^{3}$School of Physics, Trinity College Dublin, College Green, Dublin 2, D02 PN40, Ireland\\
$^{4}$Department of Civil and Environmental Engineering, Stanford University, CA, 94305 USA\\
$^{5}$Heliospace, 2448 Sixth St, Berkeley, CA, 94710, USA\\
$^{6}$School of Physics and Astronomy, The University of Edinburgh, Edinburgh, EH9 3JZ, UK\\
$^{7}$Brookhaven National Laboratory, Physics Department, Upton, NY 11973 USA\\
$^{8}$SETI Institute, 339 Bernardo Ave, Suite 200, Mountain View, CA 94043, USA\\
$^{9}$Department of Physics and Astronomy, University of New Mexico, Albuquerque, NM 87131, USA\\
$^{10}$Department of Astronomy and Carl Sagan Institute, Cornell University, Ithaca, New York 14853, USA\\
$^{11}$ECE Dept., Brigham Young University, Provo, UT, USA\\
$^{12}$Breakthrough Initiatives, 9 Rue du Laboratoire, 1911 Gare Luxembourg, Luxembourg\\
}

\date{Accepted XXX. Received YYY; in original form ZZZ}

\pubyear{\the\year{}}

\begin{document}
\label{firstpage}
\pagerange{\pageref{firstpage}--\pageref{lastpage}}
\maketitle

\begin{abstract}
We present here an overview of the Lunar Farside Transients and Technology Telescope (LFT3) mission to take advantage of the extremely clear radio frequency environment on the lunar farside. Radio observations performed from the lunar farside effectively and fully mitigate two unavoidable limitations of terrestrial-based radio telescopes: (i) the prevalence of interfering radio transmitters from human activity; and (ii) the impact of the Earth's ionosphere. However, in the era of cost-effective access to the Moon, there are many scheduled lunar missions over the next few years, and the window of opportunity to perform radio interference-free observations from the lunar farside is closing fast. LFT3 is the only mission proposed to go to the lunar farside and exploit this unique opportunity in human history. LFT3 will observe in an uncluttered radio environment to conduct unambiguous technosignature searches, transient surveys, solar physics and planetary emissions studies, spectral line observations, and cosmological science observations. LFT3 will provide an important incumbent use of the microwave spectrum for cislunar radio astronomy.

\end{abstract}

\begin{keywords}
Space vehicles: instruments -- Moon -- Telescopes -- Extraterrestrial intelligence
\end{keywords}



\section{Introduction} 
\label{sec:intro}
Despite numerous efforts to decipher its origin (for example, \citealt{Wow_2024, 2022RNAAS...6..197P}), the "Wow!" signal  \citep{wow}  remains 
among the most tantalizing unexplained radio signals to this day. This signal, seen in 1977 and rendered as ``6EQUJ5'' on the observatory printout, was unmistakably detected and aligned with the expected sidereal progression through the telescope beam, implying an origin in deep space. At the time, the world had few transmitters in air or space, allowing the signal to stand out. Today, such a signal would likely be lost in the cacophony of human-originating emissions that permeate the sky. Since it was only seen once but was consistent with a signal originating in the sky, the signal remains enigmatic; possible explanations include a novel natural phenomenon, an anomalous anthropogenic emission, or an indicator of extraterrestrial intelligence~\citep{Saide_2023,Haqq_2025,Sheikh_2025}.

To have a chance of discerning signals that would be crowded out by current human activity, one must conduct observations from a pristine radio environment shielded from the Earth. In such a place, any observed signal would be of scientific interest, and a modern radio astronomy mission there would allow for the use of technology much more advanced than that available in 1977.  That is the promise of a lunar farside telescope: recovery of the sky in which a signal like "Wow!" stands out but in an era when better means exist to measure and characterize it\footnote{The "Wow!" signal is estimated to have had a flux density of approximately 50–200 Jy (\url{https://phl.upr.edu/wow/data}), and would be detectable by LFT3 with averaging over only a few tens of kHz of bandwidth.}.

\begin{figure*}
    \centering
    \includegraphics[scale=0.55,trim={0mm 3mm 0mm 0mm}, clip]{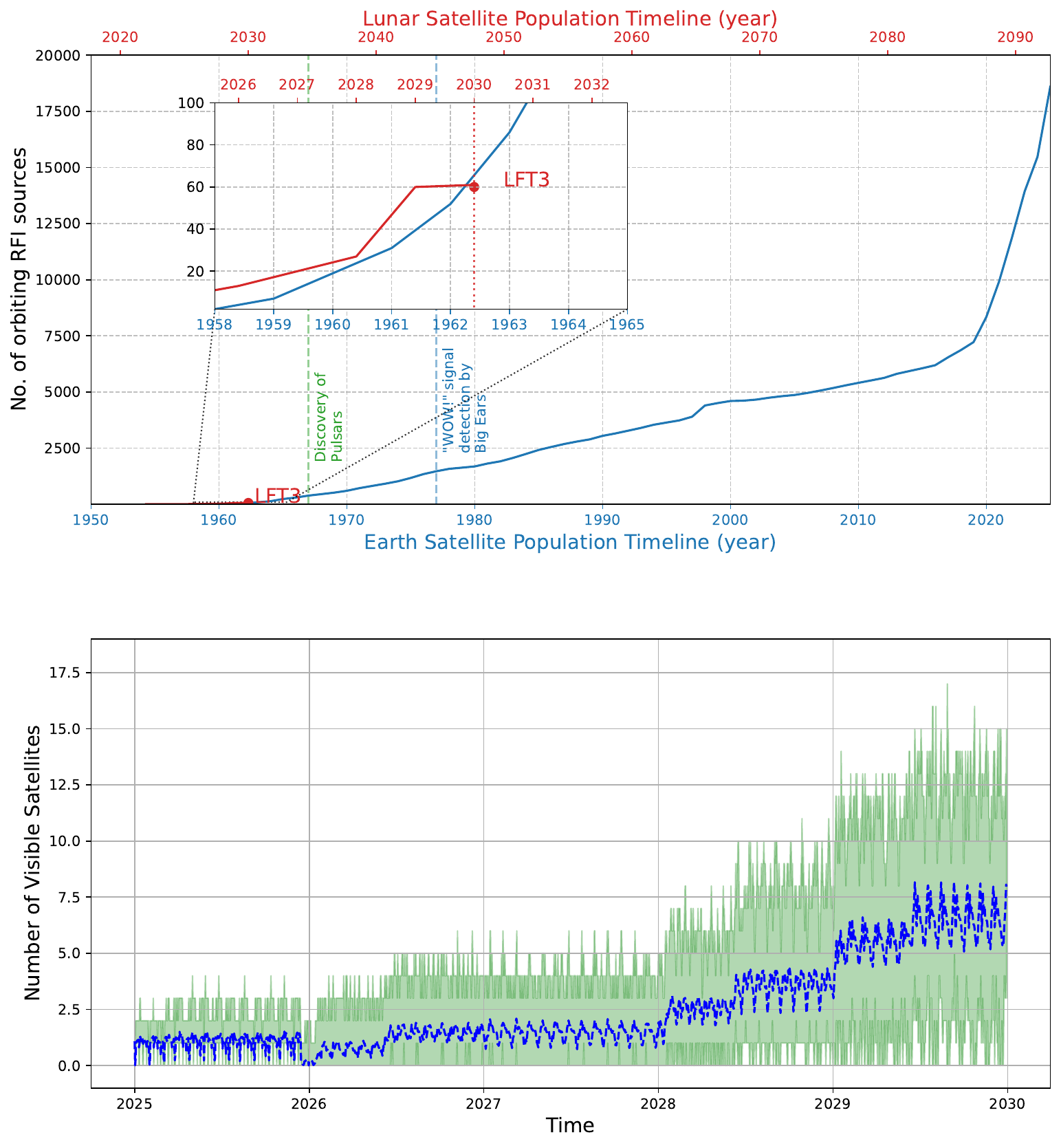}
    \caption{
        {\bf Top Panel} evolution of orbiting satellites on Earth and the Moon. In blue, we show the growth in the number of Earth-orbiting satellites, and in red we show the predicted number of lunar-orbiting satellites, with their timeline indicated on the top x-axis. From the image inset, we see that the projected RFI environment on the lunar farside in 2030 is equivalent to that of the early 1960s on Earth, more than a decade before the detection of the “Wow!” signal (shown by the dashed vertical blue line), and a few years before the discovery of pulsars (shown by vertical green dashed line). Hence, in essence, the LFT3 mission allows us to “time travel” back to a 1960s-like RFI environment, enabling technosignature searches with an instrument far more advanced than that used for the “Wow!” detection.\\
    {\bf Bottom Panel} Plot of the number of satellites above the horizon as a function of time for currently planned missions.  The green area is the minimum and maximum above the horizon at any given time over a day, while the blue line is the average number above the horizon in that day \citep{Ashe2026}.
    } 
    \label{fig:VisibilityCount}
\end{figure*}

Figure \ref{fig:VisibilityCount} (top panel) illustrates this using the number of orbiting satellites as a proxy.  The blue curve/bottom axis is a count of non-classified earth-orbiting satellites.  The red curve/top axis is a projection of lunar-orbiting satellites, matched to the historical earth-orbiting record, indicating that we still have an opportunity to make these new measurements in the RFI environment of an earlier era.

Due to the rise of technology, humanity is at a tipping point in transitioning to space-based astronomy. One push in this direction is the ever-increasing use of wireless communication devices on the ground and in orbit as in Figure \ref{fig:VisibilityCount}, which means that nowhere on our planet, not even remote locations with a very low population density, is free of significant radio frequency interference (RFI). At the same time, the increasing affordability and accessibility of space technology now makes it feasible to deploy telescopes more effectively in space.
However, in this transition to space, we risk injecting the same RFI now ubiquitous on Earth into this new domain. This is particularly true for the lunar farside, with its unique feature of always pointing away from the Earth, providing essentially perfect shielding from the Earth and its environs \citep{1975A&A....40..365A,2023ExA....56..333Y,MACCONE2019233}.
This makes it urgent to get well-designed sensors there as soon as possible to make early baseline measurements and conduct the unique science allowed by its essentially RFI-free environment at this time.  The bottom panel of Figure \ref{fig:VisibilityCount} shows the projected growth in the visibility of lunar-orbiting satellites to LFT3, and is further discussed in \S\ref{sec:context}.

\begin{figure*}
    \centering
    \includegraphics[width=0.65\linewidth]{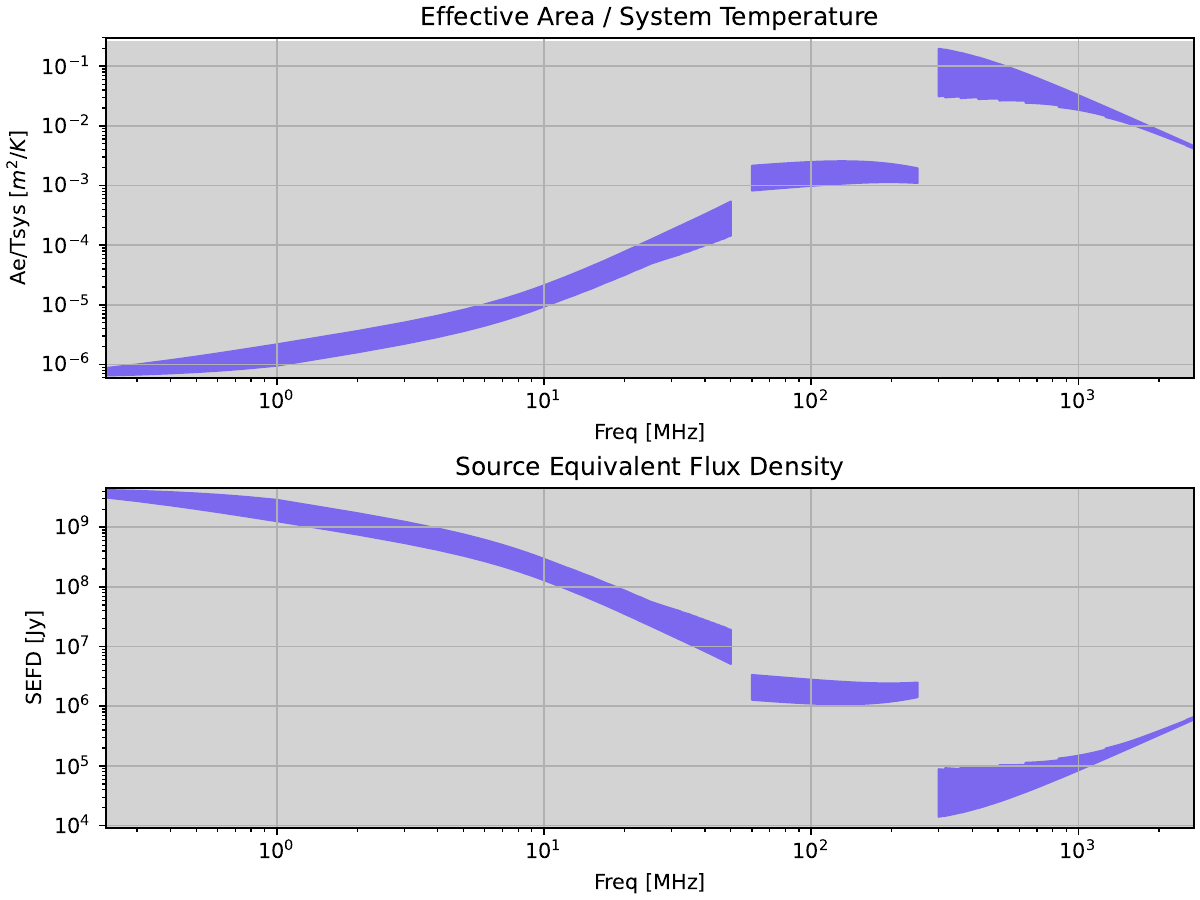}
    \caption{HF/VHF/UHF frequency bands of LFT3 and their associated sensitivity in terms of effective area over system temperature (top) and source equivalent flux density (bottom). The range stems from different fields over the varying sky temperatures due to Galactic emission and beam-size.}
    \label{fig:freqbands}
\end{figure*}

In this paper, we describe the Lunar Farside Transient and Technology Telescope (LFT3) --- a proposed mission to land near the lunar antipode within five years to deploy a radio telescope to conduct these unique-in-history measurements, within a Commercial Lunar Payload Service \citep[CLPS;][]{NASA_CLPS_Brochure_2024} budget profile of $\sim$\$150M. Figure \ref{fig:freqbands} shows the planned frequency bands and their expected sensitivity.
The scope is to cover the HF, VHF, and UHF bands (0.1 - 2700 MHz), which, from Earth, are impacted by the ionosphere and radio frequency interference \citep[e.g.][]{Zawdie_2017RS006256,Offringa_RFI,Hobbs_2020,Grigg_2025}.
The primary component (see \S~\ref{sec:payload}) is a phased UHF dual-polarization multibeam array operating at $300-2700$~MHz to conduct a complete spectral sky survey. 
The band is also sensitive to S-band communication signals from the Moon and potentially Mars, which will be helpful in demonstrating the technosignature capability. Individual antennas will cover HF ($0.1-50$~MHz) and VHF ($60-260$~MHz). Note that at HF and VHF frequencies for the scope of this mission, multiple antennas per HF or VHF bands provide little to no sensitivity benefit. A more comprehensive discussion may be found in \cite{whitepaper}.

In addition to conducting this unique survey, LFT3 serves as an important incumbent user of the broader microwave spectrum in the cislunar environment.  Incumbent use is important for spectrum management because it helps regulators understand how existing users rely on specific frequency bands before making changes or allowing new services. Incumbent use generally serves as a marker for regulators and other potential spectrum users to pay attention to bands and services. Protecting incumbent operations reduces harmful interference and supports the continuity of essential services.

In \S\ref{sec:context} we describe the context of the RFI environment on the lunar farside over the next decade, and describe the scientific opportunities of operating such a mission as soon as possible. In \S\ref{sec:science} we elaborate on the many potential scientific applications one could make with a telescope like LFT3 operating on the lunar farside. In \S\ref{sec:payload} we describe the payload and operational model for the mission, before summarising and concluding in \S\ref{sec:conclusions}.

\section{Lunar Farside for Radio Science}
\label{sec:context}
The lunar farside represents a unique opportunity in human history for quiet, high-quality observations \citep{heidmann2002,bassett2020,michaud2020lunar}
This importance was recognized in the 1970s when the International Telecommunication Union (ITU)\footnote{The ITU is the international organization that handles electromagnetic emission across national boundaries.} defined the Shielded Zone of the Moon (SZM) as “compris[ing] the area of the Moon’s surface and an adjacent volume of space which are shielded from emissions originating within a distance of 100,000 km from the center of the Earth” \citep{itu_rr_2024}.  This essential radio quietness from Earth will be exploited by LFT3. And, as mentioned previously, demonstrating incumbent use of the spectrum is an important concept in spectrum management.

This window will soon begin to close as increasing scientific and political interest in lunar exploration brings new assets to the Moon.
To characterize this growth,
Figure \ref{fig:VisibilityCount} (bottom panel) shows the number of satellites above the horizon for a projection of orbits and lifetimes for the currently planned missions. 
Beginning in mid-2028, there are significant periods of time when there will always be at least one satellite above the horizon, and typically there will be many.  Each satellite is a potential source of RFI, as shown by \cite{2024A&A...689L..10B, Grigg_2025}.

\begin{figure}
	\centering
	\includegraphics[trim = 0mm 0mm 0mm 0mm, clip, width=\linewidth]{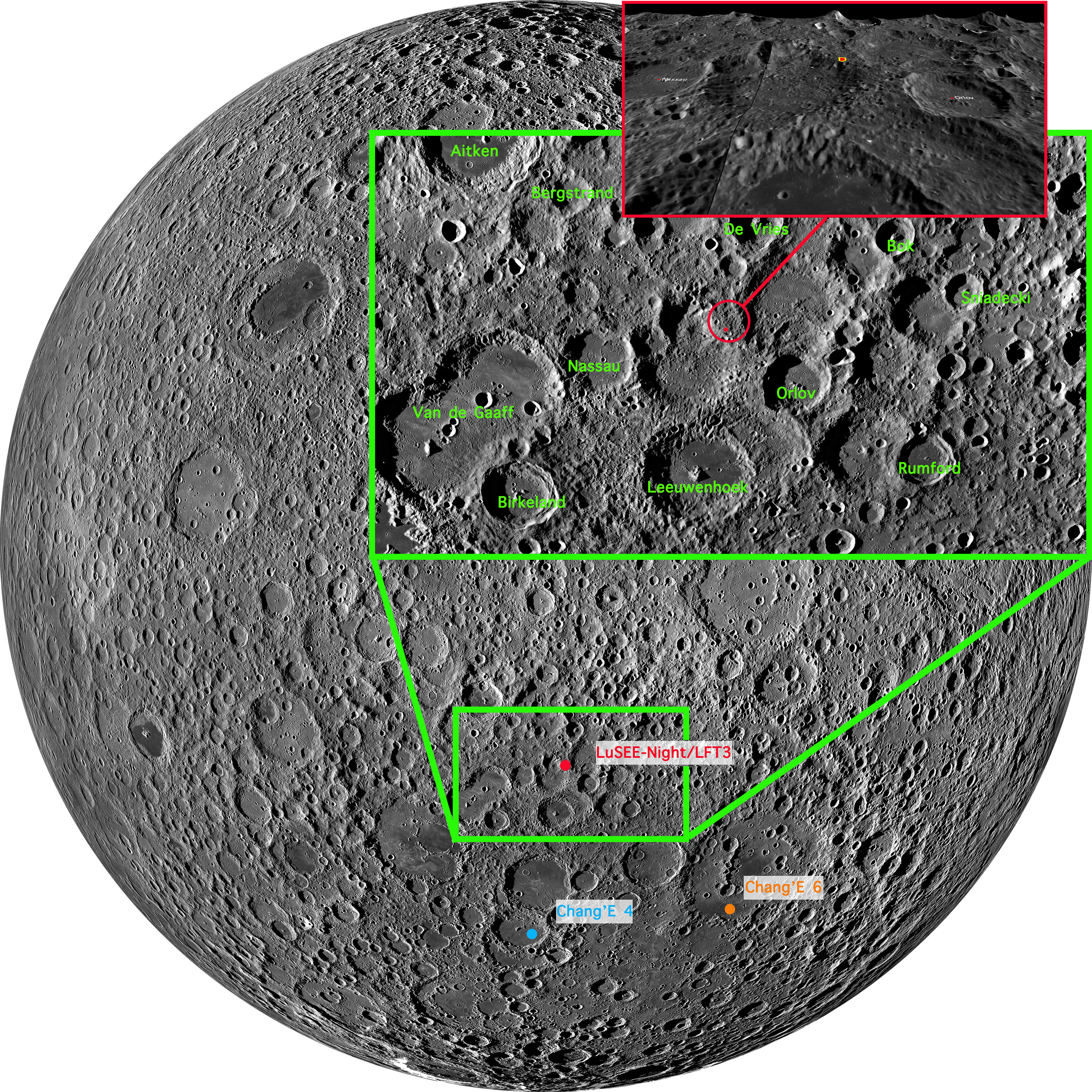}
	\caption{Lunar Reconnaissance Orbiter Camera \citep[LROC;][]{LRO_Camera2010} image of the landing sites for LFT3/LuSEE-Night, Chang'E 4 \citep{Chen2022Change4Achievements} and Chang'E 6 \citep{Yue2024Change6GeologicalContext}, zoom in green box.  The red box has a high-resolution JMARS \citep{christensen2009jmars,LRO_Camera2010} view of the proposed LFT3 landing side.}
    \label{fig:landing_site}
\end{figure}

\begin{figure*}
    \centering
    \includegraphics[trim = 0mm 0mm 0mm 0mm, clip, width=1\linewidth]{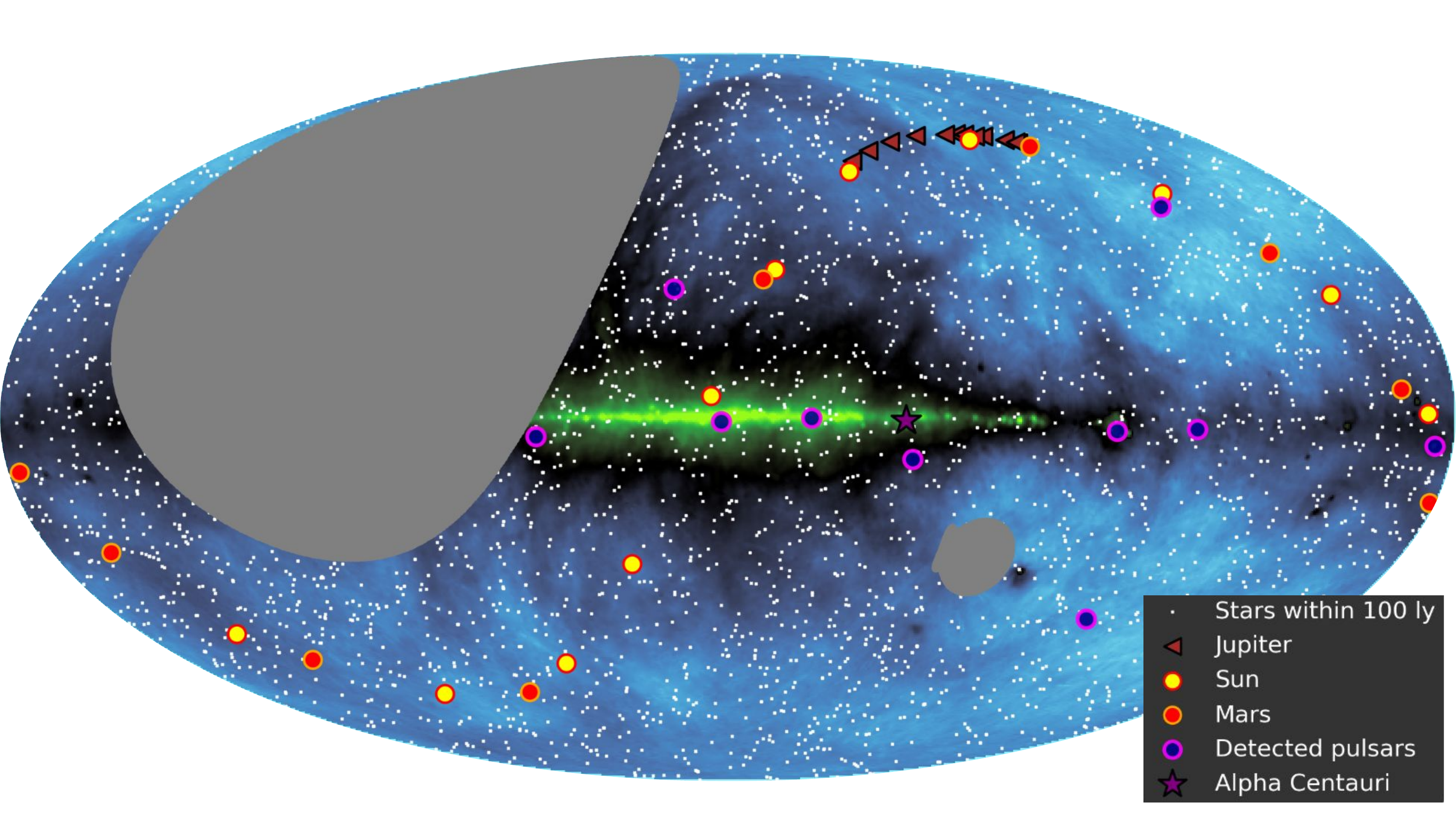}
    \caption{Graphic showing the sky visible over LFT3 over the calendar year 2028 super-imposed over a graphic of Milky Way emission at 408 MHz. The white points represent stars within 100 light years, and Alpha Centauri is shown using a purple star marker. The monthly positions of the sun, Mars and Jupiter on the ecliptic are also shown. We also show the distribution of pulsars detectable by LFT3 through orbital folding of 10 hours of data at 400 MHz (see \citealt{prabu2025lft3} for more info). }
    \label{fig:fieldofview}
\end{figure*}

The landing site coordinates within the SZM at 23.789°S, 182.137°E (Fig. \ref{fig:landing_site}) are close to the LuSEE-Night landing location, with which it will coordinate. In addition, landing near LuSEE-Night provides a known location with recent landing experience, increasing the chance of a successful landing.  Additionally, it is advantageous to study the same region of regolith from two nearby vantage points.
The sky to be observed from this location is shown in Figure~\ref{fig:fieldofview}. The features of this landing site are elaborated upon in the following, but we note here that it provides a protected location for a wide field-of-view view of the sky in the declination range from $\sim +60\deg$ to $\sim -80\deg$. Thus, during its mission lifetime, the telescope will observe most of the sky and conduct historical surveys from the solar system's most RFI-pristine environment before the presence of humanity's technology increases the radio frequency interference even at this very remote place.

In addition to the RFI considerations, which can be extensive even in remote Earth locations~\citep{Offringa_RFI,Hobbs_2020,Grigg_2025}, 
there are other very significant scientific reasons for operating a radio telescope on the farside of the Moon. The ionosphere surrounding our planet creates a conducting medium through which electromagnetic radiation is inaccessible at frequencies below about 10\,MHz and distorted below about 30\,MHz; the impacted frequencies vary in time and place and are also a function of solar activity, see e.g. \citealt{Zawdie_2017RS006256}. For this reason, low frequencies remain the last unexplored region of the electromagnetic spectrum. Due to the absence of a tangible ionosphere on the Moon, there are no technical barriers to studying physics at these new and unexplored frequencies. The primary motivations for targeting the lunar farside of getting above the ionophere and away from RFI are central to the design of the system. Going to the lunar farside allows humanity to reclaim the bands lost to radio astronomy due to persistent and pernicious RFI. 

Although the lunar farside provides an unparalleled environment for these measurements, it is important to think of this deployment in the context of other options.  The surface of the Earth will provide the vast majority of the sensitive radio telescope locations for decades to come; however, we have discussed the limitations due to RFI.
VLBI offers one very important mitigation strategy by correlating signals from widely separated antennas, suppressing interference local to any one telescope, while achieving microarcsecond-scale resolution. However, VLBI cannot completely remove RFI, especially when satellites appear in the fields of view of multiple antennas, and it cannot overcome atmospheric or ionospheric frequency limits.  Additionally, practical and computational affects limit the operation of such interferometers for widefield surveys \citep[e.g.][]{Garrett2018SETISO,radcliffe2025widefieldvlbi,herbegeorge2025sweeps}.

While a telescope in earth-orbit provides an observing platform above the ionosphere, it does not much ameliorate RFI.  Lunar orbit greatly helps reduce RFI but it does not completely eliminate it for the majority of its orbit.  Additionally, station-keeping activities in an orbiting satellite require constant monitoring and correction, meaning navigation systems cannot be depowered and reduce that source of locally-generated RFI.  The lunar farside provides the quietest stable location and, with an orbital period $\sim 30 \times$ slower than sidereal, provides additional dwell time for sensitivity.

\section{Science Case}
\label{sec:science}
\begin{figure*}
	\centering
	\includegraphics[width=0.9\linewidth]{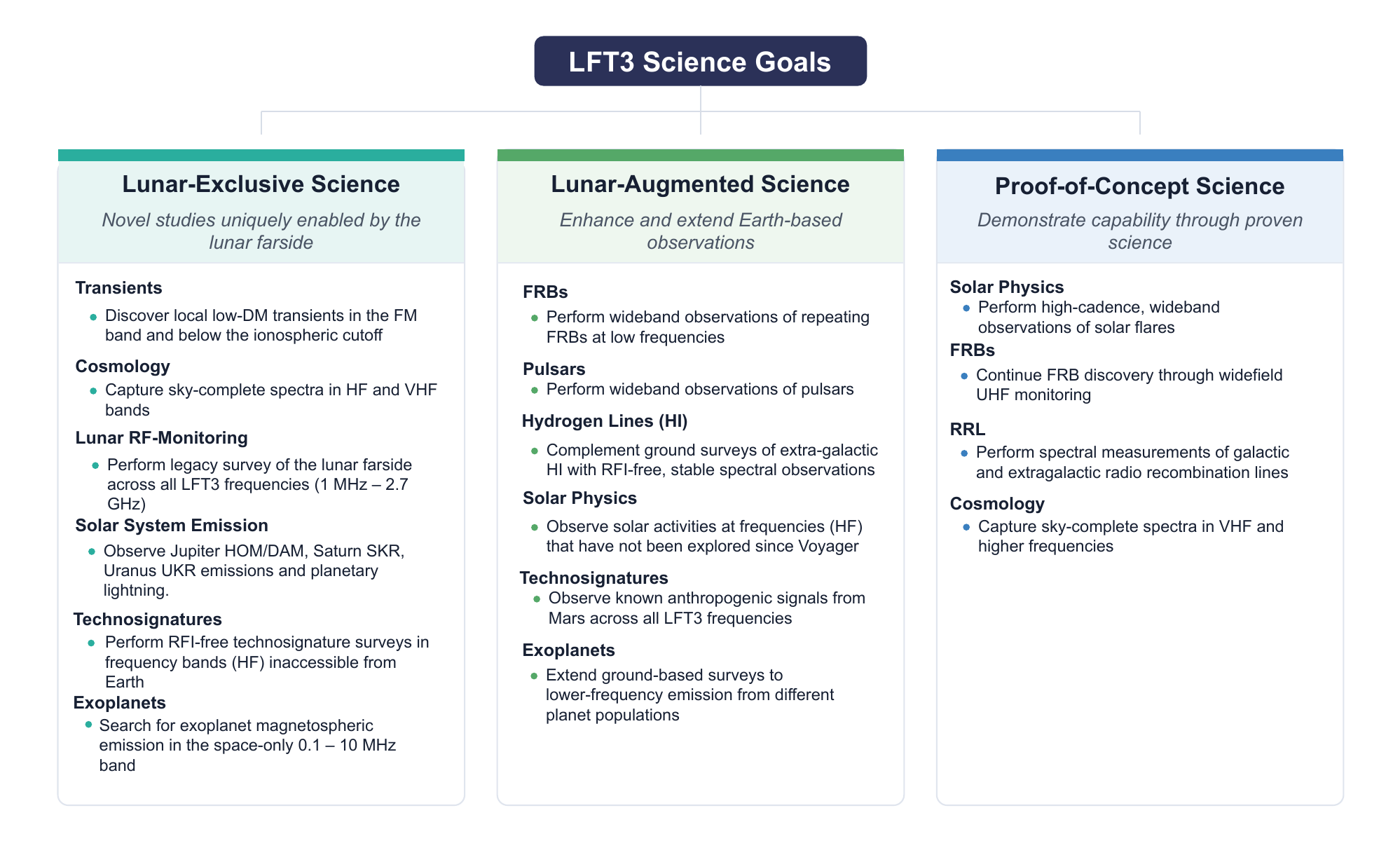}
	\caption{A high level breakdown of the LFT3 science goals discussed in this section divided into lunar-exclusive science goals, lunar-augmented science goals, and proof-of-concept science goals.}
	\label{fig:sciencegoals}
\end{figure*}

The high-level scientific goals of LFT3 taking advantage of the lunar farside and demonstrating operations from there are summarized in Figure~\ref{fig:sciencegoals}. As mentioned, {\em anything} detected there is of scientific interest and thus provides a unique opportunity. In the following, we describe the scientific applications of the LFT3 mission. These fall into 3 categories:
(1) science where LFT3 can do unique science that cannot be done, or done as well, from Earth;
(2) science where LFT3 provides useful information that could be used in addition to observations that can be done from Earth;
and (3) science that provides confidence in the performance of the telescope and lunar operations. 
The science case is not predicated on what might be done if one had a large budget to design and build a radio telescope on Earth, but rather on what one could do for an inexpensive lunar mission that advances science and our understanding of operating on the moon and can take advantage of the current unique window in time. 


\subsection{Transients}
\begin{figure*}
    \centering
    \includegraphics[width=0.9\textwidth]{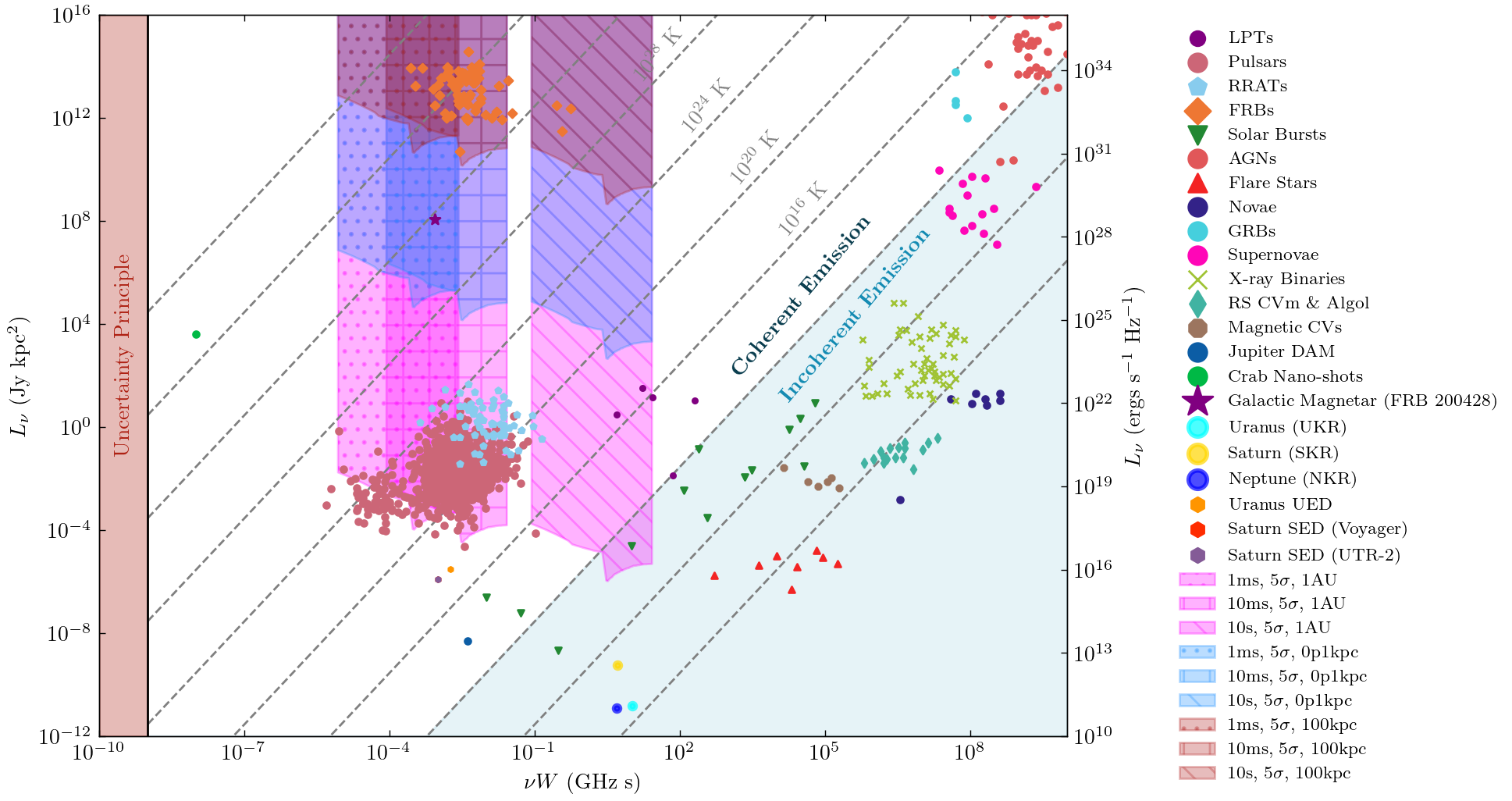} 
    \caption{Transient Parameter Space figure adapted from \citet{pietka}. This plot illustrates the sensitivity regions of a LFT3 observational setup across different pulse widths (1~ms, 10~ms, 10~s and 100~s) up to 10~kpc on a $\nu \text{W}$ vs. $L_\nu$ phase diagram. Where $\nu \text{W}$ represents the product of observed frequency and pulse width and $L_\nu$ shows the spectral luminosity. The blue, green and red hashed regions correspond to a pulse width and show the detectable signal levels above a threshold of $5\sigma$ for the LFT3 bandwidth.}
    \label{fig:transient_space}
\end{figure*}

The study of transient and variable sources in radio astronomy is one that often reveals extreme physical environments. A lunar-based telescope operating with a `new sky', in terms of spectral coverage and RFI occupancy, has the potential to discover the currently unknown. This includes the possible detection of various `low-DM' 
transient events on time scales of milliseconds to minutes. Such events may include low-DM FRB analogues, FRBs from Galactic magnetars, decimetric solar bursts, stellar flares, and any other as-yet undiscovered time-domain phenomena. In particular, LFT3 would uniquely enable the discovery of new populations of transient astrophysical sources in the very local Universe in the frequency range of $0.1-30$~MHz (below Earth's ionospheric cut-off) and in the range $87-108$~MHz (the FM band). Figure \ref{fig:transient_space} details various observational setups of LFT3 against well-known transients and variables in the transient and variable phase space of \citet{2004NewAR..48.1459C} and \citet{pietka}; highlighting the areas where LFT3 will probe. In the following, we detail the range of transient science goals, starting at the most tractable and extending to ever more challenging science targets.

\subsubsection{Solar Physics}
The primary known transient source that LFT3 will detect is the sun. 
Key solar science goals include the study of non-thermal emissions, gyrosynchrotron emission associated with coronal mass ejections (CMEs), targeted analyzes of solar bursts (Types I--V), and polarimetric measurements \citep{Dulk1985,2013POEMAS}. Many of these studies require high-bandwidth imaging on short time scales~\citep{Kansabanik_2022}, although the spectral-temporal characteristics of quiescent versus active solar states are also of interest given their impact on Earth-based technologies. 
In an effort to explore the connection between CMEs and phenomena such as Type II bursts, solar flares, and shocks, \citet{Bastian_2001} analyzed observations at 150 and 450 \, MHz with 32 \, s integration and $\sim$1 \, MHz spectral resolution. Their work yielded constraints on thermal plasma density, density filling factors, relativistic electron number density, and magnetic field strength within CMEs -- values still widely referenced today. Solar observations at varying time and frequency resolutions remain challenging. A farside lunar telescope would offer a dynamic spectral view of solar activity across a frequency range inaccessible from Earth, revisiting and extending measurements first pioneered by the Voyager missions and more recently targeted by SunRISE\footnote{\url{https://science.nasa.gov/mission/sunrise/}}.
As solar maximum approaches, such a telescope would be uniquely suited for capturing transient low-frequency events, like Type II/III bursts and solar-driven auroral emissions from planetary bodies. 

In addition, interplanetary scintillation (IPS) observations from the lunar farside were obtained. 
could be used to pinpoint solar wind and heliophysical structures by measuring rapid intensity fluctuations of the compact radio background caused by turbulence in the solar wind \citep{fallows_application_2023}. This technique could potentially reveal bulk flow velocities and turbulence levels and detect large-scale heliospheric density structures (e.g., tracking CME-driven shocks) throughout the inner heliosphere. Compared to Earth-based IPS experiments~\citep{fallows_separating_2016,kaplan_murchison_2015} that are limited by terrestrial ionospheric distortion~\citep{DEX}.


\textbf{System Implications:} The study the above solar phenomena requires high time-resolution dynamic spectra, on the order of 1--2\,ms, and spectral resolution of $\leq$0.5\,MHz. These are critical for resolving fast-evolving solar radio bursts, particularly Type III. Coherent bursts from solar events are usually $\sim$10$^3$--10$^6$~Jy \citep{Melnik2014} in VHF frequencies, often peaking at $10^{10}$ Jy at HF frequencies \citep{2011SoPh..269..335M}. However, these bursts occur over narrow frequency bands and short timescales, requiring fine temporal and spectral sampling for full characterisation.
Dynamic spectra in Stokes $I$ and $V$ would allow for polarimetric studies of coronal magnetic fields and emission mechanisms. Low-frequency IPS studies require long integration baselines (e.g. $10-100$~s) across a wide field of view to track scintillation patterns and infer bulk solar wind properties and density turbulence. The frequency range of interest spans $\sim0.1-30$~MHz, encompassing the decametric and hectometric bands relevant for Type II/III solar bursts and the IPS regime. This range lies entirely below Earth's ionospheric cutoff, making space deployment essential. A phased array configuration with beam steering would facilitate flexible observation of both targeted and wide-field events. Coordinated observations with instruments like LWA, MWA and LOFAR, and space-based platforms like SunRISE, would enable multi-instrument solar diagnostics across a broader frequency range and heliocentric baselines. These data products would be valuable across both active and quiescent solar phases, contributing to long-term solar monitoring and forecasting of geomagnetic activity. 


\subsubsection{Solar System Emission} 
After the sun, the most prominent radio source is Jupiter. Of the numerous sources of radio emission throughout the solar system, Jupiter's is the most notable. Its strong magnetic field interacts with its moons, particularly Io~\citep{Goldreich1969}, but also Europa and Ganymede~\citep{zarka_auroral_1998, Louis2023}, producing intense MJy-level decameter (DAM; 10-40~MHz) auroral radio emissions. Jupiter emits over a broad frequency range from 10~kHz to 40~MHz~ \citep{zarka_auroral_1998}, with its hectometer-wavelength (HOM; 1--5~MHz) emission modulated by solar wind activity~\citep{Desch1984}. The electron cyclotron maser instability~\citep[ECMI;][]{EMI}
is the main driver of Jupiter's emission \citep{zarka_auroral_1998}. Jupiter is the only planet in the solar system whose ECMI radio emission can be seen from the ground due to the ionospheric cutoff; auroral radio emissions from Earth, Saturn, Uranus and Neptune are all below 1 MHz~\citep{zarka_auroral_1998}. A lunar-based telescope would enable observations of Jupiter's HOM emissions, particularly around solar maximum, offering an opportunity to study solar-wind-driven dynamics on Jupiter. Simultaneous unphased observations (10-40 MHz) with LFT3 and the ground (e.g., LOFAR and NenuFAR) will allow us to measure the size of the emission beam \citep{Imai2019}. In addition, Very Long Baseline Interferometry (VLBI) observations of DAM emission would allow us to measure the source size of the emitting region \citep{Wucknitz2024}. LFT3's study of Jupiter will be a benchmark for studying exoplanet radio emission from close-in exoplanets \citep{Zarka2007,joe_nature_review}. 

LFT3 is also well positioned to study the auroral emisisons from Saturn (Saturn kilometric radiation, SKR) and Uranus (Uranus kilometric radiation, UKR), which can only be studied from space \citep{zarka_auroral_1998}. Saturn has been studied by Voyager and Cassini \citep{zarka_auroral_1998,Lamy2017} and no observations have been taken since 2017. LFT3 will enable the study of seasonal variations of the SKR. Saturn's average radio flux density at the Moon would be $\sim$6$\times$10$^3$--6$\times$10$^4$ Jy \citep{Zarka2012} and can vary by order of magnitude higher \citep{Lamy2023}. Therefore, the average SKR can be detected with $\sim$1 hour intergrations in 200 kHz channel widths. Uranus' auroral radio emission has only been studied by Voyager \citep{zarka_auroral_1998}. LFT3 will be able to study UKR again for the first time, search for variability due to solar wind fluctuations or internal processes, and study a completely different hemisphere than the one seen by Voyager \citep{zarka_auroral_1998}. Thus, LFT3 observations will be a useful benchmark to prepare for a future Uranus mission. Uranus' average radio flux density at the Moon would be $\sim$100 Jy \citep{Zarka2012}. However, strong coronal mass ejections could enhance the UKR by many orders of magnitude \citep{Lamy2012}. Therefore, the average UKR can only be studied by averaging over long timescales (days to months) and large frequency bandwiths ($\sim$500 kHz). The study of SKR and UKR will also be a benchmark for the study of radio emission from smaller exoplanets \citep{Zarka2007}.  

A lunar telescope is also well suited to search for any bright low-frequency emission generated by lightning from bodies in the solar system. Radio detections of lightning on Uranus (UED, 0.9--40~MHz; \citealt{1986Zarka_Emission}) and Saturn (SED, 2--30~MHz; \citealt{Zarka2004}) have been detected, while claims of Neptunian lightning remain unconfirmed \citep{Kaiser1991}. Lightning radio emissions from Venus ($2-30$~MHz) and Mars ($20-30$~MHz) are also predicted to occur, but have not been observed \citep{Zarka2008}. Efforts to re-detect Earth's emission from Saturn and Uranus have not been successful, and the cause remains unknown \citep{Zarka2008}. If this is due to ionospheric constraints or excess ground-based RFI (lightning and RFI have similar characteristics), then it means that LFT3 will be in a unique position to probe the crucial part of the spectrum where these emissions have been seen/predicted. Individual lightning events (30-300 ms duration) for the planets will not be detectable (maximum of 1000~Jy for Saturn lightning; \citealt{Zarka2004}). We will need to average over large bandwidths ($\sim$10s MHz) and also use burst detection techniques that can average thousands of individual bursts to find fainter signals \citep{Turner2019,Turner2021_Radio}. LFT3 is unique as it would allow for a relatively long-term observing campaign in a very low RFI environment. A confirmed re-detection or new discovery would carry significant implications for understanding planetary atmospheres and the general nature of the weather on planets in the outer Solar System.

\textbf{System Implications:} Dynamic spectra in Stokes $I$ and $V$ with high time resolution (e.g. milliseconds to hours) would enable the characterization of Solar System emission morphology and polarization. The emissions associated with lightning and auroral emissions are expected to have burst durations of milliseconds to hours~\citep{1986Zarka_Emission,Zarka1998_review}, requiring longer integration times and high spectral resolution (e.g., $\leq$100~kHz). A bandwidth covering at least $0.1-30$~MHz is necessary to capture the full spectral envelope of lightning and auroral radio emissions. Additionally, dual-polarization capability is critical for distinguishing between thermal and non-thermal processes and identifying strongly polarized bursts.

\subsubsection{Exoplanet auroral radio emission} 

Exoplanets with and without a magnetic field are expected to form, behave, and evolve very differently \citep{Brain2024}. The knowledge of a planetary magnetic field can provide robust constraints on the planet's interior structure, atmospheric escape, atmospheric dynamics and evolution, and potential habitability \citep{Lazio2019,Brain2024,joe_nature_review}. Despite decades of effort, the direct detection of magnetic fields on exoplanets has remained elusive \citep{G2015,Brain2024}. Auroral radio observations, analogous to auroral radio emissions by solar system planets, are among the best methods to detect exoplanetary magnetic fields \citep{G2015,Zarka2015SKA}, as many of the other methods are prone to false positives \citep{G2015,Turner2016a,Route2019}. This radio emission is predicted to be entirely driven by the stellar wind \citep{Zarka2007}. Therefore, the radio flux densities of exoplanets in close-in orbits are predicted to be orders of magnitude higher than Jupiter's radio flux density \citep{Griessmeier2007_AA,Griessmeier17PREVIII} and thus potentially detectable by with modern low-frequency ground-based radio telescopes \citep{Zarka2015SKA,Griessmeier17PREVIII,Turner2019}. 
To date, tentative radio detections on two hot Jupiters have been reported using LOFAR \citep{Turner2021_Radio} and NenuFAR \citep{Zhang2025}. A large-scale exoplanet survey on hot Jupiters with NenuFAR is currently ongoing. In order to study smaller planets with weaker magnetnic fields \citep{Griessmeier17PREVIII}, we need a space-based telecope below the ionospheric cutoff. 



The search for exoplanets with LFT3 would be unique and would greatly benefit the field. Our survey with LFT3 will be the second exoplanet radio search ever between $0.1-10$~MHz following the upcoming LuSEE-Night mission.  
The exoplanets, similar to the Solar System planets, that can be studied in this frequency range are smaller than Jupiter and thus cannot be studied from the ground \citep{Zarka2007,Griessmeier17PREVIII}. Although individual bursts from an exoplanet cannot be seen with LFT3, several key aspects of the emission will allow for stacking of observations. Exoplanetary radio emission occurs over long time scales (minutes to hours) and large bandwidths (tens of MHz), and the emission is periodic (for tidally locked planets the emission beam will be pointed towards the observer during the same part of the orbit) and circularly polarized \citep{Zarka2007}. Therefore, we would be able to stack the entire dataset together to search for these exoplanet signals. This could be done by using a Lomb–Scargle periodogram, as recently verified on Jupiter observations from \textit{NenuFAR} \citep{Louis2025}, or similar techniques. For close-in hot exoplanets, we can observe $\sim50-100$ orbits throughout the duration of the mission. First studying the radio-loud exoplanets \citep{Turner2021_Radio,Zhang2025} found by NenuFAR and LOFAR would allow us to verify this technique and enable the exploration of the outer parts of the magnetosphere that are not accessible from the ground. This study would place important constraints on the dynamo modeling, since this parameter space has never been explored before. Regardless of any detections, the science and technology lessons learned from LFT3 will help us better plan for larger radio telescopes (e.g., FARSIDE; \citealt{Burns2021_RSPTA}) on the Moon going forward.

\textbf{System Implications:} The exoplanet auroral emission is predicted to peak in LFT3's low-band (0.1-10 MHz) for a handful of nearby exoplanets \citep{Griessmeier17PREVIII}. As the emission is predicted to be highly circularly polarized, only Stokes $V$ is needed. Previous detections of individual bursts from the $\uptau$~Boo and HD 189733 systems are $\sim$2 Jy at most~\citep{Turner2021_Radio,Zhang2025}. Flux density predictions for some planets in the $0.1-10$~MHz frequency range extend to $\sim100$~Jy for individual bursts \citep{Griessmeier17PREVIII}. Therefore, exoplanets will only be observable by binning over large time-scales of tens of hours and large bandwidths of then of MHz. Flux density predictions for exoplanets vary by orders of magnitude between different groups \citep{Lazio2004,Lynch2018,Griessmeier17PREVIII}. Therefore, only observations can help constrain these models and narrow down their applicability. LFT3's low-band observations are complementary to ongoing ground-based searches, as LFT3 will probe different planet types. 

\subsubsection{Radio-bright Stars and Ultra Cool Dwarfs}
Radio stars, which are often young, magnetically active stars, provide insight into stellar magnetic activity, star formation, and the early stages of stellar evolution \citep{joe_nature_review}. Radio-bright stars can also be used as a means of studying star-exoplanet interactions \citep{Cuntz2000,Lanza2009,joe_nature_review}. Specifically, studying radio flare stars also has implications for exoplanetary science, as the intense radiation from these flares can affect the habitability of orbiting exoplanets \citep{Linsky2019}. Much time has also been spent observing Ultracool Dwarfs (UCDs, spectral type M7) in the radio at GHz frequencies \citep{Williams2018}. The observed properties of UCDs provide a good analog to the Jovian system \citep{hallinan_rotational_2006,Kao2023}, therefore, they are a bridge to studying giant exoplanets in the radio too \citep{Kao2016,Kao2018}.

Radio-loud stars and UCDs have been detected at mostly GHz frequencies on the scale of tens of mJy \citep{Williams2018,driessen_sydney_2024}. These stars will only be detectable over large bandwidths ($>$100 MHz) and timescales ($\sim$hours). A large-scale survey in the medium and high bands may find new star-planet interactions radio emission from nearby planet-hosting stars. Given that LFT3 will observe unique frequencies below 300 MHz, it can probe the magnetic field strengths higher in the magnetosphere of UCDs, offering significant insight into field geometry and strength not accessible from the ground. As in the other science cases, LFT3 allows observations in previously contaminated or unreachable frequencies, with the FM and sub 30~MHz bands of particular interest for monitoring already known bright sources \citep{joe_nature_review}.

\textbf{System Implications:}  Time resolution of seconds to hours and spectral resolution of $\sim$10-100\,MHz would enable the detection and classification of coherent bursts, including ECMI-driven emission. Dual-polarization capability will be important to detect and analyze strongly circularly polarized bursts, which are diagnostic of magnetic field geometry and strength in stellar objects.

\subsubsection{Pulsars}
Pulsars are highly magnetized rotating neutron stars that emit beams of electromagnetic radiation from their poles~\citep{pulsar_handbook}. They serve as precise cosmic clocks, providing insight into the solar wind, interstellar medium, gravitational waves, and the fundamental physics of matter under extreme conditions \citep{epta_inpta,ppta,agazie_nanograv_2023,sct+24,Basu2025_SKA_EOS}. Pulsars are theorized to spin rapidly when they are created, but their rotation slows over time due to the combined braking effects of electromagnetic radiation, pulsar wind, and possibly gravitational wave emission~\citep{agf+16}. 
Figure \ref{fig:transient_space} shows the \textit{average} single-pulse radio pseudo-luminosity values for the pulsar population; the brightest individual pulses can be several orders of magnitude higher~\citep{karuppusamy_giant_2010}.  LFT3 will perform single-pulse searches to detect the brightest subset of these pulses. For the Crab pulsar, for instance, there are one or two pulses per day bright enough to be detected at $10\sigma$ significance with a $\sim2$-metre diameter radio telescope with $\sim100$~MHz bandwidth \citep[][ and Johnson et al., in preparation]{2017ApJ...851...20M}. Broader bandwidth studies allow higher signal-to-noise ratios and enable studies of their emission physics and surrounding material, where higher frequencies are sensitive to emission closer to the pulsar surface~\citep{hassall_wide-band_2012}. With ground-based radio telescopes face challenges at low frequencies, especially in weaker pulsars and those that have a spectral turn-over between 100--
~MHz \citep{Stappers_2011,jankowski_spectral_2018}. Moving the radio telescope off Earth circumvents the ionospheric cut-off. One of the lowest-frequency pulsar detections made to date has been with LOFAR which detected PSR~B0809$+$74 down to 15~MHz with favorable ionospheric conditions \citep{Kondratiev_2013}.



\textbf{System Implications:} Beamformed data products are needed for pulsar observations, as is the ability to de-disperse these.
The pulsar candidates identified by LFT3 will likely be giant pulses; this is due to the luminosity distribution and pulse amplitude distributions of the Galactic pulsar population~\citep{bjb+12,k13,lbb+13,zgw+24,dwl+25}. 
For some pulsars, however, a periodicity-based search could yield detections based on the average flux densities after long integration. Figure \ref{fig:fieldofview} 
shows the distribution of known pulsars potentially detectable by LFT3 with 10-hour coherently combined integrations, assuming flux densities reported in the ATNF Pulsar Catalogue~\citep{psrcat}. The ability to coherently fold data using known ephemeredes is needed for such efforts. 


\subsubsection{Fast Radio Bursts}
Fast Radio Bursts (FRBs) are intense,
millisecond-duration radio pulses originating from extragalactic sources. 
Although Galactic magnetars have been shown to exhibit similar FRB-like emission \citep{BC_2020,chime2020sgr1935}, the exact progenitors and emission mechanisms for all FRBs is still unknown, but the consensus is that they all involve emission from neutron stars~\citep{ck21}. 
The Canadian Hydrogen Intensity Mapping Experiment (CHIME) telescope, which operates in the $400-800$~MHz range, is currently the world leader for detecting new Fast Radio Bursts, helped tremendously by its 200 \, deg$^{2}$ field-of-view~\citep{leung2021synoptic,lanman2024kko}.
A lunar telescope, capable of high-time-resolution observations, could greatly advance our understanding of fast radio bursts.
LFT3, supplemented by wide field-of-view ground stations on Earth could be used to co-observe to enable microarcsecond localization of the brightest FRBs, something which would be of unprecedented scientific value (see e.g. \citealt{nimmo_burst_2023}). 

Low-frequency observations of FRBs are crucial for understanding their propagation through the intergalactic medium, potentially unveiling the distribution of baryonic matter and offering clues about the emission mechanism and source plasma environments. LOFAR observations of FRB 20180916 have revealed potential evidence for the interaction between a neutron star and its binary companion by observing bursts at 100 MHz~\citep{pleunis2021lofar}. 


\textbf{System Implications:} The requirements for observations of bright FRBs by LFT3 are no more stringent than those needed for observing pulsars.

\subsubsection{Long-Period Transients}
Galactic long-period radio transients represent a burgeoning new subclass of transients. Distinguished by their exceptionally long periods, minute-long pulse duration, and low-frequency emission. The prototypical example is GLEAM-X J162759.5-523504.3 \citep{hurley_walker_radio_2022}. This object exhibits a period of 18.18 minutes and a pulse duration of $\sim 1$ minute at 100--200~MHz. The measured DM for this source is $\sim57~\text{pc cm}^{-3}$, indicating a Galactic origin. The duration of pulses seen from this source is orders of magnitude longer (30--60~s) than that of FRBs and have peak flux densities of $5-40$~Jy. The signal-to-noise ratio scales with the square root of the duration; this makes their discovery and cataloging one of the most feasible and impactful science goals of LFT3 (see Fig. \ref{fig:transient_space}). For pulsars, data from successive rotations can be stacked to increase detection significance. Their emission is highly linearly polarized ($\sim90\%$)~\citep{Men_2025}.
Subsequent surveys have revealed more of these objects with periods ranging from minutes to hours (\citealt{HW_2024}). In particular, they show low dispersion measures ($\lesssim$ few $10^2~\text{pc cm}^{-3}$) which make them relatively easy to distinguish when observing at low frequencies. These sources are not exceedingly rare and harbor many open questions yet to be fully explored. LFT3 has an opportunity to advance this field, still in its infancy. 

\textbf{System Implications:} The specifications needed for LPTs mostly mirror (and are less stringent) those needed for the preceding science cases. An additional consideration is the configuration of the noise baseline stability. For instance, low-frequency red noise in the data, e.g. from running median filters, should be chosen so as not to filter out the long-periods of astrophysical interest. 

\subsection{Technosignatures}

\begin{figure}
    \centering
    \includegraphics[trim = 0mm 2mm 0mm 0mm, clip, width=\linewidth]{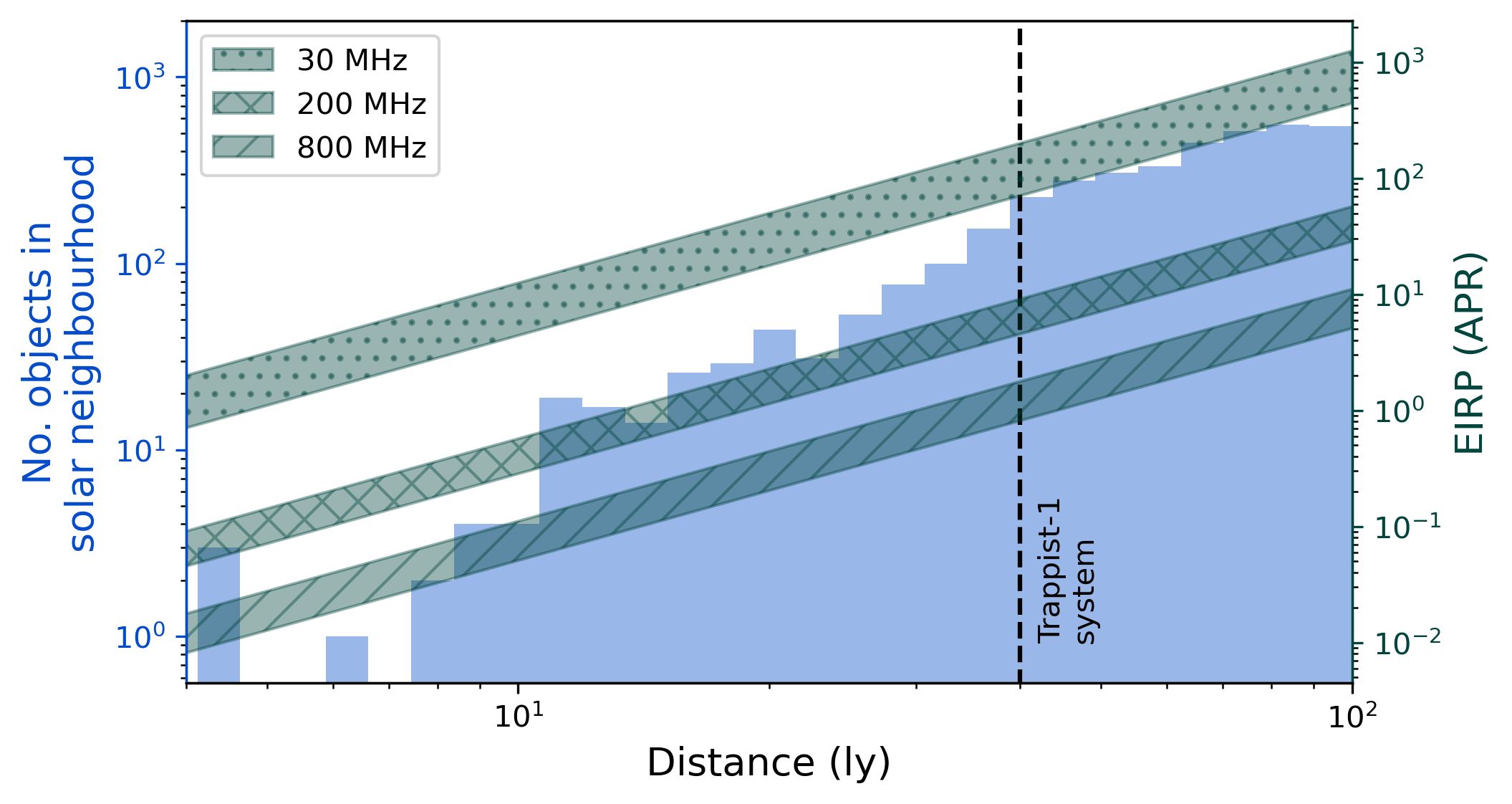}
    \caption{Distribution of objects within 100 light-years that LFT3 will be surveyed for technosignatures, covering $0.1-1000$ times the equivalent isotropic radiated power (EIRP) of the Arecibo Planetary Radar (20 TW). The distribution of targets is from the LFT3 star catalog \citep{lft3StarCatalog}. Hatched regions represent the sensitivity of each of the LFT3 bands sensitivity. 
    In this case EIRP,is assumed to have dual polarisation with a bandwidth of $1$\,MHz. 
    For reference the black dashed line shows the distance to the trappist one system.}
    \label{fig:LFT3_EIRP}
\end{figure}

One key scientific objective for LFT3 is to search for evidence of non-anthropogenic technological electromagnetic signals, or ``technosignatures'' \citep{WORDEN201798} and more generally anomalous features in frequency and/or time. We do not have \textit{a priori} information on any parameter of a technosignature, so one must search agnostically with the most sensitive receivers one can muster \citep{wright_cosmic_haystack}. 
At radio wavelengths, the discovery parameter space generally comprises frequency and time, and narrowband (e.g \citealt{Enriquez_2017,Tremblay_2024,tremblay_k218b}) or pulsed (e.g. \citealt{Suresh_2021,Gajjar_2021}) signals are the most likely to be detected. They are also among the least likely signals to be confused with an astrophysical process \citep{Hippke_2017}, but are most likely to be confused with RFI. Within that parameter space, there is a case to be made that the most likely initial detection will be a slow transient arising from some underlying cadence such as a civilization targeting many distant worlds with a beacon or some fortuitous alignment such as the conjunction of two extrasolar planets with our telescope beam \citep{Fan_2025}. 
In order to assess whether a small telescope can meaningfully search for technosignature anomalies, it is instructive to look at detectability of various transmitter strengths for the stellar distribution around us. The Arecibo planetary radar (APR), when it was operating, had been the highest effective isotropic power radiator at a level greater than 20 TW and is a key metric widely used to quantify the strength of technosignatures~\citep{Siemion_2013}. Another metric used is $1000$ times this value~\citep{Gray_2020}. Figure \ref{fig:LFT3_EIRP} shows LFT3's sensitivity towards nearby star systems. As shown, even {\em our} current APR technological capability provides a detectable signal to this telescope, let alone a stronger signal from a more advanced civilization.

For Earth-based searches, in order to further guard against interference, one usually uses a cadence of a series of on-target and off-target observations, where the ``off'' can be the ``on'' for another target.  Using a multibeam system allows ``on'' and ``off'' to be simultaneous coherent beams, making it even more efficient (i.e., \citealt{multibeam}). The use of this in a multibeam system like LFT3 is demonstrated in \cite{Huang_2023} and \cite{tremblay_k218b}. Although we note that remote sites on Earth (where on-off observing is a significant discriminant) do not provide shielding from satellites and airplanes, regardless of the on/off observing strategy. Moreover, even if the satellite signals could be subtracted from radio telescope data, it is unlikely that a technosignature detection made ``through'' the RFI would be believed.



\textbf{System Implications:} 
The telescope's frequency range covers a new parameter space in technosignature searches, especially in the HF band (0.1--50~MHz). Technosignatures may emit in a single polarization, and therefore polarization sensitivity is needed. The technosignature science case requires some in-line processing of the data to find signals of interest and for the generation of dynamic spectra or raw voltage postage stamps around signals of interest \citep[e.g,][]{Tremblay_2024}. Dedisperion is required to search for signals towards beam-formed targets. 
As there is no known technosignature signal thus far identified, frequency resolution and time resolution are open parameters that are flexible to the needs to keep data rates and processing to a minimum. The nominal specifications are 1-sec time resolution and 10-Hz frequency resolution.  Recently, it has been suggested that narrow spectral lines may be additionally broadened by coronal activity of the local star~\citep{GajjarBrown}, better matching these somewhat larger spectral bins. 
Technosignature search methods will be verified by looking toward Mars, where there are a number of space-based and ground-based communication signals transmitted back to Earth on a regular basis, as well as for lunar orbiters. Their carrier signals are typically emitted at $2200-2290$~MHz and represent a population of narrowband artificial signals.
The sky rotates $\approx30$ times slower when observing from the Moon's surface compared to Earth. This allows longer dwell times on any given patch of sky, which in turn allows prolonged studies of signal characteristics. 

\subsection{Spectral Lines} 
The most uniquely enabled spectral-line science on the lunar farside lies at the lowest radio frequencies, where ionospheric opacity and terrestrial RFI render observations from Earth impossible or severely limited. In particular, ultra–high principal quantum number radio recombination lines ($n \gtrsim 400$), low-frequency molecular transitions, and deep stacked CRRL measurements represent regimes where a farside instrument such as LFT3 does not simply improve upon ground-based capabilities, but opens entirely new observational parameter space.


\subsubsection{Extra-galactic HI}


Neutral hydrogen (H{\sc I}) emission at 21\,cm (1.420\,GHz) is a cornerstone tracer of baryonic structure, enabling studies ranging from Galactic dynamics to the large-scale distribution of matter \citep[e.g.,][]{HI4PI,Meiksin_2022,Pingel_2022}. Although this transition lies within a protected radio astronomy band, observations from Earth are increasingly limited by RFI, instrumental systematics, and ionospheric effects, particularly for experiments requiring long integrations and high spectral fidelity. These limitations become more severe for extragalactic studies, where the H{\sc I} line is redshifted to lower frequencies as $(1+z)^{-1}$, moving outside protected allocations and into heavily contaminated spectral regions.

Observations of extragalactic H{\sc I} at low to intermediate redshift provide critical insight into the gas content, kinematics, and evolution of galaxies. Large-area surveys such as WALLABY \citep{Koribalski_2020} and deeper integrations such as DINGO on the Australian SKA Pathfinder \citep{Meyer_2020,Rhee_2023} have demonstrated the power of widefield H{\sc I} mapping in the local Universe, but remain limited by the presence of RFI, bandpass stability, and the need for aggressive spectral flagging \citep{Rhee_2023}. These limitations are particularly acute when targeting low-column-density gas, faint satellite populations, or modest redshifts ($z \gtrsim 0.1$), where the H{\sc I} line is shifted into increasingly contaminated frequency bands.

The lunar farside provides a uniquely clean observing environment in which these challenges are mitigated. The absence of terrestrial RFI and ionospheric distortions enables stable, spectrally smooth observations over long integrations, improving sensitivity to weak H{\sc I} emission and reducing systematic uncertainties in baseline subtraction. In this regime, an instrument such as LFT3 would not replace ground-based surveys but would complement them by providing access to frequency ranges and spectral fidelity that are difficult to achieve from Earth, thereby enhancing studies of galaxy gas content and evolution in the nearby Universe.

\textbf{System Implications:} For this work, the data product would need to be a time-averaged power spectrum of beamformed targets or an incoherent sum. A spectral resolution of 5-10\,kHz and a time resolution of 8-10\,seconds, would be sufficient due to Doppler broadening of the 21\,cm signal. 

\subsubsection{Radio Recombination Lines}

The diffuse cold neutral medium (CNM; $T_s < 100$ K) is a key component of the interstellar medium and is traditionally probed by H{\sc I} absorption at 21 \, cm \citep{Dickey_1990}. A powerful complementary diagnostic is provided by cold radio recombination lines (RRLs), which trace partially ionized gas associated with the CNM. Due to the low ionization potential of carbon (11.4 \, eV), far-ultraviolet radiation readily ionizes carbon atoms and subsequent recombination produces CRRLs that are sensitive to the physical conditions of the diffuse ISM. These lines have been detected at low radio frequencies ($\lesssim 1.5$\,GHz) in both emission and absorption along the Galactic plane \citep[e.g.,][]{Kantharia_2001,Salas_2019,GDIGS}, and can also contribute to the spectral structure relevant for global cosmic dawn measurements \citep{2024AJ....167....2V}.

At high principal quantum numbers ($n \gtrsim 300$), corresponding to frequencies below $\sim 100$\,MHz, collisional processes dominate the level populations, driving them toward thermal equilibrium and producing CRRLs in absorption against bright background emission \citep{Tremblay_2018}. At lower $n$ (frequencies $\gtrsim 200$\,MHz), radiative processes lead to population inversion and CRRLs are observed in emission \citep{Gordon-RRL}. The transition between these regimes occurs between $\sim 100$ and $200$\,MHz, and depends sensitively on the gas density and temperature, shifting to higher frequencies in environments of higher-pressure. This behaviour makes CRRLs a sensitive probe of the thermal pressure, ionisation fraction, and excitation conditions of diffuse gas.

The lowest-frequency CRRLs, which provide the strongest constraints on low-density, cold environments, are the most difficult to access from Earth. Ionospheric effects—scaling approximately as $\lambda^2$—introduce phase distortions, spectral structure, and line broadening that limit calibration fidelity and obscure weak absorption features at frequencies below $\sim$100–500\,MHz. In addition, terrestrial RFI further contaminates these bands. The lunar farside offers a uniquely enabling environment for CRRL studies: the absence of an ionosphere and anthropogenic interference allows for stable, wide-band observations with smooth spectral baselines. This enables precise measurements of line profiles and the stacking of multiple transitions across quantum levels, significantly improving sensitivity to weak CRRL signals. As a result, a farside instrument such as LFT3 can probe the coldest and most diffuse phases of the ISM in a regime that is fundamentally limited for ground-based telescopes.

\textbf{System Implications:} The CRRLs will change width depending on the frequency of detection and the conditions of the environment. However, a resolution of 100--300\,Hz would be sufficient to study the lines across the proposed frequency bands. Additionally, an initial time resolution of 8-10 seconds is sufficient, with the option to average data over longer periods to increase sensitivity so long as the source remains within the field-of-view. We do this with nonmoving dipoles with ground-based telescopes by either employing fringe tracking or by doing short time recordings, correcting for the source position, and then time-averaging the data. If the dipoles were treated as independent entities, correlated visibilities would allow for a low-resolution map of the regions where the signals are detected. Additionally, beamforming can be used in two approaches: coherent beamforming, which maximizes sensitivity for known source positions, or incoherent beamforming, which sacrifices some sensitivity for a wider field of view and is more robust when detecting strong spectral lines. Overall, a spatial resolution of one degree in the sky would allow for follow-up by ground-based telescopes, where more detail is required. The preferred output is a time-averaged power spectrum or a spectrum converted to intensity in Jy. Either way, a calibration of the flux density values will be needed.

\subsubsection{Blind search for novel spectral features}



Several mechanisms—ranging from well-motivated to highly speculative—could give rise to previously undetected spectral features in the radio band. Among the most compelling are interactions in the dark-matter sector \citep{2025ApJ...984L..24K}, which may produce narrowband spectral signatures correlated with large-scale structure or local environments. In many models, such features trace the Galactic potential, while in others, such as dark photon dark matter \citep{2025PhRvL.134q1001A}, they may exhibit correlations with the Sun. These signals are expected to be intrinsically weak and narrowband and may lie in frequency ranges that are heavily contaminated by terrestrial RFI or distorted by ionospheric effects, potentially explaining their non-detection to date. The lunar farside provides a uniquely clean spectral environment in which to conduct blind, wide-band searches for such signals, enabling exploration of parameter space that is inaccessible from the ground.

A complementary avenue for discovery lies in low-frequency maser emission from molecular species. Transitions below $\sim 1$\,GHz can trace non-thermal population inversions and probe excitation conditions distinct from those accessible at higher frequencies \citep{Tremblay_2017}. However, many of these transitions fall within frequency bands strongly affected by terrestrial interference; for example, nitric oxide lines near 107\, MHz lie within the FM radio band and are severely affected by RFI \citep{Tremblay_2020_NO}. This limitation extends to a wider set of molecular transitions studied with ground-based facilities \citep{Jacob_2024}. In contrast, a lunar-based instrument such as LFT3 enables continuous, interference-free coverage of these frequencies, opening a new discovery space for weak maser emission and previously undetected spectral lines. Together, these considerations highlight the unique role of a lunar farside observatory in enabling blind spectral surveys across wide frequency ranges. By providing stable, RFI-free access to bands that are otherwise unusable from Earth, LFT3 offers a qualitatively new opportunity to search for novel spectral phenomena and to probe the physical and chemical processes shaping astrophysical environments.

\textbf{System Implications:}  
To reduce data volume, observations may focus on targeted frequency intervals selected according to the science objectives, with the option to define multiple narrow spectral windows to capture key transitions. This strategy avoids recording unnecessary bandwidth while retaining flexibility across different lines. Based on Galactic CO observations \citep{Dame-2001} velocity coverage of roughly $\pm 280~\mathrm{km~s^{-1}}$ is sufficient for molecular gas in the disk.

\subsection{Cosmology}
Radio observations of red-shifted atomic hydrogen have the potential to uniquely inform our knowledge of the early Universe, from the so-called Cosmic Dark Ages over 13 billion years ago, to the formation of the first stars and black holes that reionized the Universe in its last cosmic-wide change. This is because neutral hydrogen was the dominant form of baryonic matter during this period and can be detected on the basis of its 21 cm spectral line (rest frequency 1420 MHz). Of particular interest is the rise of the astrophysical structure in the cosmic fabric beginning about 13 billion years ago.  Since the relevant signal emanates from a distance of 13 giga-lightyears (Gly), it is exceedingly faint, requiring extreme sensitivity and radio-quiet conditions.  Since the signal of interest is also global, a single antenna is technically sufficient.  The lunar farside is a key proposed location for this science \citep{2021arXiv210305085B,Fialkov_2024}.


Due to strong emission from our Galaxy at the relevant frequencies, the sky is extremely bright, with brightness temperatures ranging from about 100 K to $10^6$K depending on frequency and location, while the signal of interest is very faint (in the milliKelvin range), hence the difficulty in making the measurement, which requires extremely precise knowledge of the smooth shape of the measured spectrum and removing this as a baseline. Unaccounted for low-level RFI adds uncertainty to any measurement. Precise measurements from the RFI quiet lunar farside will be invaluable in understanding that effect.

Note that LuSEE-Night, which should launch early 2027, is similarly focussed on making path-finding measurements relevant to the global signal, with all of its inherent difficulty.  LFT3 will work in collaboration with LuSEE-Night to help inform these measurements.  Both missions will fly a similar HF antenna, although by necessity the physical arrangements will differ.  Ideally, both would operate at the same time; however, this would require LuSEE-Night to remain operating for several years.  There is the possibility of using the same calibration orbiter, which is a payload delivered by LuSEE-Night.  One of the primary difficulties in this measurement is the affect of the lunar regolith on the received signal, and having two nearly identical systems in close proximity may be able to uniquely inform this issue.


Although it is highly unlikely that LFT3 would have sensitivity to detect either the Dark Ages or the Epoch of Reionization signals in the monopole, in the HF it could detect excess photons in the Rayleigh-Jeans tail of the CMB spectrum, which could again put strong limits on some dark matter models \citep{2018PhRvL.121c1103P}. In fact, there have been long-standing hints on excess emission in the radio sky at low frequencies (\cite{2018ApJ...858L...9D,2011ApJ...734....4K}) and this is something LFT3 is uniquely positioned to test.

\textbf{System Implications:} 
The expected signal is thermal; however, that thermal profile changes with cosmic time and hence frequency, so the bandwidth and binning impact the result (see, e.g. \citealt{2017PASP..129d5001D}).  
Long integrations improve the sensitivity, so FOV changes on the order of minutes are the limiting factor.
Given the noise of the Galactic plane, the Sun, and Jupiter, it is preferable to observe when they are all below the horizon.


\subsection{Polarization modulation by axion-like dark matter}
There has been a renewed interest in cosmology in ultralight dark matter candidates \cite{2022SciA....8J3618C}. This framework postulates a dark matter particle with such a small mass that its de Broglie wavelength spans roughly a kiloparsec—coinciding with the scale where structure formation issues arise. This corresponds to a particle mass around $m\sim 10^{-22}$\,eV. Due to this extremely low mass, achieving the observed dark matter energy density necessitates a very high number density of ULDM particles, implying that they behave collectively as a classical bosonic field, or a Bose-Einstein condensate. To stabilize this tiny mass against radiative corrections, it is typically assumed that the ULDM field is a pseudo-Goldstone boson resulting from a broken symmetry analogous to the QCD axion, which stems from the breaking of the Peccei-Quinn symmetry. Whether connected to QCD or not, such particles tend to couple to photons through non-renormalizable interactions, similar to those of axions. Particles of this kind, known as axion-like particles (ALPs), arise naturally in many beyond-the-Standard-Model scenarios, including string theory frameworks.

Coherent field oscillations of ALPs affect the propagation of electromagnetic waves through the ULDM condensate. It has been shown \citep{2019JCAP...02..059I,2024PhRvD.110f3013A} that the polarization angle of linearly polarized light oscillates with the same period—set by the ALP mass—which remains constant across all ALP domains in the Universe. The prediction is that the polarization angle of linearly polarized radio emission oscillates in time, which can, in principle, be detected with an instrument such as LFT3.  The observable would be small and periodic changes in linar polarization when the same patch of the sky is observed over several lunar cycles. LFT3's unique gain stability given the lunar environment and slow moon rotation allowing for long dwelling time should make it particularly sensitive to search for linearly polarized light oscillations. 

\textbf{System Implications:} 
The measurements require linear polarization sensitivity. The expected signal is weak, and the polarization angle can appear to oscillate due to other time-dependent effects such as the gain mismatch between two polarization state channels and the variable polarization leakage due to thermal expansion. The system should be designed for gain and polarization stability and the ability to discern polarization on various timescales. The stable platform on the lunar surface and the long dwell times are very beneficial.

\subsection{Exploration of VLBI Techniques}
Very Long Baseline Interferometry (VLBI) is a powerful technique of using a large separation between antennas to obtain incredible detail of astronomical objects. 
By having a telescope on the moon with a wide range of frequency coverage, we could collaborate with ground-based facilities to test Earth-Moon VLBI methods. As the earth-based telescope has to contend with the ionosphere, joint measurements would not be available for most of the HF band ($<$30\,MHz). However, the other bands have significant overlap with ground-based telescopes in both the southern and northern hemispheres.  Joint experiments with multiple facilities from VHF to UHF will be explored. Additionally, the possibility may exist to conduct limited real-time VLBI experiments at times when the communication relay satellite can see both the spacecraft and the Earth.  Given the data rates involved, this would be very limited in scope, and we are looking at partnering with potential cislunar communication companies.

\textbf{System Implications:} This goal requires clock-cycle accurate time-stamping of baseband data (relative between wavefrom acquisitions as the absolute time offset can be fitted for). It also requires long waveform acquisitions and efficient onboard quantization of waveforms to minimize the volume of transferred data. 

\subsection{Lunar Environment}
Along with advancing astronomical science, operating a telescope on the moon will also yield invaluable insights into the lunar environment and details regarding the engineering and functionality of lunar telescopes. These insights can be classified into (1) the RF environment, (2) lunar regolith, and (3)  operational/engineering impacts, as discussed below. 

\subsubsection{RF Environment and Spectral Monitoring}
Although the lunar farside should be a pristine RF environment, there is the potential of some RFI from the lunar communication orbiters that will be in place, as well as spacecraft that are at distances further than the moon, such as L2.  There is also the potential of unknown actors emitting radio frequencies, as well as additional payloads being deployed during the LFT3 operational period, for example China's orbiting interferometer called ``Discovering the Sky at the Longest Wavelengths'' (DSL).  As mentioned throughout, this is a time of unique change in the operating cislunar environment, and measuring a legacy spectral baseline as well as the potential increasing emergence of RFI is a key objective of this mission.

The dynamic spectra taken over the operational period and transmitted back to Earth will provide the key record of the baseline and evolving RF spectrum, and LFT3 represents the only mission proposed to make these measurements  across these bands.  In addition to the dynamic spectrum, LFT3 will catalog every measured RF event.  This will allow a unique record of the changing cislunar RF environment.

\subsubsection{Lunar Regolith}
Regolith is a term used to describe any loose, unconsolidated material above the bedrock of a celestial body. The lunar regolith has effectively two different science impact components for LFT3: (1) electromagnetic, (2) structural.

{\em Electromagnetic.}  The electromagnetic properties of the regolith have a huge impact on the mission, particularly at the lowest frequencies and for cosmology.  Understanding the lunar surface and its underlying topography, conductivity, dielectric constant, emission, and reflection is the key to understanding its impact on scientific results.  For cosmic dark age science, this is arguably the most important component.  The EM missions as part of NASA's Commercial Lunar Payload Service (CLPS) are essentially all leveraged to understand the lunar environment and interactions with the regolith, such as ROLSES \citep{2025arXiv250309842H}, LuSEE-Night \citep{2023URSIGASSLuSEENight, 2023arXiv230110345B} and LuSEE-Lite \citep{2023AGUFM.P31B..02B}.  LFT3 would complement those efforts and will conduct coordinated experiments with LuSEE-Night, near which LFT3 will land.  Being nearby will allow studies of the same region but from a different vantage point and will provide additional information on the effects of the regolith.

{\em Structural.} LFT3 provides a unique opportunity to investigate the structure and composition of the regolith at its landing site in the farside highlands, where the regolith layer is estimated to be about 4-5 m deep.  Characterization of the regolith at LFT3's landing site will be of significant interest to engineers and planetary scientists, especially given that only two missions, Chang'e 4 \citep{wimmer2020lnd} and Change'E 6 \citep{Ren2025ChangE6}, have successfully landed on the farside to date\footnote{Chang'E 6 returned samples of the lunar farside}. There are promising low-cost options for this characterization that will not interfere with the core radio astronomy objectives of LFT3. For example, the first \textit{Blue Ghost} mission, which landed on February 2, 2025,  included a set of six cameras called SCALPSS (Stereo Cameras for Lunar Plume-Surface Studies) to film the descent of the lander to study interactions between regolith and the lander’s exhaust plume\citep{Atkinson2025NASA}. 

The SCALPSS images are used to create digital elevation maps of the landing site via stereo photogrammetry. Observing how these maps change throughout the landing process, researchers can estimate the total volume of regolith ejected, how the rate of ejection varies over time, and the depth of the removed layers \citep{Tyrrell2022PlumeSurface}.  These data could reveal how the properties of the regolith vary with depth near the surface.  Characterizing plume-surface interactions, as with the SCALPSS experiment, could also help quantify the hazards to surrounding infrastructure posed by landing on unprepared lunar surfaces nearby.  Even something as simple as having cameras with the lander's footpads in view could reveal their penetration depth into the regolith after landing, allowing the regolith's bearing capacity to be calculated. Any of these low-cost measurements and analyses will be valuable for planning future missions and improving the quality of lunar simulations.

\subsubsection{Operational and Engineering Impact}

Beyond operational considerations for the characterization of the electromagnetic properties of the local regolith in coordination with LuSEE-Night (which are anticipated to only involve a small fraction of the observation time of LFT3) and the minor engineering impact of low-cost cameras and sensors for characterization of the regolith, the main engineering impact of the lunar environment on the design of LFT3 will be considerations for the thermal environment of the Moon and dust.  Surviving the two-week lunar night, during which solar power is unavailable and temperatures can reach -$130 ^\circ$C, will be a major concern solved in conjunction with the provider of the commercial lander housing LFT3’s hardware.  Endurance throughout the two-week lunar day, during which temperatures can reach $120 ^{\circ}$C, is a lesser concern that will be addressed with the design of the LFT3 radiators.

The top surface of the lunar regolith comprises a very fine dust with 20\% of particles by weight having a diameter below 20 \textmu m \citep{ZANON2023627}.  Lunar dust is very abrasive and becomes electrostatically charged as a result of interactions with solar wind and UV radiation.  The act of landing on the moon scatters large amounts of dust that can settle on the exterior of the lander. Although this has proven destructive to mechanical joints and pressure seals, it has been less of a problem for RF hardware .  The Chang'E 4 lander has been operational since landing on the lunar farside in 2019, with no negative effects from dust reported on the operation of its X-band antenna, lunar penetrating radar (which has channels at 60 and 500 MHz), or low-frequency radio spectrometer (which measures in the 0.1-40 MHz range) \citep{JIA2018207}.

To help increase LFT3’s resistance to lunar dust, proactive measures can be taken.   \textit{Blue Ghost} Mission 1 successfully tested a passive "clear dust repellant coating" by Voyager Space \citep{voyager2025dustcoating} as well as an electrodynamic dust shield by NASA \citep{buhler2020current}.  The low-mass dust shield draws only 2-4 W of power to remove dust particles with electric fields and has demonstrated 98\% dust removal during laboratory testing \citep{WANG20246194}.  Various passively dust-tolerant mechanisms, including seals and bearings, are also being developed \citep{fritz2024dustroadmap}.

The dust mitigation techniques employed by LFT3 will be relevant for future lunar surface missions, regardless of their specific purpose.  Alongside the CLPS missions and other future surface missions, LFT3 will play an important role in helping engineers converge on a set of practical solutions to avoid the negative effects of lunar dust adhesion.

\subsection{Science Summary}
The strength of LFT3 lies in its ability to uncover the unknown while operating in a completely unprecedented environment. As the only mission with the goal of capitalizing on this unique moment in human history, LFT3 stands at the forefront of discovery.  Table \ref{tab:scisum} provides a summary of the primary science objectives along with some of their system needs.  Additionally, the robust design is tolerant to system issues that may arise in landing or operation, within the scope of the mission. It is important to emphasize that any observation with LFT3, regardless of resolution, will be unique and scientifically valuable.  With its high-performance design, LFT3 will leave a lasting legacy, documenting the transition of the cislunar environment, potentially discovering new phenomena, and addressing long-standing questions about known phenomena. LFT3 will serve as an important incumbent of broad frequency use in the cislunar environment.

\begin{table*}
\centering
\caption{System requirements for each mission objective.
Science category 1 is Lunar-Exclusive Science, category 2 is Lunar-Augmented Science, and category 3 is Proof-of-Concept Science (see Figure \ref{fig:sciencegoals}). }
\label{tab:scisum}
\begin{tabular}{|l|l|l|l|l|l|l|}\hline
\textbf{Mission Objective} & \textbf{Receiver Band} & \textbf{Freq. Res.} & \textbf{Time Res.} & \textbf{Objects of interest} & \textbf{Stokes} & \textbf{Category} \\\hline
Technosignatures & HF/VHF/UHF & 10Hz & 1sec & catalog/blind survey & I, Q, U, V & 1,2 \\\hline
Pulsar & HF/VHF/UHF & 0.5MHz & 1ms & catalog/blind survey & I & 2,3 \\\hline
FRB Analogs & HF/VHF/UHF & 0.5MHz & 10ms & catalog/blind survey & I & 1,2 \\\hline
Radio-loud Stars & HF/VHF/UHF & 10 MHz & minutes & Gaia Stars & I, V & 1, 2\\\hline
Exoplanets & HF & 10MHz & hours & Nearby exoplanets & I,V & 1 \\\hline
Solar System & HF           & 100 kHz & 1ms     & Jupiter             & I,V & 1,2 \\\hline
             & HF           & 100 kHz & minutes & Saturn, Neptune     & I,V & 1  \\\hline         
             & HF           & 1 MHz   & 1ms     & SS Planets         & I & 1,2 \\\hline          
Studies of the Sun & HF/VHF & 0.5 MHz & 1--2ms & Sun & I,V & 3 \\\hline
 & VHF & 100 kHz & 1--2ms & Sun & I & 3 \\\hline
HI Studies & HF & 100--500Hz & 1–-10sec & Galactic Center & I & 3 \\\hline
 & VHF/UHF & 1kHz & 5--8sec & Radio Galaxies & I & 3 \\\hline
RRL & HF & 100--300Hz & 1--5sec & SNR, H\textsc{II} regions & I & 2 \\\hline
 & VHF/UHF & 100--300Hz & 1--5sec & SNR, H\textsc{II}$^{1}$ & I,V & 3 \\\hline
Cosmology & HF/VHF & 1 MHz & minutes & Non-Galactic Plane & I & 2 \\\hline
VLBI & HF/VHF/UHF &  &  & Sun, Jupiter, and AGN cores & BB & 1 \\\hline
Lunar studies & HF/VHF/UHF & 1 kHz & 1s & Moon & I & 1 \\\hline
\end{tabular}
\end{table*}




\section{Payload and Operations}
\label{sec:payload}
\begin{figure}
	\centering
	\includegraphics[width=\linewidth]{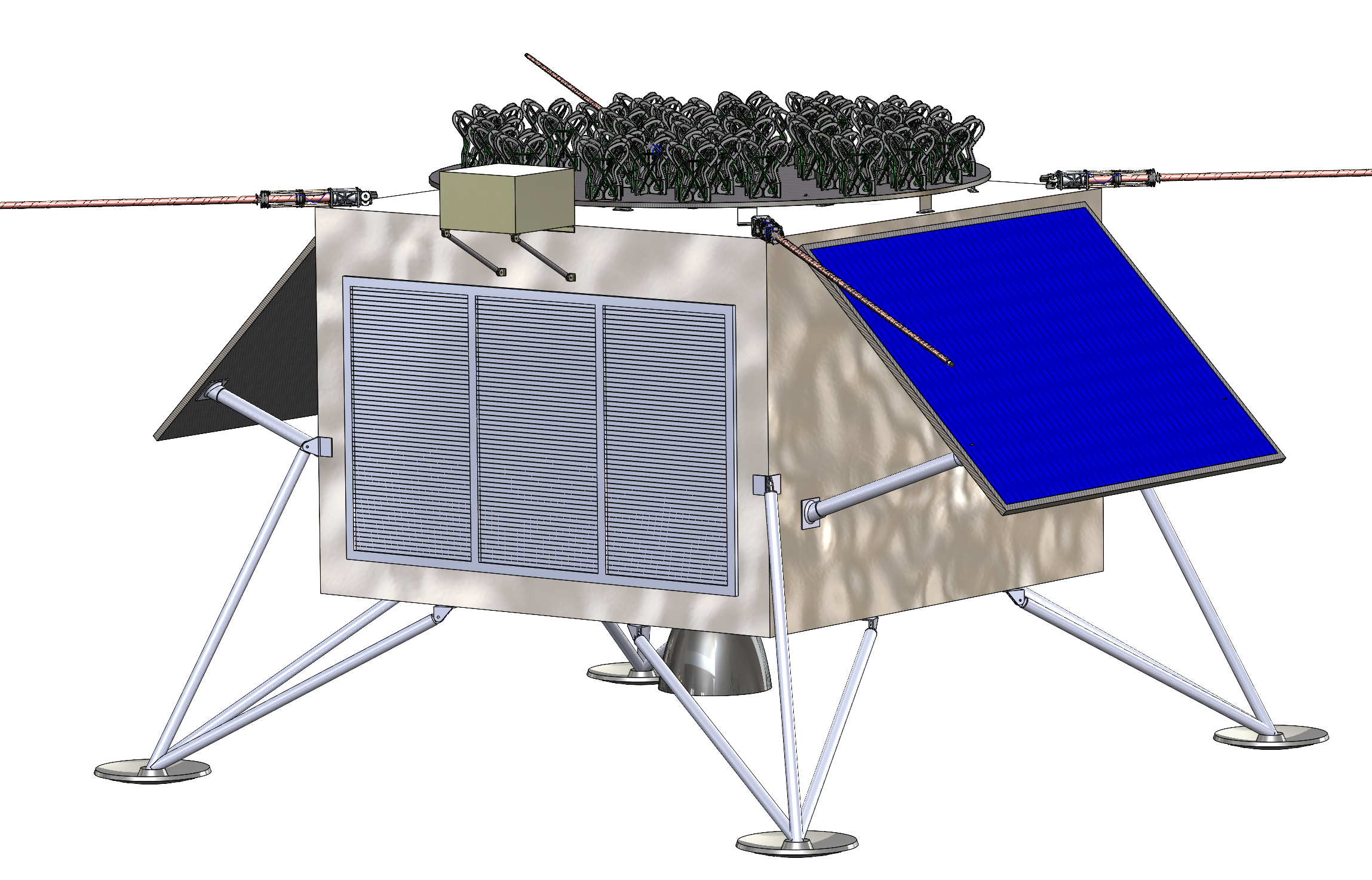}
	\caption{Rendering of the generic science lander \citep[based on][]{neal2020lunarGeophysicalNetwork} and payload, with upper left inset showing the deployed HF stasors.  The antennas are all located on the top deck, and the beamformer and processor are located in a thermal cavity directly below.   The UHF array of Vivaldi antennas is within the top deck circle, and the VHF antenna is shown schematically as a deployed block.\label{fig:render}}
\end{figure}

The previous section outlined the science case for a telescope on the lunar farside.  Ideally, one would build a large and flexible observatory, and discussions for such telescopes are underway (e.g. \citealt{burns2021lunarfarsidelowradio}, \citealt{9438165}).  However, to take advantage of this unique opportunity for early measurements, two additional constraints inform the payload design: (1) land and operate before the end of the decade (by 2030) and (2) achieve the goals for a total mission budget of $\sim$US\$150M (including launch).  The design is therefore structured around these additional objectives, which means that it has similar scope, schedule, and budget requirements to missions under the NASA Commercial Lunar Payload Service (CLPS) program \citep{nasa2026clps}, which informs the broad specifications.

The lunar lander will have a top-deck diameter of approximately 3\, m and a science payload capacity of approximately 100 kg.  One main difference from the NASA CLPS program is our ability to work with potential vendors to increase the co-design between the lander and science payload. For example, the LuSEE-Night program carries its own power system (solar panel and battery) and communication system and requires the lander to electrically passivate itself at the end of the lunar day. This simplifies lander-payload interfaces and gives LuSEE-Night full control over the RFI environment at the expense of increased payload complexity, weight, and cost. For LFT3, we intend to work much more closely with the vendors and rely on the lander's own power and communication systems.   The primary concern is electromagnetic compatibility and the need to minimize RFI effects, which will be achieved by a combination of requirements on instrument design ({\em e.g.} syncing all clocks to precisely place power supply harmonics) and operational constraints (switching off non-essential equipment when taking science data). Table \ref{tab:constraints} summarizes the scope of the payload that fits within the constraints. These numbers represent an evolutionary and realistic step-up from the parameters of LuSEE-Night.

\begin{table}
    \caption{Payload Requirements}
    \begin{tabular}{|l|l|l|} \hline
    \textbf{Parameter} & \textbf{Value} & \textbf{Notes} \\ \hline
    \textbf{Span} & $\sim$3 m & This varies with vendor. \\ \hline
    \textbf{Mass} & 100 kg & This is the science payload mass. \\ \hline
    \textbf{Power (day)} & 100 W & Balancing day/night operations. \\ \hline
    \textbf{Power (night)} & 20 W & Included in the lander mass budget. \\ \hline
    \textbf{Comms} & 100 GB/month & The nominal budget includes \\ & & 20 weeks of operation \\ \hline
    \textbf{Storage} & 20 TB & This coupled with comms \\ & & and processing scopes the system. \\ \hline
    \end{tabular}
    \label{tab:constraints}
\end{table}

\begin{figure*}
	\centering
	\includegraphics[width=0.9\linewidth]{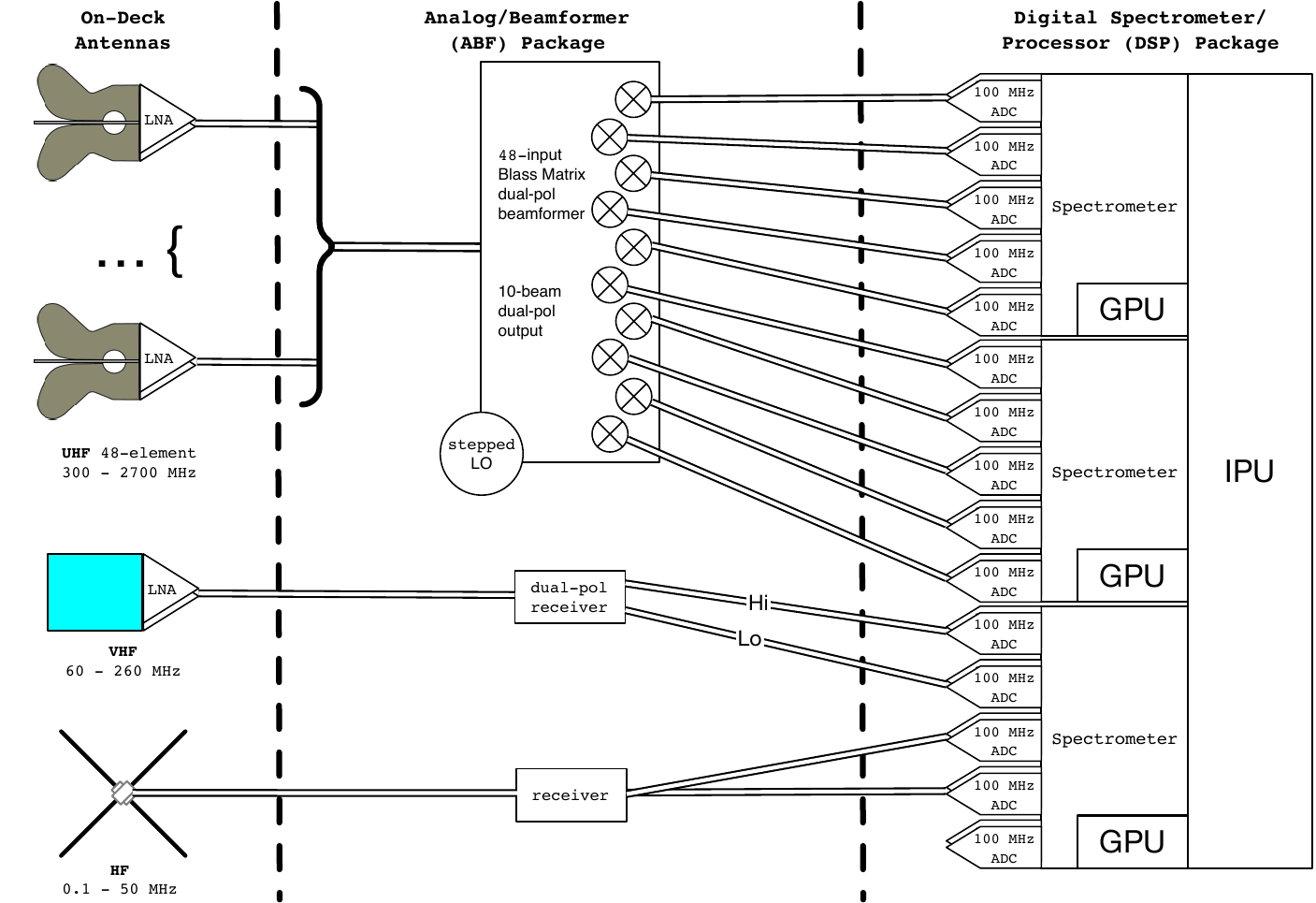}
	\caption{Block diagram of the science payload.  The three antenna bands are shown on the left.  The UHF bands go into a heterodyne and beamforming system to select the frequency sub-band and form the beams.  VHF is split into low/high 100 MHz bands. The Instrument Processing Unit (IPU) conducts the processing and saves the data for transmission to Earth.\label{fig:block}}
\end{figure*}


\begin{table}
\centering
\caption{Basic LFT3 spectrometer parameters. The spectrometers have the additional capabilities to perform technosignature searches, pulsar signal detection and folding, dedispersion and transient detection.\label{tab:spectrometer}}
\begin{tabular}{|l|l|} \hline
\textbf{Parameter} & \textbf{Value}  \\ \hline
Number of input channels & 10 \\ \hline
Max. no. of active input channels & 6 (3 dual-pol inputs) \\ \hline
Bandwidth $\gtrsim100,\mathrm{MHz}$ \\ \hline
Duty-cycle & 100\% \\ \hline
Correlation products & $XX^*$, $YY^*$, Re$(XY^*)$, Im$(XY^*)$  \\ \hline
Spectral Channels & $2^{9}$ - $2^{22}$ \\ \hline
Polyphase filterbank taps & 8 \\ \hline
ring-buffer size & 1\,second \\ \hline
Power-consumption & $<$12\,W \\ \hline
\end{tabular}
\end{table}

A rendering of the lander is shown in Figure \ref{fig:render} and a schematic of the science payload is shown in Figure \ref{fig:block}. Thermal survival is a critical top-level specification for the entire lander. 
In the Equatorial Farside region, lunar days and lunar nights each last approximately 14 Earth days. This results in survival temperatures ranging from 100 K to 400 K \citep{2012JGRE..117.0H18V}.   

\begin{figure*}
	\centering
	\includegraphics[width=\linewidth]{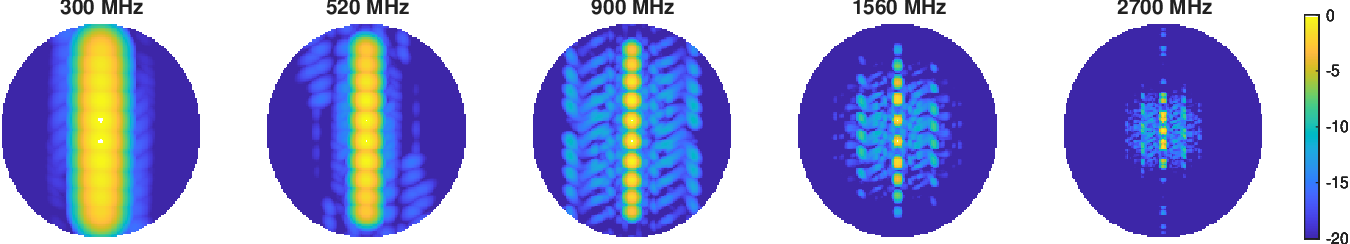}
	\caption{UHF array superimposed formed beam pattern maps over frequency. Beams are centered within 10 equal intervals over a 110 degree field of view.}
	\label{fig:midband_beam_maps}
\end{figure*}

The UHF array covers most of the lander top surface with 48 dual-polarization wideband antenna elements. The UHF array forms a fan of beams on the sky designed to sweep 70\% of the lunar sky over time (Fig. \ref{fig:midband_beam_maps}). The signals received from the antennas are amplified and combined along rows of elements with equal time delays to form an effective line array with 18 cm spacing. Line array outputs are combined in a modified wideband analog Blass matrix beamformer \cite{warnick2025modified} to form 10 dual-polarized beams. 

The 10 UHF dual-pol formed beams, two dual-pol VHF, and HF outputs (28 total ports) are downconverted and sampled in a 100 MHz subband that is scanned across the array operating bandwidth and processed by three identical spectrometers. Each spectrometer will contain 5 dual-pol inputs, of which selectable dual-pol inputs will feed the digital pipeline.  Note that generally only a subset of inputs will be processed, except in a limited set of daytime triggered opportunities.  This architecture is preferred over RF switches for flexibility, reliability, and redundancy.
Each separate dual-pol signal will be channelized and correlated into 4 independent correlation products to maintain full polarization sensitivity. Primary channelization will produce 2048 50kHz spectral channels across 100MHz using an 8-tap polyphase filterbank algorithm implemented on an FPGA. Data will be averaged and passed to the IPU for further processing, storage, and eventual transmission to earth.

Additionally, the system will maintain a concurrent ringbuffer of raw digitized data for 1 second of data that will allow a ring-buffer dump on an interesting time-domain trigger. A simple trigger will be implemented in the FPGA based on outliers in the raw channelized data. Each spectrometer will also contain a small GPU that will be enabled on demand and further process raw channelized data. The GPU will run spectral zooming to achieve a  final spectral resolution of $\sim 10\,$Hz, allowing the search for technosignatures. It will also enable a more sophisticated search for technosignatures in the raw datastream based on non-Gaussian correlations.  The system will moderate power consumption by deactivating entire spectrometers, individual signal paths, and GPUs on demand. Having three spectrometers also provides redundancy against any one spectrometer failing catastrophically. 

The LuSEE-Night spectrometer is based on the Microsemi Polarfire FPGAs family of FPGAs, which has the performance required of the LFT3's spectrometer \citep{SlosarSpectromer2026}. For GPU options, the Nvidia Jetson AGX Orin family of system-on-a-chip solutions is being considered. These have sufficient compute-power to easily handle complex spectral zoom and technosignature search tasks. One of the major issues is the ambient radiation environment on the Moon, which is harsher than that on Earth or in low earth orbit. The radiation tolerance of Jetson AGX Orin has been measured in the lab with encouraging results \citep{Slater2023}.  The electronics are contained well within the lander, which provides several layers of shielding. We estimate that with sufficient mitigation, the LFT3 should be able to demonstrate production GPU data processing from the lunar farside. 

The current best estimate of power consumption per spectrometer is 17 W, including power conversion losses. This would allow a single spectrometer to operate throughout the night. GPU use raises the power consumption dramatically (by about 20-50W, depending on load), but can be duty-cycled at night to stay within overal power budgets. During the day the system could theoretically consume up to 200W assuming a high performing GPU and with all three spectrometers running. However, the limiting factor will be thermal control rather than available power.  Potentially, short periods of high-power computing may be available--power/thermal modeling is an important design aspect to be explored further.

Operational constraints have a huge impact on the observation strategy, the derived data products, and the downloaded data and are discussed in Sect. \ref{sec:conops}.  The available full-day power is $\sim$100 W, which reduces to $\sim$20 W at night, and the expected download capacity is 100 GB/month.  The expected on-board memory will be 20 TB.  
The on-board instrument processor unit (IPU) will provide any post-processing of the dynamic spectra and coordinate memory and transmission back to Earth.  The communication back to Earth for data will be limited to the lunar day and will be episodic to handle both channel constraints, as well as operational parameters.



\subsection{Concept of Operations}
\label{sec:conops}

In this section, we outline the LFT3 science operations framework designed to efficiently support the proposed science goals by allocating appropriate mission resources (computing, observation time, beam, band, and downlink capacity) to each science objective. A more complete description can be found in \cite{prabu2025lft3}. The science goals described in this white paper are broadly grouped into three categories which correspond to the categories in Table \ref{tab:scisum}: lunar-exclusive science goals, lunar-augmented science goals, and proof-of-concept science goals. Lunar-exclusive science goals include measurements uniquely enabled by the radio-quiet, ionosphere-free environment of the lunar farside, such as unambiguous searches for technosignatures, detection of low DM transients, cosmology, and legacy RF surveys of the Moon's farside that will serve as a baseline before future RFI contamination. Lunar-augmented science goals build upon and enhance Earth-based observations with additional complementary low-frequency wideband measurements, thus improving the overall scientific understanding of known astrophysical sources. Proof-of-concept science involves replicating measurements that can also be performed from Earth, with the goal of validating LFT3's scientific capabilities and calibrating its measurements against the ones from Earth-based observatories. Examples of this will be observing bright quasars, pulsars, and hydroxyl masers. A high-level categorization of all science goals into these three groups is shown in Figure \ref{fig:sciencegoals}.


All LFT3 science operations will be conducted in two distinct modes: real-time observations and targeted observations. A real-time processing pipeline will operate continuously across all observed fields, constantly looking for transients and technosignature candidates. A real-time search system that runs throughout the 20-week mission life of LFT3 maximizes our likelihood of serendipitous discoveries of transient events. Due to their persistence in the sky, all other science targets (such as pulsars, Sun, solar system planets, etc.) will be observed as the targeted observations with appropriate scheduling (such as choice of beam to process, time and frequency resolution of the output science data, frequency tuning, and number of Stokes parameters to record) as and when the target appears within LFT3's visible sky. A complete breakdown of the observing schedule for these science targets and the volume of corresponding generated science data can be found in \cite{prabu2025lft3}.

Due to the lack of a line-of-sight link to LFT3 from Earth-based groundstations, science data products from LFT3 must be relayed to Earth through an intermediate lunar orbiting satellite. Currently, this method of downlinking data back to Earth is effectively limited to 100 GB/month. Hence, to ensure optimal use of this limited data rate, we define four different science data products that will be produced by LFT3, namely, baseband data, sky-complete spectra, high-resolution dynamic spectra and an event catalog (see Table \ref{tab:data}). For high SNR transient and technosignature events seen by LFT3, 1s of baseband data (about 2GB/s) will be downlinked back to Earth. For all other targeted science observations, only high-resolution dynamic spectra (produced at a rate of $0.0004 - 50$ MB/s based on the science target being observed) are transmitted back to Earth from the LFT3 beam containing the target, and the time/frequency resolution of these dynamic spectra is dictated by the science requirements of the target. For all the fields being observed by LFT3 during its life, a coarse resolution spectra (1MHz frequency resolution and 5 minute time integrations) called the sky complete spectra are generated. This ``sky-complete'' spectra data will map out the complete celestial sphere seen by LFT3 during its life, and contributes towards the cosmology and lunar-RF monitoring studies proposed in this work. The payload also maintains an onboard catalog of low-SNR pulse signals in the form of an event catalog, which will be used for gathering statistics of false positives (and any underlying low-SNR astrophysical transient population) seen by the instrument.

\begin{table}
    \centering
    \caption{Uncompressed volume of data products produced by LFT3 per lunar cycle (28 Earth days). Note that the total data generated is over the 100GB limit. The science data can be fit within the budget using data compression methods, which may potentially be lossy. $^{*}$The event catalog contains the statistics of low SNR pulses seen by LFT3 during the entirety of the lunar cycle. We estimate about $1.4 \times 10^9$ low SNR events to be detected per lunar cycle.}
    \begin{tabular}{|l|l|l|l|} \hline
    \textbf{Product type} & \textbf{Data Volume [GB]} & \textbf{Total integration time} \\ 
    & \textbf{(per lunar cycle)} & \textbf{(per lunar cycle)}  \\\hline
    Baseband data        & 20 &  10 \, seconds \\ \hline
    Dynamic spectra      & 255.6 & 336\, hours  \\ \hline
    Sky-complete spectra & 0.5 & 672\, hours \\  \hline
    Event Catalog ($\geq5\sigma$)       & $\le$ 1 & 28 Earth days$^{*}$ \\ \hline
    Meta-data & $\le 1$ & - \\ \hline
    \end{tabular}
    \label{tab:data}
\end{table}

The mission timeline aims to launch the LFT3 payload before 2030, with a high-level functional requirement for 20 weeks of operation.
The proposed science operations plan in \cite{prabu2025lft3} aims to meet all mission objectives within the 20-week time frame, by allocating sufficient observation times to each of the science goals and a data management plan to download the highest priority data.


\section{Conclusions}
\label{sec:conclusions}


As the only proposed mission to exploit the singular opportunity for RFI-free astrophysical observations across HF, VHF, and UHF from the lunar farside, the LFT3 mission is uniquely positioned to redefine our understanding of the radio universe from this unique vantage point in space and time. In an environment quieter than even that in which the famed "Wow!" signal was detected (and comparable to the environment at the time of the discovery of pulsars), LFT3 offers an unprecedented chance to conduct unambiguous technosignature searches and conduct a once-in-human-history spectral survey of pristine RFI skies.  LFT3 serves as an important incumbent operating over a broad frequency range for future developments of spectrum management and use from the moon.

The essence of LFT3's mission represents this return to the radio-quiet skies from the very early days of radio astronomy but armed with a next-generation high-sensitivity radio telescope. To demonstrate this, figure \ref{fig:VisibilityCount} shows the increase of satellites on the earth (in blue/bottom axis) and the projected lunar orbiters (in red/top axis). 
In doing so, it will enable the most decisive technosignature search ever undertaken by humankind. Due to the pristine RFI environment, every single detection by LFT3 will have unambiguous scientific value. In addition to the technosignature searches, LFT3 is designed to discover previously undetected populations of low-dispersion radio transients, probing low-frequency phenomena inaccessible from Earth. Alongside the payload's preliminary science of transients and technosignatures, the mission opens up new  frontiers in radio astronomy, including low frequency Very Long Baseline Interferometry (VLBI) with Earth-based observatories, and studies of low frequency observations of emission from the outer planets of our solar system (last observed by Voyager). In summary, LFT3 represents a unique time-critical opportunity to listen to the radio universe in a way that has never been possible before and, if missed now, will likely never come again.


\section*{Acknowledgements}


This article is based on work supported by Breakthrough Listen. Breakthrough Listen is managed by the Breakthrough Initiatives, sponsored by the Breakthrough Prize Foundation.

\section*{Data Availability}

 

Data associated with this article can be found in the white paper and memo series at \url{https://lft3.space/}. The compiled list of objects within 100 light-years that will be probed for anomalies by LFT3 is available from the Oxford University Research Archive \url{https://ora.ox.ac.uk/objects/uuid:2389ba0e-51d3-441c-aee0-274987a61320}.



\bibliographystyle{mnras}
\bibliography{bibfiles/bibs,bibfiles/radio-stars,bibfiles/solar,bibfiles/solarsystem-emission,bibfiles/transients} 

@ARTICLE{Siemion_2013,
       author = {{Siemion}, Andrew P.~V. and {Demorest}, Paul and {Korpela}, Eric and {Maddalena}, Ron J. and {Werthimer}, Dan and {Cobb}, Jeff and {Howard}, Andrew W. and {Langston}, Glen and {Lebofsky}, Matt and {Marcy}, Geoffrey W. and {Tarter}, Jill},
        title = "{A 1.1-1.9 GHz SETI Survey of the Kepler Field. I. A Search for Narrow-band Emission from Select Targets}",
      journal = {\apj},
         year = 2013,
        month = apr,
       volume = {767},
       number = {1},
          eid = {94},
        pages = {94},
          doi = {10.1088/0004-637X/767/1/94},
archivePrefix = {arXiv},
       eprint = {1302.0845},
 primaryClass = {astro-ph.GA},
       adsurl = {https://ui.adsabs.harvard.edu/abs/2013ApJ...767...94S}
}

@ARTICLE{Fan_2025,
       author = {{Fan}, Pinchen and {Wright}, Jason T. and {Lazio}, T. Joseph W.},
        title = "{Detecting Extraterrestrial Civilizations that Employ an Earth-level Deep Space Network}",
      journal = {\apjl},
         year = 2025,
        month = sep,
       volume = {990},
       number = {1},
          eid = {L1},
        pages = {L1},
          doi = {10.3847/2041-8213/adf6b0},
archivePrefix = {arXiv},
       eprint = {2508.15425},
 primaryClass = {astro-ph.IM},
       adsurl = {https://ui.adsabs.harvard.edu/abs/2025ApJ...990L...1F}
}

@article{LRO_Camera2010,
	author = {Robinson, M. S. and Brylow, S. M. and Tschimmel, M. and Humm, D. and Lawrence, S. J. and Thomas, P. C. and Denevi, B. W. and Bowman-Cisneros, E. and Zerr, J. and Ravine, M. A. and Caplinger, M. A. and Ghaemi, F. T. and Schaffner, J. A. and Malin, M. C. and Mahanti, P. and Bartels, A. and Anderson, J. and Tran, T. N. and Eliason, E. M. and McEwen, A. S. and Turtle, E. and Jolliff, B. L. and Hiesinger, H.},
	date = {2010/01/01},
	doi = {10.1007/s11214-010-9634-2},
	id = {Robinson2010},
	isbn = {1572-9672},
	journal = {Space Science Reviews},
	number = {1},
	pages = {81--124},
	title = {Lunar Reconnaissance Orbiter Camera (LROC) Instrument Overview},
	url = {https://doi.org/10.1007/s11214-010-9634-2,https://lroc.im-ldi.com/images/298},
	volume = {150},
	year = {2010}}

@misc{nasa2026clps,
  author       = {{National Aeronautics and Space Administration}},
  title        = {{Commercial Lunar Payload Services}},
  year         = {2026},
  month        = mar,
  day          = {24},
  howpublished = {\url{https://www.nasa.gov/commercial-lunar-payload-services/}},
  note         = {Page last updated March 24, 2026. Accessed May 15, 2026}
}

@techreport{neal2020lunarGeophysicalNetwork,
  author      = {Neal, Clive R. and Amato, M.},
  title       = {The Lunar Geophysical Network},
  institution = {NASA},
  type        = {Technical report},
  year        = {2020},
  month       = aug,
  day         = {10},
  url         = {https://ilrs.gsfc.nasa.gov/docs/2020/Lunar_Geophysical_Network_August_10-2020.pdf},
  note        = {Accessed: 2026-05-12}
}

@ARTICLE{Suresh_2021,
       author = {{Suresh}, Akshay and {Cordes}, James M. and {Chatterjee}, Shami and {Gajjar}, Vishal and {Perez}, Karen I. and {Siemion}, Andrew P.~V. and {Price}, Danny C.},
        title = "{4-8 GHz Spectrotemporal Emission from the Galactic Center Magnetar PSR J1745-2900}",
      journal = {ApJ},
         year = 2021,
        month = nov,
       volume = {921},
       number = {2},
          eid = {101},
        pages = {101},
          doi = {10.3847/1538-4357/ac1d45},
archivePrefix = {arXiv},
       eprint = {2108.05404},
 primaryClass = {astro-ph.HE},
       adsurl = {https://ui.adsabs.harvard.edu/abs/2021ApJ...921..101S}
}

@article{wright_cosmic_haystack,
	doi = {10.3847/1538-3881/aae099},
  
	url = {https://doi.org/10.3847%2F1538-3881%2Faae099},
  
	year = 2018,
	month = {nov},
  
	publisher = {American Astronomical Society},
  
	volume = {156},
  
	number = {6},
  
	pages = {260},
  
	author = {Jason T. Wright and Shubham Kanodia and Emily Lubar},
  
	title = {How Much {SETI} Has Been Done? Finding Needles in the $\less$i$\greater$n$\less$/i$\greater$-dimensional Cosmic Haystack},
  
	journal = {The Astronomical Journal}
}

@ARTICLE{tremblay_k218b,
      title={A Narrowband Technosignature Search Toward the Hycean Candidate K2-18b Using the VLA and MeerKAT}, 
      author={C. D. Tremblay and S. Chaudhary and Megan G. Li and Sofia Z. Sheikh and T. Myburgh and D. Czech and D. E. MacMahon and P. B. Demorest and R. A. Donnachie and A. P. V. Siemion and V. Gajjar and M. Lebofsky and K. Wandia and K. I. Perez. and Nikku Madhusudhan},
      journal = {The Astronomical Journal},
      year={2026},
      eprint={2602.09553},
      archivePrefix={arXiv},
      primaryClass={astro-ph.EP},
      url={https://arxiv.org/abs/2602.09553}
}

@inproceedings{warnick2025modified,
  title={Modified Wideband Blass Matrix Beamformer for the Lunar Farside Technosignatures and Transients Telescope (LFT3)},
  author={Haymore, Rebecca and Kemp, Daniel and Warnick, Karl F. and David R.\ DeBoer},
  booktitle={Proc. 2026 IEEE International Symposium on Antennas and Propagation (AP-S), Detroit, MI, 12-17 July},
  pages={},
  year={2026},
  organization={IEEE}
}

@unpublished{SlosarSpectromer2026,
  title   = {LuSEE-Night: Analog and Digital Spectrometer Subsystem},
  author  = {{LuSEE-Night Collaboration}},
  journal = {Space Science Review},
  year    = {2026},
  note    = {in preparation}
}

@INPROCEEDINGS{Slater2023,
  author={Slater, Windy S. and Rutherford, Benjamin B.W. and Mee, Jesse K. and Pinson, Ryan E. and Gruber, Matthew and Sabogal, Daniel and Troxel, Ian A.},
  booktitle={2023 IEEE Radiation Effects Data Workshop (REDW) (in conjunction with 2023 NSREC)}, 
  title={Single Event Effects and Total Ionizing Dose Radiation Testing of NVIDIA Jetson Orin AGX System on Module}, 
  year={2023},
  volume={},
  number={},
  pages={1-6},
  doi={10.1109/REDW61050.2023.10265818}}

@article{Koribalski_2020,
  author = {Koribalski, B. S. et al.},
  title = {The WALLABY survey: science goals and early results},
  journal = {Ap\&SS},
  year = {2020},
  volume = {365},
  pages = {118}
}

@article{Meyer_2020,
  author = {Meyer, M. et al.},
  title = {The DINGO survey design and science goals},
  journal = {PASA},
  year = {2020},
  volume = {37},
}

@ARTICLE{Dame-2001,
   author = {{Dame}, T.~M. and {Hartmann}, D. and {Thaddeus}, P.},
    title = "{The Milky Way in Molecular Clouds: A New Complete CO Survey}",
  journal = {\apj},
   eprint = {astro-ph/0009217},
     year = 2001,
    month = feb,
   volume = 547,
    pages = {792-813},
      doi = {10.1086/318388},
   adsurl = {http://adsabs.harvard.edu/abs/2001ApJ...547..792D}
}

@ARTICLE{2025ApJ...984L..24K,
       author = {{Keller}, Aya and {Wolff}, Nicole and {van Bibber}, Karl},
        title = "{A Model-independent Radio Telescope Dark Matter Search in the L and S Bands}",
      journal = {\apjl},
         year = 2025,
        month = may,
       volume = {984},
       number = {1},
          eid = {L24},
        pages = {L24},
          doi = {10.3847/2041-8213/adc9aa},
       adsurl = {https://ui.adsabs.harvard.edu/abs/2025ApJ...984L..24K}
}

@ARTICLE{Jacob_2024,
       author = {{Jacob}, Arshia M. and {Nandakumar}, Meera and {Roy}, Nirupam and {Menten}, Karl M. and {Neufeld}, David A. and {Faure}, Alexandre and {Tiwari}, Maitraiyee and {Pillai}, Thushara G.~S. and {Robishaw}, Timothy and {Dur{\'a}n}, Carlos A.},
        title = "{Revisiting rotationally excited CH at radio wavelengths: A case study towards W51}",
      journal = {\aap},
         year = 2024,
        month = dec,
       volume = {692},
          eid = {A164},
        pages = {A164},
          doi = {10.1051/0004-6361/202449603},
archivePrefix = {arXiv},
       eprint = {2411.08193},
 primaryClass = {astro-ph.GA},
       adsurl = {https://ui.adsabs.harvard.edu/abs/2024A&A...692A.164J}
}

@ARTICLE{Tremblay_2017,
       author = {{Tremblay}, Chenoa D. and {Hurley-Walker}, Natasha and {Cunningham}, Maria and {Jones}, Paul A. and {Hancock}, Paul J. and {Wayth}, Randall and {Jordan}, Christopher H.},
        title = "{A first look for molecules between 103 and 133 MHz using the Murchison Widefield Array}",
      journal = {\mnras},
         year = 2017,
        month = nov,
       volume = {471},
       number = {4},
        pages = {4144-4154},
          doi = {10.1093/mnras/stx1838},
archivePrefix = {arXiv},
       eprint = {1707.06009},
 primaryClass = {astro-ph.GA},
       adsurl = {https://ui.adsabs.harvard.edu/abs/2017MNRAS.471.4144T}
}

@ARTICLE{2011ApJ...734....4K,
       author = {{Kogut}, A. and {Fixsen}, D.~J. and {Levin}, S.~M. and {Limon}, M. and {Lubin}, P.~M. and {Mirel}, P. and {Seiffert}, M. and {Singal}, J. and {Villela}, T. and {Wollack}, E. and {Wuensche}, C.~A.},
        title = "{ARCADE 2 Observations of Galactic Radio Emission}",
      journal = {\apj},
         year = 2011,
        month = jun,
       volume = {734},
       number = {1},
          eid = {4},
        pages = {4},
          doi = {10.1088/0004-637X/734/1/4},
archivePrefix = {arXiv},
       eprint = {0901.0562},
 primaryClass = {astro-ph.GA},
       adsurl = {https://ui.adsabs.harvard.edu/abs/2011ApJ...734....4K}
}

@ARTICLE{2018ApJ...858L...9D,
       author = {{Dowell}, Jayce and {Taylor}, Greg B.},
        title = "{The Radio Background below 100 MHz}",
      journal = {\apjl},
         year = 2018,
        month = may,
       volume = {858},
       number = {1},
          eid = {L9},
        pages = {L9},
          doi = {10.3847/2041-8213/aabf86},
archivePrefix = {arXiv},
       eprint = {1804.08581},
 primaryClass = {astro-ph.CO},
       adsurl = {https://ui.adsabs.harvard.edu/abs/2018ApJ...858L...9D}
}

@ARTICLE{Tremblay_2020_NO,
       author = {{Tremblay}, C.~D. and {Gray}, M.~D. and {Hurley-Walker}, N. and {Green}, J.~A. and {Dawson}, J.~R. and {Dickey}, J.~M. and {Jones}, P.~A. and {Tingay}, S.~J. and {Wong}, O.~I.},
        title = "{Nitric Oxide and Other Molecules: Molecular Modeling and Low-frequency Exploration Using the Murchison Widefield Array}",
      journal = {\apj},
         year = 2020,
        month = dec,
       volume = {905},
       number = {1},
          eid = {65},
        pages = {65},
          doi = {10.3847/1538-4357/abc33a},
archivePrefix = {arXiv},
       eprint = {2010.09868},
 primaryClass = {astro-ph.GA},
       adsurl = {https://ui.adsabs.harvard.edu/abs/2020ApJ...905...65T}
}

@ARTICLE{2022SciA....8J3618C,
       author = {{Chadha-Day}, Francesca and {Ellis}, John and {Marsh}, David J.~E.},
        title = "{Axion dark matter: What is it and why now?}",
      journal = {Science Advances},
         year = 2022,
        month = feb,
       volume = {8},
       number = {8},
          eid = {eabj3618},
        pages = {eabj3618},
          doi = {10.1126/sciadv.abj3618},
archivePrefix = {arXiv},
       eprint = {2105.01406},
 primaryClass = {hep-ph},
       adsurl = {https://ui.adsabs.harvard.edu/abs/2022SciA....8J3618C}
}

@ARTICLE{2024PhRvD.110f3013A,
       author = {{Adachi}, Shunsuke and {Adkins}, Tylor and {Baccigalupi}, Carlo and {Chinone}, Yuji and {Crowley}, Kevin T. and {Errard}, Josquin and {Fabbian}, Giulio and {Feng}, Chang and {Fujino}, Takuro and {Hasegawa}, Masaya and {Hazumi}, Masashi and {Jeong}, Oliver and {Kaneko}, Daisuke and {Keating}, Brian and {Kusaka}, Akito and {Lee}, Adrian T. and {Lonappan}, Anto I. and {Minami}, Yuto and {Murata}, Masaaki and {Piccirillo}, Lucio and {Reichardt}, Christian L. and {Siritanasak}, Praween and {Spisak}, Jacob and {Takakura}, Satoru and {Teply}, Grant P. and {Yamada}, Kyohei and {Polarbear Collaboration}},
        title = "{Exploration of the polarization angle variability of the Crab Nebula with POLARBEAR and its application to the search for axionlike particles}",
      journal = {\prd},
         year = 2024,
        month = sep,
       volume = {110},
       number = {6},
          eid = {063013},
        pages = {063013},
          doi = {10.1103/PhysRevD.110.063013},
archivePrefix = {arXiv},
       eprint = {2403.02096},
 primaryClass = {astro-ph.CO},
       adsurl = {https://ui.adsabs.harvard.edu/abs/2024PhRvD.110f3013A}
}

@ARTICLE{Rhee_2023,
       author = {{Rhee}, Jonghwan and {Meyer}, Martin and {Popping}, Attila and others},
        title = "{Deep investigation of neutral gas origins (DINGO): H I stacking experiments with early science data}",
      journal = {\mnras},
         year = 2023,
        month = jan,
       volume = {518},
       number = {3},
        pages = {4646-4671},
          doi = {10.1093/mnras/stac3065},
archivePrefix = {arXiv},
       eprint = {2210.09697},
 primaryClass = {astro-ph.GA},
       adsurl = {https://ui.adsabs.harvard.edu/abs/2023MNRAS.518.4646R}
}

@ARTICLE{Pingel_2022,
       author = {{Pingel}, N.~M. and {Dempsey}, J. and {McClure-Griffiths}, N.~M. and {Dickey}, J.~M. and others},
        title = "{GASKAP-HI pilot survey science I: ASKAP zoom observations of HI emission in the Small Magellanic Cloud}",
      journal = {\pasa},
         year = 2022,
        month = feb,
       volume = {39},
          eid = {e005},
        pages = {e005},
          doi = {10.1017/pasa.2021.59},
archivePrefix = {arXiv},
       eprint = {2111.05339},
 primaryClass = {astro-ph.GA},
       adsurl = {https://ui.adsabs.harvard.edu/abs/2022PASA...39....5P}
}

@ARTICLE{HI4PI,
       author = {{HI4PI Collaboration} and {Ben Bekhti}, N. and {Fl{\"o}er}, L. and {Keller}, R. and {Kerp}, J. and {Lenz}, D. and {Winkel}, B. and {Bailin}, J. and {Calabretta}, M.~R. and {Dedes}, L. and {Ford}, H.~A. and {Gibson}, B.~K. and {Haud}, U. and {Janowiecki}, S. and {Kalberla}, P.~M.~W. and {Lockman}, F.~J. and {McClure-Griffiths}, N.~M. and {Murphy}, T. and {Nakanishi}, H. and {Pisano}, D.~J. and {Staveley-Smith}, L.},
        title = "{HI4PI: A full-sky H I survey based on EBHIS and GASS}",
      journal = {\aap},
         year = 2016,
        month = oct,
       volume = {594},
          eid = {A116},
        pages = {A116},
          doi = {10.1051/0004-6361/201629178},
archivePrefix = {arXiv},
       eprint = {1610.06175},
 primaryClass = {astro-ph.GA},
       adsurl = {https://ui.adsabs.harvard.edu/abs/2016A&A...594A.116H}
}

@INPROCEEDINGS{Meiksin_2022,
       author = {{Meiksin}, A.},
        title = "{Signatures of HI in the Early Universe: The End of the Dark Ages}",
    booktitle = {The Universe at Low Radio Frequencies},
         year = 2002,
       editor = {{Pramesh Rao}, A. and {Swarup}, G. and {Gopal-Krishna}},
       series = {IAU Symposium},
       volume = {199},
        month = jan,
        pages = {71},
       adsurl = {https://ui.adsabs.harvard.edu/abs/2002IAUS..199...71M}
}

@ARTICLE{HW_2024,
       author = {{Hurley-Walker}, N. and {McSweeney}, S.~J. and {Bahramian}, A. and {Rea}, N. and {Horv{\'a}th}, C. and {Buchner}, S. and {Williams}, A. and {Meyers}, B.~W. and {Strader}, Jay and {Aydi}, Elias and {Urquhart}, Ryan and {Chomiuk}, Laura and {Galvin}, T.~J. and {Coti Zelati}, F. and {Bailes}, Matthew},
        title = "{A 2.9 hr Periodic Radio Transient with an Optical Counterpart}",
      journal = {\apjl},
         year = 2024,
        month = dec,
       volume = {976},
       number = {2},
          eid = {L21},
        pages = {L21},
          doi = {10.3847/2041-8213/ad890e},
archivePrefix = {arXiv},
       eprint = {2408.15757},
 primaryClass = {astro-ph.SR},
       adsurl = {https://ui.adsabs.harvard.edu/abs/2024ApJ...976L..21H}
}

@ARTICLE{Men_2025,
       author = {{Men}, Yunpeng and {McSweeney}, Sam and {Hurley-Walker}, Natasha and {Barr}, Ewan and {Stappers}, Ben},
        title = "{A highly magnetized long-period radio transient exhibiting unusual emission features}",
      journal = {Science Advances},
         year = 2025,
        month = jan,
       volume = {11},
       number = {3},
          eid = {eadp6351},
        pages = {eadp6351},
          doi = {10.1126/sciadv.adp6351},
archivePrefix = {arXiv},
       eprint = {2501.10528},
 primaryClass = {astro-ph.HE},
       adsurl = {https://ui.adsabs.harvard.edu/abs/2025SciA...11P6351M}
}

@ARTICLE{2024AJ....167....2V,
       author = {{Vydula}, Akshatha K. and {Bowman}, Judd D. and {Lewis}, David and {Crawford}, Kelsie and {Kolopanis}, Matthew and {Rogers}, Alan E.~E. and {Murray}, Steven G. and {Mahesh}, Nivedita and {Monsalve}, Raul A. and {Sims}, Peter and {Samson}, Titu},
        title = "{Low-frequency Radio Recombination Lines Away from the Inner Galactic Plane}",
      journal = {\aj},
         year = 2024,
        month = jan,
       volume = {167},
       number = {1},
          eid = {2},
        pages = {2},
          doi = {10.3847/1538-3881/ad08ba},
archivePrefix = {arXiv},
       eprint = {2302.14185},
 primaryClass = {astro-ph.GA},
       adsurl = {https://ui.adsabs.harvard.edu/abs/2024AJ....167....2V}
}

@ARTICLE{2019JCAP...02..059I,
       author = {{Ivanov}, M.~M. and {Kovalev}, Y.~Y. and {Lister}, M.~L. and {Panin}, A.~G. and {Pushkarev}, A.~B. and {Savolainen}, T. and {Troitsky}, S.~V.},
        title = "{Constraining the photon coupling of ultra-light dark-matter axion-like particles by polarization variations of parsec-scale jets in active galaxies}",
      journal = {\jcap},
         year = 2019,
        month = feb,
       volume = {2019},
       number = {2},
          eid = {059},
        pages = {059},
          doi = {10.1088/1475-7516/2019/02/059},
archivePrefix = {arXiv},
       eprint = {1811.10997},
 primaryClass = {astro-ph.CO},
       adsurl = {https://ui.adsabs.harvard.edu/abs/2019JCAP...02..059I}
}

@ARTICLE{2018PhRvL.121c1103P,
       author = {{Pospelov}, Maxim and {Pradler}, Josef and {Ruderman}, Joshua T. and {Urbano}, Alfredo},
        title = "{Room for New Physics in the Rayleigh-Jeans Tail of the Cosmic Microwave Background}",
      journal = {\prl},
         year = 2018,
        month = jul,
       volume = {121},
       number = {3},
          eid = {031103},
        pages = {031103},
          doi = {10.1103/PhysRevLett.121.031103},
archivePrefix = {arXiv},
       eprint = {1803.07048},
 primaryClass = {hep-ph},
       adsurl = {https://ui.adsabs.harvard.edu/abs/2018PhRvL.121c1103P}
}

@ARTICLE{2025PhRvL.134q1001A,
       author = {{An}, Haipeng and {Ge}, Shuailiang and {Liu}, Jia and {Liu}, Mingzhe},
        title = "{In Situ Measurements of Dark Photon Dark Matter Using Parker Solar Probe: Going beyond the Radio Window}",
      journal = {\prl},
         year = 2025,
        month = may,
       volume = {134},
       number = {17},
          eid = {171001},
        pages = {171001},
          doi = {10.1103/PhysRevLett.134.171001},
archivePrefix = {arXiv},
       eprint = {2405.12285},
 primaryClass = {hep-ph},
       adsurl = {https://ui.adsabs.harvard.edu/abs/2025PhRvL.134q1001A}
}

@ARTICLE{Gray_2020,
       author = {{Gray}, Robert H.},
        title = "{The Extended Kardashev Scale}",
      journal = {\aj},
         year = 2020,
        month = may,
       volume = {159},
       number = {5},
          eid = {228},
        pages = {228},
          doi = {10.3847/1538-3881/ab792b},
       adsurl = {https://ui.adsabs.harvard.edu/abs/2020AJ....159..228G}
}

@misc{burns2021lunarfarsidelowradio,
      title={A Lunar Farside Low Radio Frequency Array for Dark Ages 21-cm Cosmology}, 
      author={{Burns}, Jack and {Hallinan}, Gregg and {Chang}, Tzu-Ching and {Anderson}, Marin and {Bowman}, Judd and {Bradley}, Richard and {Furlanetto}, Steven and {Hegedus}, Alex and {Kasper}, Justin and {Kocz}, Jonathan and {Lazio}, Joseph and {Lux}, Jim and {MacDowall}, Robert and {Mirocha}, Jordan and {Nesnas}, Issa and {Pober} Jonathan and {Polidan}, Ronald and {Rapetti}, David and {Romero-Wolf}, Andres and {Slosar}, Anže and {Stebbins}, Albert and {Teitelbaum}, Lawrence and {White}, Martin},
      year=2021,
      eprint={2103.08623},
      archivePrefix={arXiv},
      primaryClass={astro-ph.IM},
      url={https://arxiv.org/abs/2103.08623}, 
}

@INPROCEEDINGS{9438165,
  author={Bandyopadhyay, Saptarshi and Mcgarey, Patrick and Goel, Ashish and Rafizadeh, Ramin and Delapierre, Melanie and Arya, Manan and Lazio, Joseph and Goldsmith, Paul and Chahat, Nacer and Stoica, Adrian and Quadrelli, Marco and Nesnas, Issa and Jenks, Kenneth and Hallinan, Gregg},
  booktitle={2021 IEEE Aerospace Conference (50100)}, 
  title={Conceptual Design of the Lunar Crater Radio Telescope (LCRT) on the Far Side of the Moon}, 
  year={2021},
  volume={},
  number={},
  pages={1-25},
  doi={10.1109/AERO50100.2021.9438165}}

@ARTICLE{2025arXiv250309842H,
       author = {{Hibbard}, Joshua J. and {Burns}, Jack O. and {MacDowall}, Robert and {Gopalswamy}, Natchimuthuk and {Boardsen}, Scott A. and {Farrell}, William and {Bradley}, Damon and {Schulszas}, Thomas M. and {Dorigo Jones}, Johnny and {Rapetti}, David and {Turner}, Jake D.},
        title = "{Results from NASA's First Radio Telescope on the Moon: Terrestrial Technosignatures and the Low-Frequency Galactic Background Observed by ROLSES-1 Onboard the Odysseus Lander}",
      journal = {arXiv e-prints},
         year = 2025,
        month = mar,
          eid = {arXiv:2503.09842},
        pages = {arXiv:2503.09842},
          doi = {10.48550/arXiv.2503.09842},
archivePrefix = {arXiv},
       eprint = {2503.09842},
 primaryClass = {astro-ph.IM},
       adsurl = {https://ui.adsabs.harvard.edu/abs/2025arXiv250309842H}
}

@INPROCEEDINGS {2023URSIGASSLuSEENight,
  author = {{Bale}, Stuart D. and {Bassett}, Neil and {Burns}, Jack O. and {Jones}, Johnny Dorigo and {Goetz}, Keith and {Hellum-Bye}, Christian and
{Herrmann}, Sven and {Hibbard}, Joshua and {Maksimovic}, Milan and {McLean}. Ryan and  {Monsalve}, Raul and {O’Connor}, Paul and {Parsons}, Aaron and {Pulupa}, Marc and {Pund}, Rugved and {Rapetti}, David and {Rotermund}, Kaja M. and {Saliwanchik}, Ben and
{Slosar}, Anže and {Sundkvist}, David and {Suzuki}, Aritoki},
  title = {LuSEE 'Night': The Lunar Surface Electromagnetics Experiment},
  booktitle = {URSI International Union of Radio Science General Assembly and Scientific Symposium URSI GASS, Sapporo, Japan},
  year = 2023,
  month = aug
  }

@INPROCEEDINGS{2023AGUFM.P31B..02B,
       author = {{Bale}, Stuart D.},
        title = "{The Lunar Surface Electromagnetics (LuSEE) payloads for NASA's CLPS program}",
    booktitle = {AGU Fall Meeting Abstracts},
         year = 2023,
       volume = {2023},
        month = dec,
          eid = {P31B-02},
        pages = {P31B-02},
       adsurl = {https://ui.adsabs.harvard.edu/abs/2023AGUFM.P31B..02B}
}

@ARTICLE{Lazio2004,
       author = {{Lazio}, W., T. Joseph and {Farrell}, W.~M. and {Dietrick}, Jill and {Greenlees}, Elizabeth and {Hogan}, Emily and {Jones}, Christopher and {Hennig}, L.~A.},
        title = "{The Radiometric Bode's Law and Extrasolar Planets}",
      journal = {\apj},
         year = 2004,
        month = sep,
       volume = {612},
       number = {1},
        pages = {511-518},
          doi = {10.1086/422449},
       adsurl = {https://ui.adsabs.harvard.edu/abs/2004ApJ...612..511L}
}

@ARTICLE{Lynch2018,
       author = {{Lynch}, C.~R. and {Murphy}, Tara and {Lenc}, E. and {Kaplan}, D.~L.},
        title = "{The detectability of radio emission from exoplanets}",
      journal = {\mnras},
         year = 2018,
        month = aug,
       volume = {478},
       number = {2},
        pages = {1763-1775},
          doi = {10.1093/mnras/sty1138},
archivePrefix = {arXiv},
       eprint = {1804.11006},
 primaryClass = {astro-ph.EP},
       adsurl = {https://ui.adsabs.harvard.edu/abs/2018MNRAS.478.1763L}
}

@misc{itu_rr_2024,
  author = {ITU},
  title = {Radio Regulations, Edition of 2024},
  organization = {International Telecommunication Union (ITU)},
  url = {https://www.itu.int/pub/r-reg},
  year = {2024},
  note = {Includes all Appendices, Resolutions, Recommendations and ITU-R Recommendations incorporated by reference.},
}

@ARTICLE{Hippke_2017,
       author = {{Hippke}, Michael and {Forgan}, Duncan H.},
        title = "{Interstellar communication. VI. Searching X-ray spectra for narrowband communication}",
      journal = {arXiv e-prints},
         year = 2017,
        month = dec,
          eid = {arXiv:1712.06639},
        pages = {arXiv:1712.06639},
          doi = {10.48550/arXiv.1712.06639},
archivePrefix = {arXiv},
       eprint = {1712.06639},
 primaryClass = {astro-ph.IM},
       adsurl = {https://ui.adsabs.harvard.edu/abs/2017arXiv171206639H}
}

@ARTICLE{2023ExA....56..333Y,
       author = {{Yan}, Jingye and {Wu}, Ji and {Gurvits}, Leonid I. and {Wu}, Lin and {Deng}, Li and {Zhao}, Fei and {Zhou}, Li and {Lan}, Ailan and {Fan}, Wenjie and {Yi}, Min and {Yang}, Yang and {Yang}, Zhen and {Wei}, Mingchuan and {Guo}, Jinsheng and {Qiu}, Shi and {Wu}, Fan and {Hu}, Chaoran and {Chen}, Xuelei and {Rothkaehl}, Hanna and {Morawski}, Marek},
        title = "{Ultra-low-frequency radio astronomy observations from a Seleno-centric orbit}",
      journal = {Experimental Astronomy},
         year = 2023,
        month = aug,
       volume = {56},
       number = {1},
        pages = {333-353},
          doi = {10.1007/s10686-022-09887-0},
archivePrefix = {arXiv},
       eprint = {2212.09590},
 primaryClass = {astro-ph.IM},
       adsurl = {https://ui.adsabs.harvard.edu/abs/2023ExA....56..333Y}
}

@ARTICLE{2022RNAAS...6..197P,
       author = {{Perez}, Karen I. and {Farah}, Wael and {Sheikh}, Sofia Z. and {Croft}, Steve and {Siemion}, Andrew and {Pollak}, Alexander W. and {Brzycki}, Bryan and {Cruz}, Luigi F. and {Czech}, Daniel and {DeBoer}, David and {Drew}, Jamie and {Gajjar}, Vishal and {Garrett}, Michael A. and {Isaacson}, Howard and {Lebofsky}, Matt and {MacMahon}, David H.~E. and {Premnath}, Pranav H. and {de Pater}, Imke and {Price}, Danny C. and {Schoultz}, Sarah and {Wlodarczyk-Sroka}, Bart S. and {Tarter}, Jill and {Worden}, S. Pete},
        title = "{Breakthrough Listen Search for the WOW! Signal}",
      journal = {Research Notes of the American Astronomical Society},
         year = 2022,
        month = sep,
       volume = {6},
       number = {9},
          eid = {197},
        pages = {197},
          doi = {10.3847/2515-5172/ac9408},
       adsurl = {https://ui.adsabs.harvard.edu/abs/2022RNAAS...6..197P}
}

@ARTICLE{Sheikh_2025,
       author = {{Sheikh}, Sofia Z. and {Huston}, Macy J. and {Fan}, Pinchen and {Wright}, Jason T. and {Beatty}, Thomas and {Martini}, Connor and {Kopparapu}, Ravi and {Frank}, Adam},
        title = "{Earth Detecting Earth: At What Distance Could Earth's Constellation of Technosignatures Be Detected with Present-day Technology?}",
      journal = {\aj},
         year = 2025,
        month = feb,
       volume = {169},
       number = {2},
          eid = {118},
        pages = {118},
          doi = {10.3847/1538-3881/ada3c7},
archivePrefix = {arXiv},
       eprint = {2502.02614},
 primaryClass = {astro-ph.IM},
       adsurl = {https://ui.adsabs.harvard.edu/abs/2025AJ....169..118S}
}

@ARTICLE{Saide_2023,
       author = {{Saide}, Ramiro C. and {Garrett}, M.~A. and {Heeralall-Issur}, N.},
        title = "{Simulation of the Earth's radio-leakage from mobile towers as seen from selected nearby stellar systems}",
      journal = {\mnras},
         year = 2023,
        month = jun,
       volume = {522},
       number = {2},
        pages = {2393-2402},
          doi = {10.1093/mnras/stad378},
archivePrefix = {arXiv},
       eprint = {2304.13779},
 primaryClass = {astro-ph.EP},
       adsurl = {https://ui.adsabs.harvard.edu/abs/2023MNRAS.522.2393S}
}

@techreport{prabu2025lft3,
  author       = {Steve Prabu and David DeBoer and Charlie Ashe},
  title        = {LFT3 Science Operations},
  institution  = {LFT3 Project},
  type         = {Memo},
  number       = {2},
  year         = {2025},
  note         = {Memo Number 2},
}

@techreport{lft3StarCatalog,
  author       = {Steve Prabu and David DeBoer and Andrew Siemion},
  title        = {LFT3 Star Catalog},
  institution  = {LFT3 Project},
  type         = {Memo},
  number       = {3},
  year         = {2025},
  note         ={Memo Number 3}
}

@techreport{whitepaper,
  author       = {David R. DeBoer and Charlie K. Ashe and Owen A. Johnson and Evan F. Keane and Andrew C. Lesh and Ella J. Marshall and Steve Prabu and Kaia L. Reenock and Anze Slosar and Chenoa D. Tremblay and Jake D. Turner and Karl F. Warnick and Andrew P. V. Siemion and Jamie Drew and S. Pete Worden},
  title        = {Lunar Farside Technosignature and Transients Telescope (LFT3)},
  institution  = {LFT3 Project},
  type         = {Memo},
  number       = {1},
  year         = {2025},
  note         = {Memo Number 1}
}

@ARTICLE{Haqq_2025,
       author = {{Haqq-Misra}, Jacob and {Vidal}, Cl{\'e}ment and {Profitiliotis}, George},
        title = "{Projections of Earth's technosphere: Luminosity and mass as limits to growth}",
      journal = {Acta Astronautica},
         year = 2025,
        month = apr,
       volume = {229},
        pages = {831-838},
          doi = {10.1016/j.actaastro.2025.01.048},
archivePrefix = {arXiv},
       eprint = {2410.23420},
 primaryClass = {astro-ph.EP},
       adsurl = {https://ui.adsabs.harvard.edu/abs/2025AcAau.229..831H}
}

@ARTICLE{2012JGRE..117.0H18V,
       author = {{Vasavada}, Ashwin R. and {Bandfield}, Joshua L. and {Greenhagen}, Benjamin T. and {Hayne}, Paul O. and {Siegler}, Matthew A. and {Williams}, Jean-Pierre and {Paige}, David A.},
        title = "{Lunar equatorial surface temperatures and regolith properties from the Diviner Lunar Radiometer Experiment}",
      journal = {Journal of Geophysical Research (Planets)},
         year = 2012,
        month = apr,
       volume = {117},
          eid = {E00H18},
        pages = {E00H18},
          doi = {10.1029/2011JE003987},
       adsurl = {https://ui.adsabs.harvard.edu/abs/2012JGRE..117.0H18V}
}

@article{Ren2025ChangE6,
  title        = {Geological Characteristics of Chang’E-6 Landing Area in Micro-scale Unveiled by New Observation Data},
  author       = {Xin Ren and Wei Yan and Xingguo Zeng and Wangli Chen and Xingye Gao and Wei Zuo and Bin Liu and Zhoubin Zhang and Qiang Fu and Jianjun Liu and Chunlai Li},
  journal      = {Nature Communications},
  year         = {2025},
  volume       = {16},
  number       = {},
  pages        = {Article number: 59443},
  doi          = {10.1038/s41467-025-59443-5},
  url          = {https://doi.org/10.1038/s41467-025-59443-5},
  publisher    = {Springer Nature},
  month        = {May},
  day          = {6}
}

@article{wimmer2020lnd,
  title={The Lunar Lander Neutron and Dosimetry (LND) Experiment on Chang’E 4},
  author={Wimmer-Schweingruber, Robert F and Yu, Jia and Böttcher, Stephan I and Zhang, Shenyi and Burmeister, Sönke and Lohf, Henning and Guo, Jingnan and Xu, Zigong and Schuster, Björn and Seimetz, Lars and others},
  journal={arXiv preprint arXiv:2001.11028},
  year={2020},
  url={https://arxiv.org/abs/2001.11028}
}

@ARTICLE{2021arXiv210305085B,
       author = {{Burns}, Jack and {Bale}, Stuart and {Bradley}, Richard and {Ahmed}, Z. and {Allen}, S.~W. and {Bowman}, J. and {Furlanetto}, S. and {MacDowall}, R. and {Mirocha}, J. and {Nhan}, B. and {Pivovaroff}, M. and {Pulupa}, M. and {Rapetti}, D. and {Slosar}, A. and {Tauscher}, K.},
        title = "{Global 21-cm Cosmology from the Farside of the Moon}",
      journal = {arXiv e-prints},
         year = 2021,
        month = mar,
          eid = {arXiv:2103.05085},
        pages = {arXiv:2103.05085},
          doi = {10.48550/arXiv.2103.05085},
archivePrefix = {arXiv},
       eprint = {2103.05085},
 primaryClass = {astro-ph.CO},
       adsurl = {https://ui.adsabs.harvard.edu/abs/2021arXiv210305085B}
}

@ARTICLE{2023arXiv230110345B,
       author = {{Bale}, Stuart D. and {Bassett}, Neil and {Burns}, Jack O. and {Dorigo Jones}, Johnny and {Goetz}, Keith and {Hellum-Bye}, Christian and {Hermann}, Sven and {Hibbard}, Joshua and {Maksimovic}, Milan and {McLean}, Ryan and {Monsalve}, Raul and {O'Connor}, Paul and {Parsons}, Aaron and {Pulupa}, Marc and {Pund}, Rugved and {Rapetti}, David and {Rotermund}, Kaja M. and {Saliwanchik}, Ben and {Slosar}, Anze and {Sundkvist}, David and {Suzuki}, Aritoki},
        title = "{LuSEE 'Night': The Lunar Surface Electromagnetics Experiment}",
      journal = {arXiv e-prints},
         year = 2023,
        month = jan,
          eid = {arXiv:2301.10345},
        pages = {arXiv:2301.10345},
          doi = {10.48550/arXiv.2301.10345},
archivePrefix = {arXiv},
       eprint = {2301.10345},
 primaryClass = {astro-ph.IM},
       adsurl = {https://ui.adsabs.harvard.edu/abs/2023arXiv230110345B}
}

@Book{Gordon-RRL,
    title = "{Radio Recombination Lines}",
booktitle = {Astrophysics and Space Science Library},
     year = 2009,
   series = {Astrophysics and Space Science Library},
   volume = 282,
   editor = {{Gordon}, M.~A. and {Sorochenko}, R.~L.},
      doi = {10.1007/978-0-387-09604-9},
   adsurl = {http://adsabs.harvard.edu/abs/2009ASSL..282.....G}
}

@ARTICLE{Fialkov_2024,
    author = {{Fialkov}, A. and {Gessey-Jones}, T. and {Dhanda}, J.},
    title = {Cosmic Mysteries and the Hydrogen 21-cm Line: Bridging the Gap with Lunar Observations},
    journal = {Philosophical Transactions of the Royal Society A: Mathematical, Physical and Engineering Sciences},
    year = 2024,
    month = may,
    volume = {382},
    number = {2271},
    doi = {10.1098/rsta.2023.0068}
}

@article{Zawdie_2017RS006256,
author = {Zawdie, K. A. and Drob, D. P. and Siskind, D. E. and Coker, C.},
title = {Calculating the absorption of HF radio waves in the ionosphere},
journal = {Radio Science},
volume = {52},
number = {6},
pages = {767-783},
doi = {https://doi.org/10.1002/2017RS006256},
url = {https://agupubs.onlinelibrary.wiley.com/doi/abs/10.1002/2017RS006256},
eprint = {https://agupubs.onlinelibrary.wiley.com/doi/pdf/10.1002/2017RS006256},
year = {2017}
}

@ARTICLE{2017PASP..129d5001D,
       author = {{DeBoer}, David R. and {Parsons}, Aaron R. and {Aguirre}, James E. and {Alexander}, Paul and {Ali}, Zaki S. and {Beardsley}, Adam P. and {Bernardi}, Gianni and {Bowman}, Judd D. and {Bradley}, Richard F. and {Carilli}, Chris L. and {Cheng}, Carina and {de Lera Acedo}, Eloy and {Dillon}, Joshua S. and {Ewall-Wice}, Aaron and {Fadana}, Gcobisa and {Fagnoni}, Nicolas and {Fritz}, Randall and {Furlanetto}, Steve R. and {Glendenning}, Brian and {Greig}, Bradley and {Grobbelaar}, Jasper and {Hazelton}, Bryna J. and {Hewitt}, Jacqueline N. and {Hickish}, Jack and {Jacobs}, Daniel C. and {Julius}, Austin and {Kariseb}, MacCalvin and {Kohn}, Saul A. and {Lekalake}, Telalo and {Liu}, Adrian and {Loots}, Anita and {MacMahon}, David and {Malan}, Lourence and {Malgas}, Cresshim and {Maree}, Matthys and {Martinot}, Zachary and {Mathison}, Nathan and {Matsetela}, Eunice and {Mesinger}, Andrei and {Morales}, Miguel F. and {Neben}, Abraham R. and {Patra}, Nipanjana and {Pieterse}, Samantha and {Pober}, Jonathan C. and {Razavi-Ghods}, Nima and {Ringuette}, Jon and {Robnett}, James and {Rosie}, Kathryn and {Sell}, Raddwine and {Smith}, Craig and {Syce}, Angelo and {Tegmark}, Max and {Thyagarajan}, Nithyanandan and {Williams}, Peter K.~G. and {Zheng}, Haoxuan},
        title = "{Hydrogen Epoch of Reionization Array (HERA)}",
      journal = {\pasp},
         year = 2017,
        month = apr,
       volume = {129},
       number = {974},
        pages = {045001},
          doi = {10.1088/1538-3873/129/974/045001},
archivePrefix = {arXiv},
       eprint = {1606.07473},
 primaryClass = {astro-ph.IM},
       adsurl = {https://ui.adsabs.harvard.edu/abs/2017PASP..129d5001D}
}

@ARTICLE{GDIGS,
       author = {{Anderson}, L.~D. and {Luisi}, Matteo and {Liu}, Bin and {Wenger}, Trey V. and {Balser}, Dana. S. and {Bania}, T.~M. and {Haffner}, L.~M. and {Linville}, Dylan J. and {Mascoop}, J.~L.},
        title = "{The GBT Diffuse Ionized Gas Survey (GDIGS): Survey Overview and First Data Release}",
      journal = {\apjs},
         year = 2021,
        month = jun,
       volume = {254},
       number = {2},
          eid = {28},
        pages = {28},
          doi = {10.3847/1538-4365/abef65},
archivePrefix = {arXiv},
       eprint = {2103.10466},
 primaryClass = {astro-ph.GA},
       adsurl = {https://ui.adsabs.harvard.edu/abs/2021ApJS..254...28A}
}

@ARTICLE{multibeam,
       author = {{Luan}, Xiao-Hang and {Tao}, Zhen-Zhao and {Zhao}, Hai-Chen and {Huang}, Bo-Lun and {Li}, Shi-Yu and {Liu}, Cong and {Wang}, Hong-Feng and {Liu}, Wen-Fei and {Zhang}, Tong-Jie and {Gajjar}, Vishal and {Werthimer}, Dan},
        title = "{Multibeam Blind Search of Targeted SETI Observations toward 33 Exoplanet Systems with FAST}",
      journal = {\aj},
         year = 2023,
        month = mar,
       volume = {165},
       number = {3},
          eid = {132},
        pages = {132},
          doi = {10.3847/1538-3881/acb706},
archivePrefix = {arXiv},
       eprint = {2301.10890},
 primaryClass = {astro-ph.EP},
       adsurl = {https://ui.adsabs.harvard.edu/abs/2023AJ....165..132L}
}

@ARTICLE{Wow_2024,
       author = {{M{\'e}ndez}, Abel and {Ortiz Ceballos}, Kevin and {Zuluaga}, Jorge I.},
        title = "{Arecibo Wow! I: An Astrophysical Explanation for the Wow! Signal}",
      journal = {arXiv e-prints},
         year = 2024,
        month = aug,
          eid = {arXiv:2408.08513},
        pages = {arXiv:2408.08513},
          doi = {10.48550/arXiv.2408.08513},
archivePrefix = {arXiv},
       eprint = {2408.08513},
 primaryClass = {astro-ph.HE},
       adsurl = {https://ui.adsabs.harvard.edu/abs/2024arXiv240808513M}
}

@ARTICLE{Hobbs_2020,
       author = {{Hobbs}, George and {Manchester}, Richard N. and {Dunning}, Alex and {Jameson}, Andrew and {Roberts}, Paul and {George}, Daniel and {Green}, J.~A. and {Tuthill}, John and {Toomey}, Lawrence and {Kaczmarek}, Jane F. and {Mader}, Stacy and {Marquarding}, Malte and {Ahmed}, Azeem and {Amy}, Shaun W. and {Bailes}, Matthew and {Beresford}, Ron and {Bhat}, N.~D.~R. and {Bock}, Douglas C. -J. and {Bourne}, Michael and {Bowen}, Mark and {Brothers}, Michael and {Cameron}, Andrew D. and {Carretti}, Ettore and {Carter}, Nick and {Castillo}, Santy and {Chekkala}, Raji and {Cheng}, Wan and {Chung}, Yoon and {Craig}, Daniel A. and {Dai}, Shi and {Dawson}, Joanne and {Dempsey}, James and {Doherty}, Paul and {Dong}, Bin and {Edwards}, Philip and {Ergesh}, Tuohutinuer and {Gao}, Xuyang and {Han}, JinLin and {Hayman}, Douglas and {Indermuehle}, Balthasar and {Jeganathan}, Kanapathippillai and {Johnston}, Simon and {Kanoniuk}, Henry and {Kesteven}, Michael and {Kramer}, Michael and {Leach}, Mark and {Mcintyre}, Vince and {Moss}, Vanessa and {Os{\l}owski}, Stefan and {Phillips}, Chris and {Pope}, Nathan and {Preisig}, Brett and {Price}, Daniel and {Reeves}, Ken and {Reilly}, Les and {Reynolds}, John and {Robishaw}, Tim and {Roush}, Peter and {Ruckley}, Tim and {Sadler}, Elaine and {Sarkissian}, John and {Severs}, Sean and {Shannon}, Ryan and {Smart}, Ken and {Smith}, Malcolm and {Smith}, Stephanie and {Sobey}, Charlotte and {Staveley-Smith}, Lister and {Tzioumis}, Anastasios and {van Straten}, Willem and {Wang}, Nina and {Wen}, Linqing and {Whiting}, Matthew},
        title = "{An ultra-wide bandwidth (704 to 4 032 MHz) receiver for the Parkes radio telescope}",
      journal = {\pasa},
         year = 2020,
        month = apr,
       volume = {37},
          eid = {e012},
        pages = {e012},
          doi = {10.1017/pasa.2020.2},
archivePrefix = {arXiv},
       eprint = {1911.00656},
 primaryClass = {astro-ph.IM},
       adsurl = {https://ui.adsabs.harvard.edu/abs/2020PASA...37...12H}
}

@ARTICLE{Offringa_RFI,
   author = {{Offringa}, A.~R. and {Wayth}, R.~B. and {Hurley-Walker}, N. and 
	{Kaplan}, D.~L. and {Barry}, N. and others},
    title = "{The Low-Frequency Environment of the Murchison Widefield Array: Radio-Frequency Interference Analysis and Mitigation}",
  journal = {\pasa},
archivePrefix = "arXiv",
   eprint = {1501.03946},
 primaryClass = "astro-ph.IM",
     year = 2015,
    month = mar,
   volume = 32,
      eid = {e008},
    pages = {e008},
      doi = {10.1017/pasa.2015.7},
   adsurl = {http://adsabs.harvard.edu/abs/2015PASA...32....8O}
}

@ARTICLE{Huang_2023,
       author = {{Huang}, Bo-Lun and {Tao}, Zhen-Zhao and {Zhang}, Tong-Jie},
        title = "{A Solution to Continuous RFI in Narrowband Radio SETI with FAST: The MultiBeam Point-source Scanning Strategy}",
      journal = {\aj},
         year = 2023,
        month = dec,
       volume = {166},
       number = {6},
          eid = {245},
        pages = {245},
          doi = {10.3847/1538-3881/ad06b1},
archivePrefix = {arXiv},
       eprint = {2307.11368},
 primaryClass = {astro-ph.IM},
       adsurl = {https://ui.adsabs.harvard.edu/abs/2023AJ....166..245H}
}

@ARTICLE{pietka,
       author = {{Pietka}, M. and {Fender}, R.~P. and {Keane}, E.~F.},
        title = "{The variability time-scales and brightness temperatures of radio flares from stars to supermassive black holes}",
      journal = {\mnras},
         year = 2015,
        month = feb,
       volume = {446},
       number = {4},
        pages = {3687-3696},
          doi = {10.1093/mnras/stu2335},
archivePrefix = {arXiv},
       eprint = {1411.1067},
 primaryClass = {astro-ph.HE},
       adsurl = {https://ui.adsabs.harvard.edu/abs/2015MNRAS.446.3687P}
}

@ARTICLE{Louis2023,
       author = {{Louis}, C.~K. and {Louarn}, P. and {Collet}, B. and {Cl{\'e}ment}, N. and {Al Saati}, S. and {Szalay}, J.~R. and {Hue}, V. and {Lamy}, L. and {Kotsiaros}, S. and {Kurth}, W.~S. and {Jackman}, C.~M. and {Wang}, Y. and {Blanc}, M. and {Allegrini}, F. and {Connerney}, J.~E.~P. and {Gershman}, D.},
        title = "{Source of Radio Emissions Induced by the Galilean Moons Io, Europa and Ganymede: In Situ Measurements by Juno}",
      journal = {Journal of Geophysical Research (Space Physics)},
         year = 2023,
        month = dec,
       volume = {128},
       number = {12},
          eid = {e2023JA031985},
        pages = {e2023JA031985},
          doi = {10.1029/2023JA031985},
archivePrefix = {arXiv},
       eprint = {2308.05541},
 primaryClass = {astro-ph.EP},
       adsurl = {https://ui.adsabs.harvard.edu/abs/2023JGRA..12831985L}
}

@ARTICLE{Goldreich1969,
       author = {{Goldreich}, P. and {Lynden-Bell}, D.},
        title = "{Io, a jovian unipolar inductor}",
      journal = {\apj},
         year = 1969,
        month = apr,
       volume = {156},
        pages = {59-78},
          doi = {10.1086/149947},
       adsurl = {https://ui.adsabs.harvard.edu/abs/1969ApJ...156...59G}
}

@ARTICLE{Kao2023,
       author = {{Kao}, Melodie M. and {Mioduszewski}, Amy J. and {Villadsen}, Jackie and {Shkolnik}, Evgenya L.},
        title = "{Resolved imaging confirms a radiation belt around an ultracool dwarf}",
      journal = {\nat},
         year = 2023,
        month = jul,
       volume = {619},
       number = {7969},
        pages = {272-275},
          doi = {10.1038/s41586-023-06138-w},
archivePrefix = {arXiv},
       eprint = {2302.12841},
 primaryClass = {astro-ph.EP},
       adsurl = {https://ui.adsabs.harvard.edu/abs/2023Natur.619..272K}
}

@ARTICLE{Imai2019,
       author = {{Imai}, Masafumi and {Lecacheux}, Alain and {Clarke}, Tracy E. and {Higgins}, Charles A. and {Panchenko}, Mykhaylo and {Zakharenko}, Vyacheslav V. and {Brazhenko}, Anatolii I. and {Frantsuzenko}, Anatolii V. and {Ivantyshin}, Oleg N. and {Konovalenko}, Alexandr A. and {Koshovyy}, Volodymyr V.},
        title = "{Concurrent Jovian S-Burst Beaming as Observed From LWA1, NDA, and Ukrainian Radio Telescopes}",
      journal = {Journal of Geophysical Research (Space Physics)},
         year = 2019,
        month = jul,
       volume = {124},
       number = {7},
        pages = {5302-5316},
          doi = {10.1029/2018JA026445},
       adsurl = {https://ui.adsabs.harvard.edu/abs/2019JGRA..124.5302I}
}

@INPROCEEDINGS{Wucknitz2024,
       author = {{Wucknitz}, O. and {Bassa}, C.~G. and {Bondonneau}, L. and {Girard}, J. and {Griessmeier}, J.-M. and {Keane}, E.~F. and {Koller}, J. and {Lamy}, L. and {Loh}, A. and {Louis}, C.~K. and {McCauley}, J. and {McKay}, D. and {Steinbergs}, J. and {Taylor}, C. and {Taylor}, G.~B. and {Vocks}, C. and {Zarka}, P.},
        title = "{Intercontinental decametric VLBI: Jupiter DAM observations with KAIRA, LOFAR, LWA and NenuFAR}",
    booktitle = {Proceedings of the 16th EVN Symposium},
         year = 2024,
       editor = {{Ros}, E. and {Benke}, P. and {Dzib}, S.~A. and {Rottmann}, I. and {Zensus}, J.~A.},
        month = sep,
        pages = {199-202},
       adsurl = {https://ui.adsabs.harvard.edu/abs/2024evn..conf..199W}
}

@ARTICLE{Kao2018,
       author = {{Kao}, Melodie M. and {Hallinan}, Gregg and {Pineda}, J. Sebastian and {Stevenson}, David and {Burgasser}, Adam},
        title = "{The Strongest Magnetic Fields on the Coolest Brown Dwarfs}",
      journal = {\apjs},
         year = 2018,
        month = aug,
       volume = {237},
       number = {2},
          eid = {25},
        pages = {25},
          doi = {10.3847/1538-4365/aac2d5},
archivePrefix = {arXiv},
       eprint = {1808.02485},
 primaryClass = {astro-ph.SR},
       adsurl = {https://ui.adsabs.harvard.edu/abs/2018ApJS..237...25K}
}

@ARTICLE{Kao2016,
       author = {{Kao}, Melodie M. and {Hallinan}, Gregg and {Pineda}, J. Sebastian and {Escala}, Ivanna and {Burgasser}, Adam and {Bourke}, Stephen and {Stevenson}, David},
        title = "{Auroral Radio Emission from Late L and T Dwarfs: A New Constraint on Dynamo Theory in the Substellar Regime}",
      journal = {\apj},
         year = 2016,
        month = feb,
       volume = {818},
       number = {1},
          eid = {24},
        pages = {24},
          doi = {10.3847/0004-637X/818/1/24},
archivePrefix = {arXiv},
       eprint = {1511.03661},
 primaryClass = {astro-ph.SR},
       adsurl = {https://ui.adsabs.harvard.edu/abs/2016ApJ...818...24K}
}

@ARTICLE{Zarka2007,
   author = {{Zarka}, P.},
    title = "{Plasma interactions of exoplanets with their parent star and associated radio emissions}",
  journal = {P$\&$SS},
     year = 2007,
    month = apr,
   volume = 55,
    pages = {598-617},
      doi = {10.1016/j.pss.2006.05.045},
   adsurl = {http://adsabs.harvard.edu/abs/2007P%26SS...55..598Z}
}

@ARTICLE{Zarka2008,
       author = {{Zarka}, P. and {Farrell}, W. and {Fischer}, G. and {Konovalenko}, A.},
        title = "{Ground-Based and Space-Based Radio Observations of Planetary Lightning}",
      journal = {\ssr},
         year = 2008,
        month = jun,
       volume = {137},
       number = {1-4},
        pages = {257-269},
          doi = {10.1007/s11214-008-9366-8},
       adsurl = {https://ui.adsabs.harvard.edu/abs/2008SSRv..137..257Z}
}

@ARTICLE{Kaiser1991,
       author = {{Kaiser}, M.~L. and {Zarka}, P. and {Desch}, M.~D. and {Farrell}, W.~M.},
        title = "{Restrictions on the characteristics of Neptunian lightning}",
      journal = {\jgr},
         year = 1991,
        month = oct,
       volume = {96},
        pages = {19043-19047},
          doi = {10.1029/91JA01599},
       adsurl = {https://ui.adsabs.harvard.edu/abs/1991JGR....9619043K}
}

@ARTICLE{Zhang2025,
       author = {{Zhang}, X. and {Zarka}, P. and {Girard}, J.~N. and {Tasse}, C. and {Loh}, A. and {Mauduit}, E. and {Mertens}, F.~G. and {Bonnassieux}, E. and {Louis}, C.~K. and {Grie{\ss}meier}, J-M. and {Turner}, J.~D. and {Lamy}, L. and {Strugarek}, A. and {Corbel}, S. and {Cecconi}, B. and {Konovalenko}, O. and {Zakharenko}, V. and {Ulyanov}, O. and {Tokarsky}, P. and {Tagger}, M.},
        title = "{A circularly polarized low-frequency radio burst from the exoplanetary system HD 189733}",
      journal = {arXiv e-prints},
         year = 2025,
        month = jun,
          eid = {arXiv:2506.07912},
        pages = {arXiv:2506.07912},
          doi = {10.48550/arXiv.2506.07912},
archivePrefix = {arXiv},
       eprint = {2506.07912},
 primaryClass = {astro-ph.EP},
       adsurl = {https://ui.adsabs.harvard.edu/abs/2025arXiv250607912Z}
}

@ARTICLE{1975A&A....40..365A,
       author = {{Alexander}, J.~K. and {Kaiser}, M.~L. and {Novaco}, J.~C. and {Grena}, F.~R. and {Weber}, R.~R.},
        title = "{Scientific instrumentation of the Radio-Astronomy-Explorer-2 satellite.}",
      journal = {\aap},
         year = 1975,
        month = may,
       volume = {40},
       number = {4},
        pages = {365-371},
       adsurl = {https://ui.adsabs.harvard.edu/abs/1975A&A....40..365A}
}

@ARTICLE{Turner2021_Radio,
       author = {{Turner}, Jake D. and {Zarka}, Philippe and {Grie{\ss}meier}, Jean-Mathias and {Lazio}, Joseph and {Cecconi}, Baptiste and {Emilio Enriquez}, J. and {Girard}, Julien N. and {Jayawardhana}, Ray and {Lamy}, Laurent and {Nichols}, Jonathan D. and {de Pater}, Imke},
        title = "{The search for radio emission from the exoplanetary systems 55 Cancri, {\ensuremath{\upsilon}} Andromedae, and {\ensuremath{\tau}} Bo{\"o}tis using LOFAR beam-formed observations}",
      journal = {A$\&$A},
         year = 2021,
        month = jan,
       volume = {645},
          eid = {A59},
        pages = {A59},
          doi = {10.1051/0004-6361/201937201},
archivePrefix = {arXiv},
       eprint = {2012.07926},
 primaryClass = {astro-ph.EP},
       adsurl = {https://ui.adsabs.harvard.edu/abs/2021A&A...645A..59T}
}

@ARTICLE{Turner2019,
       author = {{Turner}, Jake D. and {Grie{\ss}meier}, Jean-Mathias and
         {Zarka}, Philippe and {Vasylieva}, Iaroslavna},
        title = "{The search for radio emission from exoplanets using LOFAR beam-formed observations: Jupiter as an exoplanet}",
      journal = {\aap},
         year = "2019",
        month = "Apr",
       volume = {624},
          eid = {A40},
        pages = {A40},
          doi = {10.1051/0004-6361/201832848},
archivePrefix = {arXiv},
       eprint = {1802.07316},
 primaryClass = {astro-ph.EP},
   shorthand = {T19},
       adsurl = {https://ui.adsabs.harvard.edu/abs/2019A&A...624A..40T}
}

@inproceedings{Griessmeier17PREVIII,
   author    = {J-M. Grie{\ss}meier},
   title     = {The search for radio emission from giant exoplanets},
  editor    = {G. Fischer and G. Mann and M. Panchenko and P. Zarka},
   booktitle = {Planetary Radio Emissions VIII},
   year      = {2017},
       pages     = {285-300},
   publisher = {Austrian Academy of Sciences Press, Vienna},
}

@ARTICLE{Griessmeier2007_AA,
   author = {{Grie{\ss}meier}, J.-M. and {Zarka}, P. and {Spreeuw}, H.},
    title = "{Predicting low-frequency radio fluxes of known extrasolar planets}",
  journal = {A$\&$A},
archivePrefix = "arXiv",
   eprint = {0806.0327},
     year = 2007,
    month = nov,
   volume = 475,
    pages = {359-368},
      doi = {10.1051/0004-6361:20077397},
   adsurl = {http://adsabs.harvard.edu/abs/2007A%26A...475..359G}
}

@ARTICLE{Route2019,
       author = {{Route}, Matthew},
        title = "{The Rise of ROME. I. A Multiwavelength Analysis of the Star-Planet Interaction in the HD 189733 System}",
      journal = {ApJ},
         year = "2019",
        month = "Feb",
       volume = {872},
       number = {1},
          eid = {79},
        pages = {79},
          doi = {10.3847/1538-4357/aafc25},
archivePrefix = {arXiv},
       eprint = {1901.02048},
 primaryClass = {astro-ph.SR},
       adsurl = {https://ui.adsabs.harvard.edu/abs/2019ApJ...872...79R}
}

@ARTICLE{Turner2016a,
   author = {{Turner}, J.~D. and {Christie}, D. and {Arras}, P. and {Johnson}, R.~E. and 
	{Schmidt}, C.},
    title = "{Investigation of the environment around close-in transiting exoplanets using CLOUDY}",
  journal = {MNRAS},
archivePrefix = "arXiv",
   eprint = {1603.01229},
 primaryClass = "astro-ph.EP",
     year = 2016,
    month = jun,
   volume = 458,
    pages = {3880-3891},
      doi = {10.1093/mnras/stw556},
   adsurl = {http://adsabs.harvard.edu/abs/2016MNRAS.458.3880T}
}

@ARTICLE{Burns2021_RSPTA,
       author = {{Burns}, Jack O.},
        title = "{Transformative science from the lunar farside: observations of the dark ages and exoplanetary systems at low radio frequencies}",
      journal = {Philosophical Transactions of the Royal Society of London Series A},
         year = 2021,
        month = jan,
       volume = {379},
       number = {2188},
          eid = {20190564},
        pages = {20190564},
          doi = {10.1098/rsta.2019.0564},
archivePrefix = {arXiv},
       eprint = {2003.06881},
 primaryClass = {astro-ph.IM},
       adsurl = {https://ui.adsabs.harvard.edu/abs/2021RSPTA.37990564B}
}

@ARTICLE{Louis2025,
       author = {{Louis}, C.~K. and {Loh}, A. and {Zarka}, P. and {Lamy}, L. and {Mauduit}, E. and {Girard}, J.~N. and {N\textbackslash'enon}, Q.},
        title = "{Detection method for periodic radio emissions from an exoplanet's magnetosphere or a star-planet interaction}",
      journal = {arXiv e-prints},
         year = 2025,
        month = mar,
          eid = {arXiv:2503.18733},
        pages = {arXiv:2503.18733},
          doi = {10.48550/arXiv.2503.18733},
archivePrefix = {arXiv},
       eprint = {2503.18733},
 primaryClass = {astro-ph.EP},
       adsurl = {https://ui.adsabs.harvard.edu/abs/2025arXiv250318733L}
}

@ARTICLE{Lazio2019,
       author = {{Lazio}, Joseph and {Hallinan}, G. and {Airapetian}, A. and
         {Brain}, D.~A. and {Clarke}, T.~E. and {Dolch}, T. and {Dong}, C.~F. and
         {Driscoll}, P.~E. and {Fares}, R. and {Griessmeier}, J. -M. and
         {Farrell}, W.~M. and {Kasper}, J.~C. and {Murphy}, T. and
         {Rogers}, L.~A. and {Shkolnik}, E. and {Stanley}, S. and
         {Strugarek}, A. and {Turner}, N.~J. and {Wolszczan}, A. and
         {Zarka}, P. and {Knapp}, M. and {Lynch}, C.~R. and {Turner}, J.~D.},
        title = "{Magnetic Fields of Extrasolar Planets: Planetary Interiors and Habitability}",
      journal = {\baas},
         year = "2019",
        month = "May",
       volume = {51},
       number = {3},
          eid = {135},
        pages = {135},
archivePrefix = {arXiv},
       eprint = {1803.06487},
 primaryClass = {astro-ph.EP},
       adsurl = {https://ui.adsabs.harvard.edu/abs/2019BAAS...51c.135L}
}

@ARTICLE{Zarka2015SKA,
   author = {{Zarka}, P. and {Lazio}, J. and {Hallinan}, G.},
    title = "{Magnetospheric Radio Emissions from Exoplanets with the SKA}",
  journal = {Advancing Astrophysics with the Square Kilometre Array (AASKA14)},
     year = 2015,
    month = apr,
      eid = {120},
    pages = {120},
   adsurl = {http://adsabs.harvard.edu/abs/2015aska.confE.120Z}
}

@INPROCEEDINGS{G2015,
   author = {{Grie{\ss}meier}, J.-M.},
    title = "{Detection Methods and Relevance of Exoplanetary Magnetic Fields}",
booktitle = {Astrophysics and Space Science Library},
     year = 2015,
   series = {Astrophysics and Space Science Library},
   volume = 411,
   editor = {{Lammer}, H. and {Khodachenko}, M.},
    pages = {213},
      doi = {10.1007/978-3-319-09749-7_11},
   adsurl = {http://adsabs.harvard.edu/abs/2015ASSL..411..213G}
}

@article{WORDEN201798,
title = {Breakthrough Listen – A new search for life in the universe},
journal = {Acta Astronautica},
volume = {139},
pages = {98-101},
year = {2017},
issn = {0094-5765},
doi = {https://doi.org/10.1016/j.actaastro.2017.06.008},
url = {https://www.sciencedirect.com/science/article/pii/S0094576517303144},
author = {S. Pete Worden and Jamie Drew and Andrew Siemion and Dan Werthimer and David DeBoer and Steve Croft and David MacMahon and Matt Lebofsky and Howard Isaacson and Jack Hickish and Danny Price and Vishal Gajjar and Jason T. Wright}
}

@article{Yue2024Change6GeologicalContext,
  author  = {Yue, Zongyu and Gou, Sheng and Sun, Shujuan and Yang, Wei and Chen, Yi and Wang, Yexin and Lin, Honglei and Di, Kaichang and Lin, Yangting and Li, Xianhua and Wu, Fuyuan},
  title   = {Geological context of the Chang'e-6 landing area and implications for sample analysis},
  journal = {The Innovation},
  volume  = {5},
  number  = {5},
  pages   = {100663},
  year    = {2024},
  doi     = {10.1016/j.xinn.2024.100663},
  url     = {https://doi.org/10.1016/j.xinn.2024.100663},
  note    = {Published online June 24, 2024}
}

@article{Chen2022Change4Achievements,
  title   = {Overview of the Latest Scientific Achievements of Chang'E-4 Mission of China's Lunar Exploration Project},
  author  = {Chen, Yuesong and Tang, Yuhua and Fan, Yu and Yan, Jun and Wang, Chi and Zou, Yongliao},
  journal = {Chinese Journal of Space Science},
  volume  = {42},
  number  = {4},
  pages   = {519},
  year    = {2022},
  doi     = {10.11728/cjss2022.04.yg30},
  url     = {https://www.cjss.ac.cn/en/article/id/faeec271-94ed-49a8-9775-b7e1ed319159}
}

@techreport{NASA_CLPS_Brochure_2024,
  author      = {{NASA}},
  title       = {{Commercial Lunar Payload Services: CLPS Brochure 2024}},
  institution = {{National Aeronautics and Space Administration}},
  year        = {2024},
  url         = {https://www.nasa.gov/wp-content/uploads/2024/02/np-2023-12-019-jsc-clps-artemis-brochure-2024-web-2-14-24.pdf},
  note        = {NASA publication NP-2023-12-019-JSC}
}

@ARTICLE{2024A&A...689L..10B,
       author = {{Bassa}, C.~G. and {Di Vruno}, F. and {Winkel}, B. and {J{\'o}zsa}, G.~I.~G. and {Brentjens}, M.~A. and {Zhang}, X.},
        title = "{Bright unintended electromagnetic radiation from second-generation Starlink satellites}",
      journal = {\aap},
         year = 2024,
        month = sep,
       volume = {689},
          eid = {L10},
        pages = {L10},
          doi = {10.1051/0004-6361/202451856},
archivePrefix = {arXiv},
       eprint = {2409.11767},
 primaryClass = {astro-ph.IM},
       adsurl = {https://ui.adsabs.harvard.edu/abs/2024A&A...689L..10B}
}

@ARTICLE{Zarka2004,
       author = {{Zarka}, P. and {Farrell}, W.~M. and {Kaiser}, M.~L. and {Blanc}, E. and {Kurth}, W.~S.},
        title = "{Study of solar system planetary lightning with LOFAR}",
      journal = {\planss},
         year = 2004,
        month = dec,
       volume = {52},
       number = {15},
        pages = {1435-1447},
          doi = {10.1016/j.pss.2004.09.011},
       adsurl = {https://ui.adsabs.harvard.edu/abs/2004P&SS...52.1435Z}
}

@article{Desch1984,
author = {Desch, M. D. and Barrow, C. H.},
title = {Direct evidence for solar wind control of Jupiter’s hectometer-wavelength radio emission},
journal = {Journal of Geophysical Research: Space Physics},
volume = {89},
number = {A8},
pages = {6819-6823},
doi = {https://doi.org/10.1029/JA089iA08p06819},
url = {https://agupubs.onlinelibrary.wiley.com/doi/abs/10.1029/JA089iA08p06819},
eprint = {https://agupubs.onlinelibrary.wiley.com/doi/pdf/10.1029/JA089iA08p06819},
year = {1984}
}

@ARTICLE{Brain2024,
       author = {{Brain}, David A. and {Kao}, Melodie M. and {O'Rourke}, Joseph G.},
        title = "{Exoplanet Magnetic Fields}",
      journal = {arXiv e-prints},
         year = 2024,
        month = apr,
          eid = {arXiv:2404.15429},
        pages = {arXiv:2404.15429},
          doi = {10.48550/arXiv.2404.15429},
archivePrefix = {arXiv},
       eprint = {2404.15429},
 primaryClass = {astro-ph.EP},
       adsurl = {https://ui.adsabs.harvard.edu/abs/2024arXiv240415429B}
}

@ARTICLE{Lanza2009,
       author = {{Lanza}, A.~F.},
        title = "{Stellar coronal magnetic fields and star-planet interaction}",
      journal = {A$\&$A},
         year = "2009",
        month = "Oct",
       volume = {505},
       number = {1},
        pages = {339-350},
          doi = {10.1051/0004-6361/200912367},
archivePrefix = {arXiv},
       eprint = {0906.1738},
 primaryClass = {astro-ph.SR},
       adsurl = {https://ui.adsabs.harvard.edu/abs/2009A&A...505..339L}
}

@article{Cuntz2000,
	Adsurl = {http://adsabs.harvard.edu/abs/2000ApJ...533L.151C},
	Author = {{Cuntz}, M. and {Saar}, S.~H. and {Musielak}, Z.~E.},
	Doi = {10.1086/312609},
	Journal = {ApJl},
	Month = apr,
	Pages = {L151-L154},
	Title = {{On Stellar Activity Enhancement Due to Interactions with Extrasolar Giant Planets}},
	Volume = 533,
	Year = 2000}

@ARTICLE{Bastian_2001,
       author = {{Bastian}, T.~S. and {Pick}, M. and {Kerdraon}, A. and {Maia}, D. and {Vourlidas}, A.},
        title = "{The Coronal Mass Ejection of 1998 April 20: Direct Imaging at Radio Wavelengths}",
      journal = {\apjl},
         year = 2001,
        month = sep,
       volume = {558},
       number = {1},
        pages = {L65-L69},
          doi = {10.1086/323421},
       adsurl = {https://ui.adsabs.harvard.edu/abs/2001ApJ...558L..65B}
}

@ARTICLE{Kansabanik_2022,
       author = {{Kansabanik}, Devojyoti and {Oberoi}, Divya and {Mondal}, Surajit},
        title = "{Tackling the Unique Challenges of Low-frequency Solar Polarimetry with the Square Kilometre Array Low Precursor: The Algorithm}",
      journal = {\apj},
         year = 2022,
        month = jun,
       volume = {932},
       number = {2},
          eid = {110},
        pages = {110},
          doi = {10.3847/1538-4357/ac6758},
archivePrefix = {arXiv},
       eprint = {2204.04578},
 primaryClass = {astro-ph.SR},
       adsurl = {https://ui.adsabs.harvard.edu/abs/2022ApJ...932..110K}
}

@ARTICLE{Tremblay_2024,
       author = {{Tremblay}, C.~D. and {Varghese}, S.~S. and {Hickish}, J. and {Demorest}, P.~B. and {Ng}, C. and {Siemion}, A.~P.~V. and {Czech}, D. and {Donnachie}, R.~A. and {Farah}, W. and {Gajjar}, V. and {Lebofsky}, M. and {MacMahon}, D.~H.~E. and {Myburgh}, T. and {Ruzindana}, M. and {Bright}, J.~S. and {Erickson}, A. and {Lacker}, K.},
        title = "{COSMIC: An Ethernet-based Commensal, Multimode Digital Backend on the Karl G. Jansky Very Large Array for the Search for Extraterrestrial Intelligence}",
      journal = {\aj},
         year = 2024,
        month = jan,
       volume = {167},
       number = {1},
          eid = {35},
        pages = {35},
          doi = {10.3847/1538-3881/ad0fe0},
archivePrefix = {arXiv},
       eprint = {2310.09414},
 primaryClass = {astro-ph.IM},
       adsurl = {https://ui.adsabs.harvard.edu/abs/2024AJ....167...35T}
}

@inproceedings{christensen2009jmars,
  title={JMARS—A Planetary GIS},
  author={Christensen, Philip R. and Engle, Eric and Anwar, Syed and Dickenshied, Stephanie and Noss, Daniel and Gorelick, Noel and Weiss-Malik, Michael},
  booktitle={AGU Fall Meeting Abstracts},
  volume={2009},
  pages={IN22A--06},
  year={2009}
}

@webpage{wow,
        author = {Jerry Ehman},
        month = {2},
        title = {The Big Ear Wow! Signal},
        url = {http://www.bigear.org/wow20th.htm},
        urldate = {2/20/2008},
        year = {2008}}

@ARTICLE{BC_2020,
       author = {{Bochenek}, C.~D. and {Ravi}, V. and {Belov}, K.~V. and {Hallinan}, G. and {Kocz}, J. and {Kulkarni}, S.~R. and {McKenna}, D.~L.},
        title = "{A fast radio burst associated with a Galactic magnetar}",
      journal = {NAT},
         year = 2020,
        month = nov,
       volume = {587},
       number = {7832},
        pages = {59-62},
          doi = {10.1038/s41586-020-2872-x},
archivePrefix = {arXiv},
       eprint = {2005.10828},
 primaryClass = {astro-ph.HE},
       adsurl = {https://ui.adsabs.harvard.edu/abs/2020Natur.587...59B}
}

@ARTICLE{Tremblay_2018,
       author = {{Tremblay}, Chenoa D. and {Jordan}, Christopher H. and {Cunningham}, Maria and {Jones}, Paul A. and {Hurley-Walker}, Natasha},
        title = "{Low-Frequency Carbon Recombination Lines in the Orion Molecular Cloud Complex}",
      journal = {\pasa},
         year = 2018,
        month = may,
       volume = {35},
          eid = {e018},
        pages = {e018},
          doi = {10.1017/pasa.2018.13},
archivePrefix = {arXiv},
       eprint = {1803.08199},
 primaryClass = {astro-ph.GA},
       adsurl = {https://ui.adsabs.harvard.edu/abs/2018PASA...35...18T}
}

@ARTICLE{Salas_2019,
       author = {{Salas}, P. and {Oonk}, J.~B.~R. and {Emig}, K.~L. and {Pabst}, C. and {Toribio}, M.~C. and {R{\"o}ttgering}, H.~J.~A. and {Tielens}, A.~G.~G.~M.},
        title = "{Carbon radio recombination lines from gigahertz to megahertz frequencies towards Orion A}",
      journal = {\aap},
         year = 2019,
        month = jun,
       volume = {626},
          eid = {A70},
        pages = {A70},
          doi = {10.1051/0004-6361/201834532},
archivePrefix = {arXiv},
       eprint = {1905.02799},
 primaryClass = {astro-ph.GA},
       adsurl = {https://ui.adsabs.harvard.edu/abs/2019A&A...626A..70S}
}

@ARTICLE{Kantharia_2001,
       author = {{Kantharia}, N.~G. and {Anantharamaiah}, K.~R.},
        title = "{Carbon recombination lines from the Galactic plane at 34.5 \& 328 MHz}",
      journal = {Journal of Astrophysics and Astronomy},
         year = 2001,
        month = mar,
       volume = {22},
       number = {1},
        pages = {51-80},
          doi = {10.1007/BF02933590},
archivePrefix = {arXiv},
       eprint = {astro-ph/0104364},
 primaryClass = {astro-ph},
       adsurl = {https://ui.adsabs.harvard.edu/abs/2001JApA...22...51K}
}

@ARTICLE{Dickey_1990,
       author = {{Dickey}, John M. and {Lockman}, Felix J.},
        title = "{H I in the galaxy.}",
      journal = {\araa},
         year = 1990,
        month = jan,
       volume = {28},
        pages = {215-261},
          doi = {10.1146/annurev.aa.28.090190.001243},
       adsurl = {https://ui.adsabs.harvard.edu/abs/1990ARA&A..28..215D}
}

@ARTICLE{Stappers_2011,
       author = {{Stappers}, B.~W. and {Hessels}, J.~W.~T. and {Alexov}, A. and {Anderson}, K. and {Coenen}, T. and {Hassall}, T. and {Karastergiou}, A. and {Kondratiev}, V.~I. and {Kramer}, M. and {van Leeuwen}, J. and {Mol}, J.~D. and {Noutsos}, A. and {Romein}, J.~W. and {Weltevrede}, P. and {Fender}, R. and {Wijers}, R.~A.~M.~J. and {B{\"a}hren}, L. and {Bell}, M.~E. and {Broderick}, J. and {Daw}, E.~J. and {Dhillon}, V.~S. and {Eisl{\"o}ffel}, J. and {Falcke}, H. and {Griessmeier}, J. and {Law}, C. and {Markoff}, S. and {Miller-Jones}, J.~C.~A. and {Scheers}, B. and {Spreeuw}, H. and {Swinbank}, J. and {Ter Veen}, S. and {Wise}, M.~W. and {Wucknitz}, O. and {Zarka}, P. and {Anderson}, J. and {Asgekar}, A. and {Avruch}, I.~M. and {Beck}, R. and {Bennema}, P. and {Bentum}, M.~J. and {Best}, P. and {Bregman}, J. and {Brentjens}, M. and {van de Brink}, R.~H. and {Broekema}, P.~C. and {Brouw}, W.~N. and {Br{\"u}ggen}, M. and {de Bruyn}, A.~G. and {Butcher}, H.~R. and {Ciardi}, B. and {Conway}, J. and {Dettmar}, R. -J. and {van Duin}, A. and {van Enst}, J. and {Garrett}, M. and {Gerbers}, M. and {Grit}, T. and {Gunst}, A. and {van Haarlem}, M.~P. and {Hamaker}, J.~P. and {Heald}, G. and {Hoeft}, M. and {Holties}, H. and {Horneffer}, A. and {Koopmans}, L.~V.~E. and {Kuper}, G. and {Loose}, M. and {Maat}, P. and {McKay-Bukowski}, D. and {McKean}, J.~P. and {Miley}, G. and {Morganti}, R. and {Nijboer}, R. and {Noordam}, J.~E. and {Norden}, M. and {Olofsson}, H. and {Pandey-Pommier}, M. and {Polatidis}, A. and {Reich}, W. and {R{\"o}ttgering}, H. and {Schoenmakers}, A. and {Sluman}, J. and {Smirnov}, O. and {Steinmetz}, M. and {Sterks}, C.~G.~M. and {Tagger}, M. and {Tang}, Y. and {Vermeulen}, R. and {Vermaas}, N. and {Vogt}, C. and {de Vos}, M. and {Wijnholds}, S.~J. and {Yatawatta}, S. and {Zensus}, A.},
        title = "{Observing pulsars and fast transients with LOFAR}",
      journal = {\aap},
         year = 2011,
        month = jun,
       volume = {530},
          eid = {A80},
        pages = {A80},
          doi = {10.1051/0004-6361/201116681},
archivePrefix = {arXiv},
       eprint = {1104.1577},
 primaryClass = {astro-ph.IM},
       adsurl = {https://ui.adsabs.harvard.edu/abs/2011A&A...530A..80S}
}

@ARTICLE{Enriquez_2017,
       author = {{Enriquez}, J. Emilio and {Siemion}, Andrew and {Foster}, Griffin and {Gajjar}, Vishal and {Hellbourg}, Greg and {Hickish}, Jack and {Isaacson}, Howard and {Price}, Danny C. and {Croft}, Steve and {DeBoer}, David and {Lebofsky}, Matt and {MacMahon}, David H.~E. and {Werthimer}, Dan},
        title = "{The Breakthrough Listen Search for Intelligent Life: 1.1-1.9 GHz Observations of 692 Nearby Stars}",
      journal = {\apj},
         year = 2017,
        month = nov,
       volume = {849},
       number = {2},
          eid = {104},
        pages = {104},
          doi = {10.3847/1538-4357/aa8d1b},
archivePrefix = {arXiv},
       eprint = {1709.03491},
 primaryClass = {astro-ph.EP},
       adsurl = {https://ui.adsabs.harvard.edu/abs/2017ApJ...849..104E}
}

@ARTICLE{Gajjar_2021,
       author = {{Gajjar}, Vishal and {Perez}, Karen I. and {Siemion}, Andrew P.~V. and {Foster}, Griffin and {Brzycki}, Bryan and {Chatterjee}, Shami and {Chen}, Yuhong and {Cordes}, James M. and {Croft}, Steve and {Czech}, Daniel and {DeBoer}, David and {DeMarines}, Julia and {Drew}, Jamie and {Gowanlock}, Michael and {Isaacson}, Howard and {Lacki}, Brian C. and {Lebofsky}, Matt and {MacMahon}, David H.~E. and {Morrison}, Ian S. and {Ng}, Cherry and {de Pater}, Imke and {Price}, Danny C. and {Sheikh}, Sofia Z. and {Suresh}, Akshay and {Webb}, Claire and {Pete Worden}, S.},
        title = "{The Breakthrough Listen Search For Intelligent Life Near the Galactic Center. I.}",
      journal = {\aj},
         year = 2021,
        month = jul,
       volume = {162},
       number = {1},
          eid = {33},
        pages = {33},
          doi = {10.3847/1538-3881/abfd36},
archivePrefix = {arXiv},
       eprint = {2104.14148},
 primaryClass = {astro-ph.HE},
       adsurl = {https://ui.adsabs.harvard.edu/abs/2021AJ....162...33G}
}

@ARTICLE{GajjarBrown,
       author = {{Gajjar}, Vishal and {Brown}, Grayce C.},
        title = "{Exo-IPM Scattering as a Hidden Gatekeeper of Narrowband Technosignatures}",
      journal = "Submitted",
         year = 2025,
        month = oct
}

@article{MACCONE2019233,
title = {Moon Farside Protection, Moon Village and PAC (Protected Antipode Circle)},
journal = {Acta Astronautica},
volume = {154},
pages = {233-237},
year = {2019},
issn = {0094-5765},
doi = {https://doi.org/10.1016/j.actaastro.2018.02.012},
url = {https://www.sciencedirect.com/science/article/pii/S0094576517316478},
author = {Claudio Maccone}
}

@misc{michaud2020lunar,
      title={Lunar Opportunities for SETI}, 
      author={Eric J. Michaud and Andrew P. V. Siemion and Jamie Drew and S. Pete Worden},
      year={2020},
      eprint={2009.12689},
      archivePrefix={arXiv},
      primaryClass={astro-ph.IM}
}

@ARTICLE{heidmann2002,
       author = {{Heidmann}, J.},
        title = {A New IAA Cosmic Study: Establishing a Radio Observatory on the Moon Farside},
      journal = {Acta Astronautica},
         year = 2002,
        month = jan,
       volume = {50},
       number = {1},
        pages = {59-63},
          doi = {10.1016/S0094-5765(01)00139-4},
       adsurl = {https://ui.adsabs.harvard.edu/abs/2002AcAau..50...59H}
}

@ARTICLE{chime2020sgr1935,
       author = {{CHIME/FRB Collaboration} and {Andersen}, B.~C. and {Bandura}, K.~M. and {Bhardwaj}, M. and {Bij}, A. and {Boyce}, M.~M. and {Boyle}, P.~J. and {Brar}, C. and {Cassanelli}, T. and {Chawla}, P. and {Chen}, T. and {Cliche}, J. -F. and {Cook}, A. and {Cubranic}, D. and {Curtin}, A.~P. and {Denman}, N.~T. and {Dobbs}, M. and {Dong}, F.~Q. and {Fandino}, M. and {Fonseca}, E. and {Gaensler}, B.~M. and {Giri}, U. and {Good}, D.~C. and {Halpern}, M. and {Hill}, A.~S. and {Hinshaw}, G.~F. and {H{\"o}fer}, C. and {Josephy}, A. and {Kania}, J.~W. and {Kaspi}, V.~M. and {Landecker}, T.~L. and {Leung}, C. and {Li}, D.~Z. and {Lin}, H. -H. and {Masui}, K.~W. and {McKinven}, R. and {Mena-Parra}, J. and {Merryfield}, M. and {Meyers}, B.~W. and {Michilli}, D. and {Milutinovic}, N. and {Mirhosseini}, A. and {M{\"u}nchmeyer}, M. and {Naidu}, A. and {Newburgh}, L.~B. and {Ng}, C. and {Patel}, C. and {Pen}, U. -L. and {Pinsonneault-Marotte}, T. and {Pleunis}, Z. and {Quine}, B.~M. and {Rafiei-Ravandi}, M. and {Rahman}, M. and {Ransom}, S.~M. and {Renard}, A. and {Sanghavi}, P. and {Scholz}, P. and {Shaw}, J.~R. and {Shin}, K. and {Siegel}, S.~R. and {Singh}, S. and {Smegal}, R.~J. and {Smith}, K.~M. and {Stairs}, I.~H. and {Tan}, C.~M. and {Tendulkar}, S.~P. and {Tretyakov}, I. and {Vanderlinde}, K. and {Wang}, H. and {Wulf}, D. and {Zwaniga}, A.~V.},
        title = "{A bright millisecond-duration radio burst from a Galactic magnetar}",
      journal = {\nat},
         year = 2020,
        month = nov,
       volume = {587},
       number = {7832},
        pages = {54-58},
          doi = {10.1038/s41586-020-2863-y},
archivePrefix = {arXiv},
       eprint = {2005.10324},
 primaryClass = {astro-ph.HE},
       adsurl = {https://ui.adsabs.harvard.edu/abs/2020Natur.587...54C}
}

@ARTICLE{lanman2024kko,
       author = {{Lanman}, Adam E. and {Andrew}, Shion and {Lazda}, Mattias and {Shah}, Vishwangi and {Amiri}, Mandana and {Balasubramanian}, Arvind and {Bandura}, Kevin and {Boyle}, P.~J. and {Brar}, Charanjot and {Carlson}, Mark and {Cliche}, Jean-Fran{\c{c}}ois and {Gusinskaia}, Nina and {Hendricksen}, Ian T. and {Kaczmarek}, J.~F. and {Landecker}, Tom and {Leung}, Calvin and {Mckinven}, Ryan and {Mena-Parra}, Juan and {Milutinovic}, Nikola and {Nimmo}, Kenzie and {Pearlman}, Aaron B. and {Renard}, Andre and {Rahman}, Mubdi and {Shaw}, J. Richard and {Siegel}, Seth R. and {Smegal}, Rick J. and {Cassanelli}, Tomas and {Chatterjee}, Shami and {Curtin}, Alice P. and {Dobbs}, Matt and {Dong}, Fengqiu Adam and {Halpern}, Mark and {Hopkins}, Hans and {Kaspi}, Victoria M. and {Khairy}, Kholoud and {Masui}, Kiyoshi W. and {Meyers}, Bradley W. and {Michilli}, Daniele and {Petroff}, Emily and {Pinsonneault-Marotte}, Tristan and {Pleunis}, Ziggy and {Rafiei-Ravandi}, Masoud and {Shin}, Kaitlyn and {Smith}, Kendrick and {Vanderlinde}, Keith and {Zegmott}, Tarik J.},
        title = "{CHIME/FRB Outriggers: KKO Station System and Commissioning Results}",
      journal = {\aj},
         year = 2024,
        month = aug,
       volume = {168},
       number = {2},
          eid = {87},
        pages = {87},
          doi = {10.3847/1538-3881/ad5838},
archivePrefix = {arXiv},
       eprint = {2402.07898},
 primaryClass = {astro-ph.IM},
       adsurl = {https://ui.adsabs.harvard.edu/abs/2024AJ....168...87L}
}

@ARTICLE{leung2021synoptic,
       author = {{Leung}, Calvin and {Mena-Parra}, Juan and {Masui}, Kiyoshi and {Bandura}, Kevin and {Bhardwaj}, Mohit and {Boyle}, P.~J. and {Brar}, Charanjot and {Bruneault}, Mathieu and {Cassanelli}, Tomas and {Cubranic}, Davor and {Kaczmarek}, Jane F. and {Kaspi}, Victoria and {Landecker}, Tom and {Michilli}, Daniele and {Milutinovic}, Nikola and {Patel}, Chitrang and {Pleunis}, Ziggy and {Rahman}, Mubdi and {Renard}, Andre and {Sanghavi}, Pranav and {Stairs}, Ingrid H. and {Scholz}, Paul and {Vanderlinde}, Keith and {Chime/Frb Collaboration}},
        title = "{A Synoptic VLBI Technique for Localizing Nonrepeating Fast Radio Bursts with CHIME/FRB}",
      journal = {\aj},
         year = 2021,
        month = feb,
       volume = {161},
       number = {2},
          eid = {81},
        pages = {81},
          doi = {10.3847/1538-3881/abd174},
archivePrefix = {arXiv},
       eprint = {2008.11738},
 primaryClass = {astro-ph.IM},
       adsurl = {https://ui.adsabs.harvard.edu/abs/2021AJ....161...81L}
}

@ARTICLE{pleunis2021lofar,
       author = {{Pleunis}, Z. and {Michilli}, D. and {Bassa}, C.~G. and {Hessels}, J.~W.~T. and {Naidu}, A. and {Andersen}, B.~C. and {Chawla}, P. and {Fonseca}, E. and {Gopinath}, A. and {Kaspi}, V.~M. and {Kondratiev}, V.~I. and {Li}, D.~Z. and {Bhardwaj}, M. and {Boyle}, P.~J. and {Brar}, C. and {Cassanelli}, T. and {Gupta}, Y. and {Josephy}, A. and {Karuppusamy}, R. and {Keimpema}, A. and {Kirsten}, F. and {Leung}, C. and {Marcote}, B. and {Masui}, K.~W. and {Mckinven}, R. and {Meyers}, B.~W. and {Ng}, C. and {Nimmo}, K. and {Paragi}, Z. and {Rahman}, M. and {Scholz}, P. and {Shin}, K. and {Smith}, K.~M. and {Stairs}, I.~H. and {Tendulkar}, S.~P.},
        title = "{LOFAR Detection of 110-188 MHz Emission and Frequency-dependent Activity from FRB 20180916B}",
      journal = {\apjl},
         year = 2021,
        month = apr,
       volume = {911},
       number = {1},
          eid = {L3},
        pages = {L3},
          doi = {10.3847/2041-8213/abec72},
archivePrefix = {arXiv},
       eprint = {2012.08372},
 primaryClass = {astro-ph.HE},
       adsurl = {https://ui.adsabs.harvard.edu/abs/2021ApJ...911L...3P}
}

@article{ZANON2023627,
title = {Current Lunar dust mitigation techniques and future directions},
journal = {Acta Astronautica},
volume = {213},
pages = {627-644},
year = {2023},
issn = {0094-5765},
doi = {https://doi.org/10.1016/j.actaastro.2023.09.031},
url = {https://www.sciencedirect.com/science/article/pii/S0094576523004939},
author = {Philipp Zanon and Michelle Dunn and Geoffrey Brooks},
}

@article{JIA2018207,
title = {The scientific objectives and payloads of Chang’E−4 mission},
journal = {Planetary and Space Science},
volume = {162},
pages = {207-215},
year = {2018},
note = {Lunar Reconnaissance Orbiter – Seven Years of Exploration and Discovery},
issn = {0032-0633},
doi = {https://doi.org/10.1016/j.pss.2018.02.011},
url = {https://www.sciencedirect.com/science/article/pii/S0032063317300211},
author = {Yingzhuo Jia and Yongliao Zou and Jinsong Ping and Changbin Xue and Jun Yan and Yuanming Ning},
}

@article{WANG20246194,
title = {Electrodynamic dust shield efficiency characterisation under UV in vacuum for lunar application},
journal = {Advances in Space Research},
volume = {74},
number = {11},
pages = {6194-6204},
year = {2024},
issn = {0273-1177},
doi = {https://doi.org/10.1016/j.asr.2024.07.082},
url = {https://www.sciencedirect.com/science/article/pii/S0273117724008019},
author = {Ya-Chun Wang and Fabrice Cipriani and Fredrik Leffe Johansson and Matthias Sperl and Masato Adachi}
}

@misc{voyager2025dustcoating,
  author       = {{Voyager Technologies}},
  title        = {Voyager’s Clear Dust-Repellent Coating Lands on the Moon},
  howpublished = {\url{https://voyagertechnologies.com/press-releases/voyagers-clear-dust-repellent-coating-lands-on-the-moon/}},
  year         = {2025},
  month        = {apr},
  note         = {Press Release. Accessed: 2025-06-20}
}

@inproceedings{buhler2020current,
  author    = {Buhler, C. R. and Johansen, M. and Dupuis, M. and Hogue, M. and Phillips, J. and Malissa, J. and Wang, J. and Calle, C. I.},
  title     = {{Current State of the Electrodynamic Dust Shield for Mitigation}},
  booktitle = {Lunar Dust 2020},
  year      = {2020},
  series    = {LPI Contribution},
  number    = {2243},
  eid       = {5027},
  publisher = {Lunar and Planetary Institute},
  url       = {https://www.hou.usra.edu/meetings/lunardust2020/pdf/5027.pdf}
}

@inproceedings{fritz2024dustroadmap,
  author    = {Fritz, Amy and Breeding, Shawn and Tamasy, Gabor},
  title     = {{NASA Dust Mitigation Technology Roadmap}},
  booktitle = {Lunar Surface Innovation Consortium (LSIC) Fall Meeting},
  year      = {2024},
  month     = {nov},
  publisher = {Johns Hopkins University Applied Physics Laboratory},
  address   = {Las Vegas, NV},
  note      = {Document ID: 20240013978},
  url       = {https://ntrs.nasa.gov/api/citations/20240013978/downloads/NASA%20Lunar%20Dust%20Mitigation%20Roadmap%20Fall%202024.pdf}
}

@inproceedings{Tyrrell2022PlumeSurface,
  author    = {Olivia K. Tyrrell and Ryan J. Thompson and Paul M. Danehy and Christopher J. Dupuis and Michelle M. Munk and Chi P. Nguyen and Robert W. Maddock and Timothy W. Fahringer and William C. Krolick and Andrew B. Weaver and Jeffrey S. West and Michael S. Manginelli and William K. Witherow},
  title     = {Design of a Lunar Plume-Surface Interaction Measurement System},
  booktitle = {AIAA SciTech Forum},
  year      = {2022},
  month     = {January},
  address   = {San Diego, CA, USA},
  publisher = {American Institute of Aeronautics and Astronautics},
  note      = {NASA Technical Report Server ID: 20210024671},
  url       = {https://ntrs.nasa.gov/citations/20210024671}
}

@misc{Atkinson2025NASA,
  author       = {Atkinson, Joe},
  title        = {NASA Cameras on Blue Ghost Capture First-of-its-Kind Moon Landing Footage},
  howpublished = {Web Page},
  organization = {NASA},
  year         = {2025},
  month        = {June},
  url          = {https://www.nasa.gov/general/nasa-cameras-on-blue-ghost-capture-first-of-its-kind-\\moon-landing-footage/},
}

@ARTICLE{sct+24,
       author = {{Susarla}, S.~C. and {Chalumeau}, A. and {Tiburzi}, C. and {Keane}, E.~F. and {Verbiest}, J.~P.~W. and {Hazboun}, J.~S. and {Krishnakumar}, M.~A. and {Iraci}, F. and {Shaifullah}, G.~M. and {Golden}, A. and {Bak Nielsen}, A. -S. and {Donner}, J. and {Grie{\ss}meier}, J. -M. and {Keith}, M.~J. and {Os{\l}owski}, S. and {Porayko}, N.~K. and {Serylak}, M. and {Anderson}, J.~M. and {Br{\"u}ggen}, M. and {Ciardi}, B. and {Dettmar}, R. -J. and {Hoeft}, M. and {K{\"u}nsem{\"o}ller}, J. and {Schwarz}, D. and {Vocks}, C.},
        title = "{Exploring the time variability of the solar wind using LOFAR pulsar data}",
      journal = {\aap},
         year = 2024,
        month = dec,
       volume = {692},
          eid = {A18},
        pages = {A18},
          doi = {10.1051/0004-6361/202450680},
archivePrefix = {arXiv},
       eprint = {2409.09838},
 primaryClass = {astro-ph.SR},
       adsurl = {https://ui.adsabs.harvard.edu/abs/2024A&A...692A..18S}
}

@ARTICLE{epta_inpta,
       author = {{EPTA Collaboration} and {InPTA Collaboration} and {Antoniadis}, J. and {Arumugam}, P. and {Arumugam}, S. and {Babak}, S. and {Bagchi}, M. and {Bak Nielsen}, A. -S. and {Bassa}, C.~G. and {Bathula}, A. and {Berthereau}, A. and {Bonetti}, M. and {Bortolas}, E. and {Brook}, P.~R. and {Burgay}, M. and {Caballero}, R.~N. and {Chalumeau}, A. and {Champion}, D.~J. and {Chanlaridis}, S. and {Chen}, S. and {Cognard}, I. and {Dandapat}, S. and {Deb}, D. and {Desai}, S. and {Desvignes}, G. and {Dhanda-Batra}, N. and {Dwivedi}, C. and {Falxa}, M. and {Ferdman}, R.~D. and {Franchini}, A. and {Gair}, J.~R. and {Goncharov}, B. and {Gopakumar}, A. and {Graikou}, E. and {Grie{\ss}meier}, J. -M. and {Guillemot}, L. and {Guo}, Y.~J. and {Gupta}, Y. and {Hisano}, S. and {Hu}, H. and {Iraci}, F. and {Izquierdo-Villalba}, D. and {Jang}, J. and {Jawor}, J. and {Janssen}, G.~H. and {Jessner}, A. and {Joshi}, B.~C. and {Kareem}, F. and {Karuppusamy}, R. and {Keane}, E.~F. and {Keith}, M.~J. and {Kharbanda}, D. and {Kikunaga}, T. and {Kolhe}, N. and {Kramer}, M. and {Krishnakumar}, M.~A. and {Lackeos}, K. and {Lee}, K.~J. and {Liu}, K. and {Liu}, Y. and {Lyne}, A.~G. and {McKee}, J.~W. and {Maan}, Y. and {Main}, R.~A. and {Mickaliger}, M.~B. and {Ni{\c{t}}u}, I.~C. and {Nobleson}, K. and {Paladi}, A.~K. and {Parthasarathy}, A. and {Perera}, B.~B.~P. and {Perrodin}, D. and {Petiteau}, A. and {Porayko}, N.~K. and {Possenti}, A. and {Prabu}, T. and {Quelquejay Leclere}, H. and {Rana}, P. and {Samajdar}, A. and {Sanidas}, S.~A. and {Sesana}, A. and {Shaifullah}, G. and {Singha}, J. and {Speri}, L. and {Spiewak}, R. and {Srivastava}, A. and {Stappers}, B.~W. and {Surnis}, M. and {Susarla}, S.~C. and {Susobhanan}, A. and {Takahashi}, K. and {Tarafdar}, P. and {Theureau}, G. and {Tiburzi}, C. and {van der Wateren}, E. and {Vecchio}, A. and {Venkatraman Krishnan}, V. and {Verbiest}, J.~P.~W. and {Wang}, J. and {Wang}, L. and {Wu}, Z.},
        title = "{The second data release from the European Pulsar Timing Array. III. Search for gravitational wave signals}",
      journal = {\aap},
         year = 2023,
        month = oct,
       volume = {678},
          eid = {A50},
        pages = {A50},
          doi = {10.1051/0004-6361/202346844},
archivePrefix = {arXiv},
       eprint = {2306.16214},
 primaryClass = {astro-ph.HE},
       adsurl = {https://ui.adsabs.harvard.edu/abs/2023A&A...678A..50E}
}

@ARTICLE{ppta,
       author = {{Reardon}, Daniel J. and {Zic}, Andrew and {Shannon}, Ryan M. and {Hobbs}, George B. and {Bailes}, Matthew and {Di Marco}, Valentina and {Kapur}, Agastya and {Rogers}, Axl F. and {Thrane}, Eric and {Askew}, Jacob and {Bhat}, N.~D. Ramesh and {Cameron}, Andrew and {Cury{\l}o}, Ma{\l}gorzata and {Coles}, William A. and {Dai}, Shi and {Goncharov}, Boris and {Kerr}, Matthew and {Kulkarni}, Atharva and {Levin}, Yuri and {Lower}, Marcus E. and {Manchester}, Richard N. and {Mandow}, Rami and {Miles}, Matthew T. and {Nathan}, Rowina S. and {Os{\l}owski}, Stefan and {Russell}, Christopher J. and {Spiewak}, Ren{\'e}e and {Zhang}, Songbo and {Zhu}, Xing-Jiang},
        title = "{Search for an Isotropic Gravitational-wave Background with the Parkes Pulsar Timing Array}",
      journal = {\apjl},
         year = 2023,
        month = jul,
       volume = {951},
       number = {1},
          eid = {L6},
        pages = {L6},
          doi = {10.3847/2041-8213/acdd02},
archivePrefix = {arXiv},
       eprint = {2306.16215},
 primaryClass = {astro-ph.HE},
       adsurl = {https://ui.adsabs.harvard.edu/abs/2023ApJ...951L...6R}
}

@UNPUBLISHED{Basu2025_SKA_EOS,
       author = {{Basu}, A. and {Graber}, V. and {Lower}, M.~E. and {Antonelli}, M. and {Antonopoulou}, D. and {Bagchi}, M. and {Char}, P. and {Freire}, P.~C.~C. and {Haskell}, B. and {Hu}, H. and {Jones}, D.~I. and {Mukhopadhyay}, B. and {Oertel}, M. and {Rea}, N. and {Sagun}, V. and {Shaw}, B. and {Singha}, J. and {Stappers}, B.~W. and {Thongmeearkom}, T. and {Watts}, A. and {Weltevrede}, P. and {{The SKA Pulsar Science Working Group}}},
        title = "{Probing neutron star interiors and the properties of cold, ultra-dense matter with the {SKA}}",
      journal = {Open Journal of Astrophysics},
         note = {Submitted},
         year = {2025}
}

@ARTICLE{agf+16,
       author = {{Archibald}, R.~F. and {Gotthelf}, E.~V. and {Ferdman}, R.~D. and {Kaspi}, V.~M. and {Guillot}, S. and {Harrison}, F.~A. and {Keane}, E.~F. and {Pivovaroff}, M.~J. and {Stern}, D. and {Tendulkar}, S.~P. and {Tomsick}, J.~A.},
        title = "{A High Braking Index for a Pulsar}",
      journal = {\apjl},
         year = 2016,
        month = mar,
       volume = {819},
       number = {1},
          eid = {L16},
        pages = {L16},
          doi = {10.3847/2041-8205/819/1/L16},
archivePrefix = {arXiv},
       eprint = {1603.00305},
 primaryClass = {astro-ph.HE},
       adsurl = {https://ui.adsabs.harvard.edu/abs/2016ApJ...819L..16A}
}

@ARTICLE{psrcat,
       author = {{Manchester}, R.~N. and {Hobbs}, G.~B. and {Teoh}, A. and {Hobbs}, M.},
        title = "{The Australia Telescope National Facility Pulsar Catalogue}",
      journal = {\aj},
         year = 2005,
        month = apr,
       volume = {129},
       number = {4},
        pages = {1993-2006},
          doi = {10.1086/428488},
archivePrefix = {arXiv},
       eprint = {astro-ph/0412641},
 primaryClass = {astro-ph},
       adsurl = {https://ui.adsabs.harvard.edu/abs/2005AJ....129.1993M}
}

@ARTICLE{lbb+13,
       author = {{Levin}, L. and {Bailes}, M. and {Barsdell}, B.~R. and {Bates}, S.~D. and {Bhat}, N.~D.~R. and {Burgay}, M. and {Burke-Spolaor}, S. and {Champion}, D.~J. and {Coster}, P. and {D'Amico}, N. and {Jameson}, A. and {Johnston}, S. and {Keith}, M.~J. and {Kramer}, M. and {Milia}, S. and {Ng}, C. and {Possenti}, A. and {Stappers}, B. and {Thornton}, D. and {van Straten}, W.},
        title = "{The High Time Resolution Universe Pulsar Survey -VIII. The Galactic millisecond pulsar population}",
      journal = {\mnras},
         year = 2013,
        month = sep,
       volume = {434},
       number = {2},
        pages = {1387-1397},
          doi = {10.1093/mnras/stt1103},
archivePrefix = {arXiv},
       eprint = {1306.4190},
 primaryClass = {astro-ph.SR},
       adsurl = {https://ui.adsabs.harvard.edu/abs/2013MNRAS.434.1387L}
}

@INPROCEEDINGS{k13,
       author = {{Keane}, E.~F.},
        title = "{Radio pulsar variability}",
    booktitle = {Neutron Stars and Pulsars: Challenges and Opportunities after 80 years},
         year = 2013,
       editor = {{van Leeuwen}, Joeri},
       series = {IAU Symposium},
       volume = {291},
        month = mar,
        pages = {295-300},
          doi = {10.1017/S1743921312023927},
archivePrefix = {arXiv},
       eprint = {1210.5397},
 primaryClass = {astro-ph.SR},
       adsurl = {https://ui.adsabs.harvard.edu/abs/2013IAUS..291..295K}
}

@ARTICLE{bjb+12,
       author = {{Burke-Spolaor}, S. and {Johnston}, S. and {Bailes}, M. and {Bates}, S.~D. and {Bhat}, N.~D.~R. and {Burgay}, M. and {Champion}, D.~J. and {D'Amico}, N. and {Keith}, M.~J. and {Kramer}, M. and {Levin}, L. and {Milia}, S. and {Possenti}, A. and {Stappers}, B. and {van Straten}, W.},
        title = "{The High Time Resolution Universe Pulsar Survey - V. Single-pulse energetics and modulation properties of 315 pulsars}",
      journal = {\mnras},
         year = 2012,
        month = jun,
       volume = {423},
       number = {2},
        pages = {1351-1367},
          doi = {10.1111/j.1365-2966.2012.20998.x},
archivePrefix = {arXiv},
       eprint = {1203.6068},
 primaryClass = {astro-ph.SR},
       adsurl = {https://ui.adsabs.harvard.edu/abs/2012MNRAS.423.1351B}
}

@ARTICLE{zgw+24,
       author = {{Zhang}, S.~B. and {Geng}, J.~J. and {Wang}, J.~S. and {Yang}, X. and {Kaczmarek}, J. and {Tang}, Z.~F. and {Johnston}, S. and {Hobbs}, G. and {Manchester}, R. and {Wu}, X.~F. and {Jiang}, P. and {Huang}, Y.~F. and {Zou}, Y.~C. and {Dai}, Z.~G. and {Zhang}, B. and {Li}, D. and {Yang}, Y.~P. and {Dai}, S. and {Chang}, C.~M. and {Pan}, Z.~C. and {Lu}, J.~G. and {Wei}, J.~J. and {Li}, Y. and {Wu}, Q.~W. and {Qian}, L. and {Wang}, P. and {Wang}, S.~Q. and {Feng}, Y. and {Staveley-Smith}, L.},
        title = "{RRAT J1913+1330: An Extremely Variable and Puzzling Pulsar}",
      journal = {\apj},
         year = 2024,
        month = sep,
       volume = {972},
       number = {1},
          eid = {59},
        pages = {59},
          doi = {10.3847/1538-4357/ad6602},
archivePrefix = {arXiv},
       eprint = {2306.02855},
 primaryClass = {astro-ph.HE},
       adsurl = {https://ui.adsabs.harvard.edu/abs/2024ApJ...972...59Z}
}

@ARTICLE{dwl+25,
       author = {{Dang}, Shi-jun and {Wu}, Zi-wei and {Lu}, Ji-guang and {Jiang}, Peng and {Li}, Wei and {Liu}, Yu-lan and {Cai}, Yan-qing and {Yuan}, Jian-ping and {Wang}, Na},
        title = "{FAST Observation of Polarized Radio Emission of Rotating Radio Transient J2325‑0530}",
      journal = {\apj},
         year = 2025,
        month = jul,
       volume = {988},
       number = {1},
          eid = {11},
        pages = {11},
          doi = {10.3847/1538-4357/ade23a},
       adsurl = {https://ui.adsabs.harvard.edu/abs/2025ApJ...988...11D}
}

@ARTICLE{ck21,
       author = {{Caleb}, Manisha and {Keane}, Evan},
        title = "{A Decade and a Half of Fast Radio Burst Observations}",
      journal = {Universe},
         year = 2021,
        month = nov,
       volume = {7},
       number = {11},
          eid = {453},
        pages = {453},
          doi = {10.3390/universe7110453},
       adsurl = {https://ui.adsabs.harvard.edu/abs/2021Univ....7..453C}
}

@ARTICLE{bassett2020,
       author = {{Bassett}, Neil and {Rapetti}, David and {Burns}, Jack O. and {Tauscher}, Keith and {MacDowall}, Robert},
        title = "{Characterizing the radio quiet region behind the lunar farside for low radio frequency experiments}",
      journal = {Advances in Space Research},
         year = 2020,
        month = sep,
       volume = {66},
       number = {6},
        pages = {1265-1275},
          doi = {10.1016/j.asr.2020.05.050},
archivePrefix = {arXiv},
       eprint = {2003.03468},
 primaryClass = {astro-ph.IM},
       adsurl = {https://ui.adsabs.harvard.edu/abs/2020AdSpR..66.1265B}
}

@ARTICLE{Grigg_2025,
       author = {{Grigg}, D. and {Tingay}, S.~J. and {Sokolowski}, M.},
        title = "{The growing impact of unintended Starlink broadband emission on radio astronomy in the SKA-Low frequency range}",
      journal = {\aap},
         year = 2025,
        month = jul,
       volume = {699},
          eid = {A307},
        pages = {A307},
          doi = {10.1051/0004-6361/202554787},
archivePrefix = {arXiv},
       eprint = {2506.02831},
 primaryClass = {astro-ph.IM},
       adsurl = {https://ui.adsabs.harvard.edu/abs/2025A&A...699A.307G}
}

@article{Garrett2018SETISO,
  title={SETI surveys of the nearby and distant universe employing wide-field radio interferometry techniques},
  author={Michael A. Garrett},
  journal={arXiv: Instrumentation and Methods for Astrophysics},
  year={2018},
  url={https://api.semanticscholar.org/CorpusID:73613963}
}

@misc{radcliffe2025widefieldvlbi,
  author        = {Radcliffe, Jack F. and McKean, J. P. and Herb{\'e}-George, C. and Coetzer, L. and Matsepane, T.},
  title         = {Latest Developments in Wide-Field {VLBI}},
  year          = {2025},
  eprint        = {2504.19579},
  archivePrefix = {arXiv},
  primaryClass  = {astro-ph.IM},
  url           = {https://arxiv.org/abs/2504.19579}
}

@article{herbegeorge2025sweeps,
  author  = {Herb{\'e}-George, C{\'e}lestin and McKean, J. P. and Morganti, Raffaella and Radcliffe, Jack F.},
  title   = {Synoptic Wide-field {EVN}--e-{MERLIN} Public Survey ({SWEEPS}) -- {I}. First Steps Towards Commensal Surveys with {VLBI}},
  journal = {Monthly Notices of the Royal Astronomical Society: Letters},
  volume  = {537},
  number  = {1},
  pages   = {L49--L54},
  year    = {2025},
  doi     = {10.1093/mnrasl/slae115},
  eprint  = {2412.02746},
  archivePrefix = {arXiv},
  primaryClass  = {astro-ph.IM}
}

@unpublished{Ashe2026,
  author = {Ashe, C. and Marshall, E. and DeBoer, D. and Keane, E.},
  title = {Modeling RFI on the Lunar Farside},
  note = {in preparation},
  year = {2026}
}

@INCOLLECTION{Williams2018,
       author = {{Williams}, Peter K.~G.},
        title = "{Radio Emission from Ultracool Dwarfs}",
    booktitle = {Handbook of Exoplanets},
         year = 2018,
       editor = {{Deeg}, Hans J. and {Belmonte}, Juan Antonio},
          eid = {171},
        pages = {171},
          doi = {10.1007/978-3-319-55333-7_171},
       adsurl = {https://ui.adsabs.harvard.edu/abs/2018haex.bookE.171W}
}

@BOOK{Linsky2019,
    author = {{Linsky}, Jeffrey},
    title = {Host Stars and their Effects on Exoplanet Atmospheres},
    year = 2019,
    volume = {955},
    doi = {10.1007/978-3-030-11452-7},
    adsurl = {https://ui.adsabs.harvard.edu/abs/2019LNP...955.....L},
    publisher = {Springer},
    series = {Springer Praxis Books}
}

@article{zarka_auroral_1998,
	title = {Auroral radio emissions at the outer planets: {Observations} and theories},
	volume = {103},
	copyright = {Copyright 1998 by the American Geophysical Union.},
	issn = {2156-2202},
	shorttitle = {Auroral radio emissions at the outer planets},
	url = {https://onlinelibrary.wiley.com/doi/abs/10.1029/98JE01323},
	doi = {10.1029/98JE01323},
	language = {en},
	number = {E9},
	urldate = {2024-02-28},
	journal = {Journal of Geophysical Research: Planets},
	author = {Zarka, Philippe},
	year = {1998},
	note = {\_eprint: https://onlinelibrary.wiley.com/doi/pdf/10.1029/98JE01323},
	pages = {20159--20194},
}

@ARTICLE{EMI,
    author = {{Zhelezniakov}, V.~V. and {Zlotnik}, E. Ia.},
    title = "{Cyclotron Wave Instability in the Corona and Origin of Solar Radio Emission with Fine Structure. I: Bernstein Modes and Plasma Waves in a Hybrid Band}",
    year = 1975,
    month = aug,
    journal = {Solar Physics},
    volume = {43},
    number = {2},
    pages = {431-451},
    doi = {10.1007/BF00152366},
    adsurl = {https://ui.adsabs.harvard.edu/abs/1975SoPh...43..431Z}
}




%
%


\bsp	
\label{lastpage}
\end{document}